\documentclass[11pt]{article}
\usepackage[utf8]{inputenc}
\usepackage[T1]{fontenc}
\usepackage[english]{babel}
\usepackage[margin=2.3cm]{geometry}
\usepackage{amsmath,amssymb,amsthm,mathtools}
\usepackage{bm}
\usepackage{enumitem}
\usepackage[protrusion=false,expansion=false]{microtype}
\usepackage[hidelinks]{hyperref}
\usepackage{bbm}
\usepackage{graphicx}
\usepackage{float}
\usepackage{xcolor}
\usepackage{etoolbox}
\usepackage{setspace}
\appto\normalsize{%
  \setlength{\abovedisplayskip}{5pt plus 2pt minus 2pt}%
  \setlength{\belowdisplayskip}{5pt plus 2pt minus 2pt}%
  \setlength{\abovedisplayshortskip}{0pt plus 2pt}%
  \setlength{\belowdisplayshortskip}{3pt plus 2pt minus 1pt}}

\theoremstyle{plain}
\newtheorem{theorem}{Theorem}[section]
\newtheorem{proposition}[theorem]{Proposition}
\newtheorem{lemma}[theorem]{Lemma}
\newtheorem{corollary}[theorem]{Corollary}
\theoremstyle{definition}
\newtheorem{definition}[theorem]{Definition}
\theoremstyle{remark}
\newtheorem{remark}[theorem]{Remark}
\newtheorem{remarks}[theorem]{Remarks}
\newtheorem{example}[theorem]{Example}
\newtheorem{application}[theorem]{\bf Application}

\newcommand{\C}{\mathbb{C}}
\newcommand{\D}{\mathcal{D}}
\newcommand{\R}{\mathbb{R}}
\newcommand{\N}{\mathbb{N}}
\newcommand{\T}{\partial\mathbb{D}}
\newcommand{\B}{\mathcal{B}}
\newcommand{\HH}{\mathcal{H}}
\newcommand{\Ucal}{\mathcal{U}}
\newcommand{\Mcal}{\mathcal{M}}
\newcommand{\Hs}{\mathcal{H}_P}
\newcommand{\Ho}{\mathcal{H}_Q}
\newcommand{\Id}{\mathbbm{1}}
\newcommand{\id}{\mathrm{id}}
\newcommand{\Tr}{\operatorname{Tr}}
\newcommand{\Lind}{\mathcal{L}}
\newcommand{\Wcal}{\mathcal{W}}
\newcommand{\Acal}{\mathcal{A}}
\newcommand{\Ccal}{\mathcal{C}}
\newcommand{\Vcal}{\mathcal{V}}

\newcommand{\Sp}{\operatorname{Sp}}
\newcommand{\opnorm}[1]{\lVert #1 \rVert}
\newcommand{\trnorm}[1]{\lVert #1 \rVert_1}

\newcommand{\Span}{\operatorname{span}}
\newcommand{\Ocal}{\mathcal{O}}

\newcommand{\Kcal}{\mathcal{K}}
\newcommand{\dbra}[1]{\langle\!\langle #1|}
\newcommand{\dket}[1]{|#1\rangle\!\rangle}
\newcommand{\Diag}{\operatorname{Diag}}
\newcommand{\E}{\mathbb{E}}
\newcommand{\Ecal}{\mathcal{E}}
\newcommand{\Prob}{\mathbb{P}}
\newcommand{\ran}{\operatorname{ran}}
\newcommand{\HSip}[2]{\langle\!\langle #1,#2\rangle\!\rangle}

\usepackage{authblk}
\begin{document}

\title{ \bfseries Reset Quantum Channels}

\author{Thomas Argoud\footnote{thomas.argoud1@univ-grenoble-alpes.fr}}
\author{Alain Joye\footnote{alain.joye@univ-grenoble-alpes.fr}}

\affil{Univ. Grenoble Alpes,
CNRS, Institut Fourier,
F-38000 Grenoble France
}

\date{}

\maketitle

\begin{abstract}
\noindent

We introduce Reset Quantum Channels (RQCs), a class of quantum channels acting on a Hilbert space structured as a direct sum of an arbitrary number of subspaces. Within each subspace, the state evolves unitarily; across subspaces, it is transferred by
measure-and-prepare resets along given states. In this general setting, we show that the determination of their invariant states reduces to a stochastic eigenvalue problem, whose size is set by the number of reset states rather
than by the dimension of the Hilbert space, and we discuss the uniqueness of these states. We further address the primitivity and the spectrum of these channels, which we compute explicitly when the reset states
commute with the unitary evolutions. Applications include a small system
coupled to several thermal reservoirs. We also consider random unitary evolutions and analyze the averages of the invariant states in this setup. Finally, we show that a suitable continuous-time limit yields Lindblad generators of reset type on the algebra of block-diagonal operators.
\end{abstract}

\tableofcontents
\medskip

\section{Introduction}

Quantum channels, represented mathematically by completely positive trace-preserving (CPTP) maps, describe the most general physical transformations that the state  of a quantum system (non-negative operator of unit trace) can undergo, whether the system is isolated or part of a larger composite system, see e.g. \cite{AF, BP, H, KLW}. In the case of open quantum systems, one often gives up describing the environment with which the system of interest interacts, and instead focuses on the reduced dynamics of the latter. When this reduction is not derived from an explicit microscopic model of the full composite system, for instance through a weak coupling limit, the choice of the quantum channel may retain some phenomenological freedom, depending on the physical features one wishes to describe, see \cite{BP, AJP, AL} for example. \medskip

In this paper, we introduce and analyze a  versatile class of quantum channels, which we call Reset Quantum Channels (RQCs for short). Let us motivate informally their construction by the following physical situation: 

Consider a Hilbert space $\HH$ structured as the direct sum of several subspaces; to keep things simple here, we consider two subspaces only, $\Hs$ and $\Ho$ with $\HH=\Hs\oplus \Ho$, and we denote by $P$ and $Q$ the orthogonal projectors onto these subspaces. We assume these projectors are further decomposed as $P=\sum_{i=1}^{N_P}P_i$ and $Q=\sum_{i=1}^{N_Q}Q_i$, by means of pairwise orthogonal projectors.
On  $\Hs$ and $\Ho$, we are given unitary operators $W_P$ and $W_Q$, and two families of quantum states $\{\pi_i\}_{i=1}^{N_Q}$ and $\{\tau_i\}_{i=1}^{N_P}$, with supports in $\Hs$ and $\Ho$ respectively. The action of the RQC  on a state $\sigma$ of the system is given by a weighted mixture of the following elementary maps: 

i) A restriction of the state to one of the subspaces followed by the unitarily implemented evolution within the subspace
$$
\sigma \mapsto  W_P P \sigma P W_P^*, \ \ \text{or } \ \ \sigma \mapsto  W_Q Q \sigma Q W_Q^*.
$$

ii) A reset of the part of the state that is supported within a piece of one subspace to a state supported in the other subspace
$$
\sigma \mapsto \Tr (\sigma P_i)\tau_i, \ \ \text{or } \ \ \sigma \mapsto \Tr (\sigma Q_i)\pi_i.
$$

By construction, these elementary maps have opposite effects with respect to the distinguished subspaces: the maps i) preserve the subspaces and generate a unitary evolution inside them, while the maps ii) transfer the components of the state from one subspace to the other along certain given reset states. With independent weights $\alpha$ and $\beta$ in $[0,1]$, the mixture defines a {\it bona fide} RQC $\Phi$, see Proposition \ref{prop:cptp}, acting on any operator $\sigma\in \B(\HH)$ as
\begin{align}\label{defphiintro}
    \Phi(\sigma)= \alpha W_P P \sigma P W_P^*+ \beta W_Q Q \sigma Q W_Q^* + (1-\alpha)\sum_{i=1}^{N_P}\Tr (\sigma P_i)\tau_i+ (1-\beta)\sum_{i=1}^{N_Q}\Tr (\sigma Q_i)\pi_i.
\end{align}
 Within each subspace, the unitary evolution and the transfer to the other subspace are assigned complementary weights $\alpha$ and $1-\alpha$ in $\Hs$, $\beta$ and $1-\beta$ in $\Ho$. Note that $\Phi$ destroys the coherences between the subspaces $\Hs$ and $\Ho$ since it annihilates contributions of the form $P\sigma Q$ and $Q\sigma P$. This means that the transfer between the subspaces is classical, whereas quantum coherence survives within each subspace. 

Note that in case $\alpha=\beta=1$, $\Phi$ acts as two independent unitarily implemented evolutions in the subspaces $\Hs$ and $\Ho$, whereas for $\alpha=\beta=0$, $\Phi$ becomes a Measure-and-Prepare or Entanglement Breaking Channel \cite{H0, HSR, AKL+}, which is reducible to a classical Markov process, as recalled in Remarks \ref{rem:ebc} ii), iii) below. In the generic case $0<\alpha, \beta <1$, the RQC $\Phi$ realizes the dynamical competition between coherent evolution within the subspaces and classical resetting across them. 

Motivated by the study of non-equilibrium steady states on existing quantum devices, related discrete-time channels with resets appear in \cite{WYP+, MTP}. There, the channel is driven by a single unitary, the underlying Hilbert space carrying no direct sum structure, and the reset involves one state, or two states entering with the populations of a given orthogonal decomposition of the identity.

\subsection{Main Results}\label{sec:mainresults}

At the level of generality of \eqref{defphiintro}, we show in Corollary \ref{prop:stoch} that determining the invariant state of $\Phi$ reduces to finding an invariant positive vector of a 
stochastic matrix of size $\min(N_P,N_Q)$ only, instead of dealing with a spectral problem in $\B(\HH)$, reflecting the classical nature of the transfers between subspaces. 
We further discuss its uniqueness in Section \ref{sec:unique}, conditions ensuring primitivity of $\Phi$ in Proposition \ref{prop:primitive}, and the spectrum of $\Phi$ in Section \ref{sec:spectrum}.

\medskip 

RQCs can also model a "small system" in $\Hs$ in contact  with a  "reservoir" in $\Ho$ at equilibrium in a state $\tau$, by taking $\tau_i=\tau$ for all $1\leq i\leq N_P$, and $[W_Q,\tau]=0$. 
This will be the case, for instance, if the reservoir is governed by a Hamiltonian with spectral decomposition $H_Q=\sum_{i=1}^{N_Q}E_i Q_i$ in $\Ho$ so that $W_Q=e^{-iH_Q}$ and $\tau=e^{-\theta H_Q}/Z_Q$ is a Gibbs state for some inverse temperature $\theta>0$; but no Gibbs structure is required. 
In the system-reservoir setting, the RQC \eqref{defphiintro} acts on $\sigma\in \B(\HH)$ as
\begin{align}\label{phisysres}
   \Phi(\sigma)= \alpha W_P P \sigma P W_P^*+ \beta W_Q Q \sigma Q W_Q^* + (1-\alpha)\Tr (\sigma P)\tau + (1-\beta)\sum_{i=1}^{N_Q}\Tr (\sigma Q_i)\pi_i.
\end{align}
By construction, the reset from $\Hs$ to $\Ho$ is fully along the state $\tau$, which is invariant under the dynamics in $\Ho$. In that sense, the small system cannot change the state of the reservoir. Hence we expect that under the repeated action of $\Phi$, the $Q$-part of the state converges  to $\tau$, so that the resets to $\Hs$ along the different states $\pi_i$ have populations $\Tr(Q_i\tau)$, or  weights $\Tr (Q_i) e^{-\theta E_i}/Z_Q$ in the Gibbs case, consistent with the effect of a reservoir in an equilibrium state $\tau$. Indeed, this intuition is justified by Application \ref{ex:NP1} showing that for the choices above and $0\leq\alpha,\beta<1$, $\Phi$ in \eqref{phisysres} admits a unique invariant state $\varrho$ of the form
\begin{align}\label{eq:ressys}
    \varrho = \frac{1-\beta}{2-\alpha-\beta}(1-\alpha)(\id_P-\alpha\Wcal_P)^{-1}\big(\sum_i\Tr (Q_i\tau)\pi_i\big)
+\frac{1-\alpha}{2-\alpha-\beta}\tau,
\end{align}
where $\Wcal_P(\cdot)=W_P \cdot W_P^*$. Hence, the $Q$-part of the invariant state is indeed $\tau$, with a prefactor $\frac{1-\alpha}{2-\alpha-\beta}$ representing the asymptotic population in the subspace $\Ho$ under the iterated dynamics: after one iteration, this population is
\begin{align}
    \Tr(\Phi(\sigma)Q)=\beta \Tr(\sigma Q)+(1-\alpha)(1-\Tr(\sigma Q))=(\beta+\alpha-1)\Tr(\sigma Q)+(1-\alpha),
\end{align}
which yields the result by iteration for $\alpha+\beta >0$; see also Lemma~\ref{lem:Sigma} and Remark~\ref{rem:35}. The $P$-part of the invariant state is given by the non trivial effect $(1-\alpha)(\id_P-\alpha\Wcal_P)^{-1}$ of the dynamics within $\Hs$ acting on the convex combination of reset states $\pi_i$ weighted by the populations $\Tr (Q_i\tau)$, multiplied by the prefactor $\frac{1-\beta}{2-\alpha-\beta}$. See Remark \ref{expdyn} for a probabilistic interpretation.

\medskip

We further address the properties of the RQC when the unitary dynamics are random in Section \ref{sec:randhaar}. The corresponding random RQC is denoted by $\Phi_\omega$, and its invariant state by $\varrho(\Phi_\omega)$, assumed to be unique. In the system-reservoir setting, we determine the expectation of the invariant state $\E[\varrho(\Phi_\omega)]$ and 
compare it to the invariant state $\varrho(\E[\Phi_\omega])$ of the expectation of the RQC $\E[\Phi_\omega]$. In the simpler case \eqref{phisysres}, \eqref{eq:ressys} with $W_Q$ deterministic, we show that both states agree if $W_P$ has uniformly distributed random eigenvalues and, in case $W_P$ is Haar distributed, that they are a distance of order $1/(\dim \Hs)^2$ apart from one another, see Corollary \ref{cor:approx}.

\medskip

Another distinctive property of RQCs is that, in weak limit regimes obtained for suitable asymptotics of their parameters, they give rise to continuous time evolutions driven by Lindblad operators associated with Quantum Reset Models, see e.g. \cite{HDB, RTL+, BBH+, HJ, HJ2} and references therein. More precisely, in Section \ref{sec:continuous} we consider  the scaling $n=t/h\in \N$, where $t>0$ is fixed and $h\to 0$, and assume $\Phi\equiv \Phi_h$ is characterized by parameters 
$$\alpha(h)=1-\gamma_Ph+O(h^2), \ \ \beta(h)=1-\gamma_Q h +O(h^2), \ \ W_\#(h)=e^{-iH_\#h}, \ \#=P,Q.$$
We assume $\sigma$ is block-diagonal with respect to $\Hs\oplus\Ho$, which we write as $\sigma=\sigma_P+\sigma_Q$  where $\sigma_P=P\sigma P$, $\sigma_Q=Q\sigma Q$. Then, $\lim_{h\to 0}\Phi_h^{t/h}(\sigma)=e^{t\Lind}(\sigma)$, and
$$\Lind(\sigma)=-i[H_P+ H_Q,\sigma]
+\gamma_P\Big(\sum_i\Tr(P_i\sigma)\,\tau_i-\sigma_P\Big)
+\gamma_Q\Big(\sum_j\Tr(Q_j\sigma)\,\pi_j-\sigma_Q\Big),$$
where we recognize the characteristic reset terms in the dissipator.
The restriction to block-diagonal operators matters, since the formula for $\Lind$, applied
to all of $\B(\HH)$, does not define a Lindblad generator, see the discussion in Section \ref{sec:continuous}.

\medskip 

 While we restrict the presentation above to two subspaces for clarity, the construction and our results hold, under suitable irreducibility assumptions, for an arbitrary number of subspaces, as shown in Section \ref{sec:rreservoir}. This allows us in particular to analyze the case of a small system in contact with several reservoirs and, more generally, to consider arbitrary transfer patterns between subspaces.

\bigskip

\section{RQC in a Simple Setup}
\label{sec:model}

We start by defining formally the notion of Reset Quantum Channel (RQC) $\Phi$ acting on $\B(\HH)$, the set of operators on a Hilbert space $\HH$ with no structure.  This simplified framework serves as  a gentle introduction to  RQCs and to the mathematical tools and concepts used in their analysis. The subsequent sections turn to the main setting where the Hilbert spaces carry additional structures and the associated RQCs exhibit richer behavior.\medskip

Let $\HH$ be a finite-dimensional Hilbert space, and let $\Id$ be the identity operator on $\HH$. We denote by $\D(\HH)=\{\sigma\in \B(\HH)\ | \ \sigma\geq 0, \ \Tr(\sigma)=1\}$ the set of states on $\HH$. The ingredients required to define a RQC are as follows. \\

\noindent{\bf Data:}
We assume the existence of \\
i)  a unitary operator $W\in \B(\HH)$.\\
ii) a set of nonzero orthogonal projectors $\{P_i\}_{i=1}^{N}$, $N\geq 1$, such that $\sum_{i=1}^{N}P_i=\Id$ and $P_iP_j=\delta_{ij}P_i.$\\
iii) a set of states $\{\pi_i\}_{i=1}^N\subset\D(\HH)$.\\
iv) a weight $\alpha\in [0,1]$. \\

This data will be denoted by $\{W,\{P_i\}, \{\pi_i\},\alpha\}$.

\begin{definition}
    The Simple Reset Quantum Channel $\Phi\in\B(\B(\HH))$ associated with the data above is given by
    \begin{align} \label{eq:phi0}
    \Phi(X)=\alpha W X W^* +(1-\alpha)\sum_{i=1}^N \Tr (P_i X)\pi_i, \ \ \forall X\in \B(\HH).
\end{align}
\end{definition}
The RQC is a convex combination of a unitary conjugation $X\mapsto \Wcal(X)= W X W^*$ and of an Entanglement Breaking Channel or a Holevo Measure-and-Prepare Channel, see \cite{H0, HSR, AKL+} on $\B(\HH)$
\begin{align}\label{eq:ebc}
   X\mapsto  \Mcal (X)=\sum_{i=1}^N \Tr (P_i X)\pi_i.
\end{align}

Both maps $\Wcal$ and $\Mcal$ are CPTP, so the same is true for $\Phi$:
\begin{proposition}
\label{prop:cptp0} For all choices of data $\{W, \{P_i\}, \{\pi_i\}, \alpha\}$, the RQC 
$\Phi\in \B(\B(\HH))$ defined by \eqref{eq:phi0} is completely positive and trace-preserving.
\end{proposition}

\begin{remark}\label{rk:samepi}
The number of reset states can sometimes be reduced. If $\pi_i=\pi_j$ for some $i\neq j$, then
\[
\Tr(P_iX)\,\pi_i+\Tr(P_jX)\,\pi_j=\Tr\big((P_i+P_j)X\big)\,\pi_i ,
\]
where $P_i+P_j$ is again an orthogonal projector which is orthogonal to every $P_k$ with $k\neq i,j$. Hence $\{P_k\}_{k\neq i,j}\cup\{P_i+P_j\}$ is an admissible family, and $\Phi$ is a simple RQC with $N-1$ reset states. Hence, $N$ can always be taken equal to the number of distinct reset states. In particular, if $\pi_i=\pi$ for all $1\leq i\leq N$, then using $\sum_{i=1}^N P_i=\Id$,
\[
(1-\alpha)\sum_{i=1}^N\Tr(P_iX)\,\pi_i
=(1-\alpha)\Tr\Big(\Big(\sum_{i=1}^N P_i\Big)X\Big)\,\pi
=(1-\alpha)\Tr(X)\,\pi ,
\]
so that $\Phi(X)=\alpha\Wcal(X)+(1-\alpha)\Tr(X)\pi$ is independent of the family $\{P_i\}$ and corresponds to $N=1$, with $P_1=\Id$.
\end{remark}

Note that a channel $\Phi$ on $\B(\HH)$ is an Entanglement Breaking Channel (EBC) if for any given finite dimensional Hilbert space $\Kcal$, any state $\Gamma\in \D(\HH\otimes \Kcal)$ is mapped by $\Phi\otimes \id$ to a separable state, where $\id$ denotes the identity channel. It is shown in \cite{HSR} that the set of EBCs coincides with maps of the form \eqref{eq:ebc}, where the set of projectors $\{P_i\}$ is generalized to a set $\{F_i\}$ which yields a positive operator valued measure (POVM) i.e. $F_i\geq 0$ for all $1\leq i \leq N$ and  $\sum_i F_i=\Id$.
\begin{remark}\label{rem:povm}
 Actually, the results of this paper remain true, mutatis mutandis, if the projectors are replaced by nonzero operators $P_i\geq 0$ that yield a POVM, here and in the more general definitions below. The few places where the arguments have to be modified are indicated in the text. For simplicity, we restrict ourselves to orthogonal projectors.
\end{remark}

\medskip
The depolarizing channel on $\B(\HH)$
 $   X\mapsto \alpha X+ (1-\alpha)\Tr(X) \Id/\dim(\HH)$
is an example of RQC corresponding to the data $\{\Id, \{\Id\}, \{\Id/\dim(\HH)\}, \alpha\}$. Since it is known that for $\alpha<1$ large enough, the depolarizing channel is not an EBC, the set of RQCs is not contained in the set of EBCs.

Also, $\Phi$ is in general not unital: $\Phi(\Id)=\alpha \Id+(1-\alpha)\sum_{i}\Tr (P_i)\pi_i$ equals $\Id$ if and only if $\alpha =1$ or $\sum_{i}\Tr(P_i)\pi_i=\Id$, which is generally not true.
\medskip

An interpretation of the action of the RQC $\Phi$ \eqref{eq:phi0} on a state $\sigma\in \D(\HH)$ is as follows:  with probability $\alpha$ the state $\sigma$ evolves coherently under the unitary conjugation $\Wcal(\cdot)$, or a reset event occurs with probability $1-\alpha$. During the reset event, the populations $\Tr (P_i \sigma)$ determine the probability of resetting or re-preparing the system to the state $\pi_i$. The reset state $\Mcal(\sigma)$ is obtained by averaging over the possible post-reset states. Hence  $\Phi(\cdot)=\alpha \Wcal(\cdot)+(1-\alpha)\Mcal(\cdot)$ is the corresponding statistical mixture of these alternatives. This actually yields a quantum trajectory interpretation for the discrete time evolution induced by $\Phi$: at each time step, the state $\sigma$ is mapped to $W\sigma W^*$ with probability $\alpha$, and to $\pi_i$ with probability $(1-\alpha)\Tr (P_i \sigma)$, $1\leq i\leq N$. This defines a discrete time Markov process on $\D(\HH)$ with initial state $\sigma$, and it follows by induction  that the expected state after $n$ steps is precisely $\Phi^n(\sigma)$.

\medskip

Let us turn to the spectral properties of $\Phi$ which are closely linked to the properties of the discrete time dynamics induced by $\Phi$ on $\D(\HH)$. Since $\HH$ is finite dimensional, recall that $\B(\HH)$ equipped with the Hilbert-Schmidt inner product $ \HSip{X}{Y}=\Tr X^* Y$ is a Hilbert space. This allows us to define $\Phi^*$, the adjoint of $\Phi$ with respect to this scalar product. It reads
\begin{align}
    \label{eq:phistar0}
    \Phi^*(X)=\alpha\,W^*XW
+(1-\alpha)\sum_{i=1}^{N}P_i\Tr(\pi_iX), \ \ \forall X\in \B(\HH).
\end{align}
As a CP map, $\Phi$ satisfies $(\Phi(X))^*=\Phi(X^*)$, which one can also check explicitly. Hence $\Sp(\Phi)=\overline{\Sp(\Phi)}$. Together with the general relation between spectra of adjoints $\Sp(\Phi)=\overline{\Sp(\Phi^*)}$, we deduce $\Sp(\Phi)=\Sp(\Phi^*)$.
Since $\Phi$ is trace-preserving, we have that $\Phi^*(\Id)=\Id$, as one can check on \eqref{eq:phistar0}, so that $1\in\Sp(\Phi^*)=\Sp(\Phi)$. Also, any CPTP $\Phi$ is a contraction for the trace norm, 
so that $\Sp(\Phi)\subset \bar {\mathbb D}$, where ${\mathbb D}$ is the open unit disc in $\C$ centered at the origin. Moreover, being a quantum channel, $\Phi$ always admits a fixed point which can be normalized to be a state, see e.g. \cite{EH-K, Schrader, Wolf, Watrous}. 

Invariant states play a central role for the dynamics, and it is therefore of interest to establish conditions under which they are unique. As mentioned in the introduction, the spectral analysis will be partially reduced to that of a classical Markov process.\\

\subsection{Conjugation Operators and Resolvents}

When addressing the spectral properties of $\Phi$, we will make repeated use of certain properties of unitary conjugation operators that we state here. We call a state $\sigma\in \D(\HH)$ faithful, if $\sigma>0$. \medskip

\begin{lemma}
\label{lem:resolvent}
The operator $\Wcal$ is unitary on the Hilbert space $(\B(\HH), \HSip{\cdot}{\cdot})$ and for $\gamma\in[0,1[$, 
the operator $\id-\gamma\Wcal$ is invertible with
$$(\id-\gamma\Wcal)^{-1}=\sum_{k\ge0}\gamma^k\Wcal^k.$$
Moreover, for  $\lambda\notin\gamma\,\Sp(\Wcal)$, the resolvent
$(\lambda\, \id-\gamma\Wcal)^{-1}$ satisfies, for all $A\in\B(\HH)$:
\begin{equation}
\Tr\big[(\lambda\,\id-\gamma\Wcal)^{-1}(A)\big]=\frac{\Tr(A)}{\lambda-\gamma}.
\label{eq:traceid0}
\end{equation}
\end{lemma}

\begin{proof} 
For $A,B\in\B(\HH)$,
$\HSip{\Wcal(A)}{B}=\Tr(W A^*W^*B)=\Tr(A^*W^*BW)=\HSip{A}{\Wcal^*(B)}$, so the adjoint of $\Wcal$ acts as $\Wcal^*(B)=W^* B W$. Hence $\Wcal^*\Wcal = \Wcal\Wcal^*=\id$ since $W^*W=WW^*=\Id$, i.e.\ $\Wcal$ is unitary and $\opnorm{\Wcal}=1$. Thus $\opnorm{\gamma\Wcal}=\gamma<1$ and the Neumann series
converges in operator norm to the inverse of $\id-\gamma\Wcal$ (with $\Wcal^0=\id$). For $\lambda\notin\gamma\,\Sp(\Wcal)$, set $B=(\lambda\,\id-\gamma\Wcal)^{-1}(A)$, so that $A=\lambda B-\gamma\Wcal(B)$. Taking the trace and
using $\Tr\Wcal(B)=\Tr(B)$ gives $\Tr(A)=(\lambda-\gamma)\Tr(B)$. Since $\Wcal(\Id)=\Id$, $\gamma\in\gamma\,\Sp(\Wcal)$, so that $\lambda\neq\gamma$, which yields \eqref{eq:traceid0}.  
\end{proof}

\begin{corollary}
    For $\gamma\in[0,1[$ the map $(1-\gamma)(\id-\gamma\Wcal)^{-1}$ on $\B(\HH)$ is a quantum channel which preserves faithfulness. %
\label{cor:tracechannel}
\end{corollary}

\begin{proof}
For $\lambda=1$, \eqref{eq:traceid0} yields $\Tr[(1-\gamma)(\id-\gamma\Wcal)^{-1}(A)]=\Tr(A)$, so the map is trace-preserving. Further, it is completely positive because $(\id-\gamma\Wcal)^{-1}$ %
is a converging positive linear combination of unitary conjugations, so that it is a quantum channel. Finally, note that unitary conjugations preserve faithfulness, and  positive linear combinations of faithful operators are faithful. Since $(\id-\gamma\Wcal)^{-1}=\sum_k\gamma^k\Wcal^k$ is such a combination, the quantum channel preserves faithfulness. 
\end{proof}
\begin{remarks}\label{expdyn}
    0) We call $(1-\gamma)(\id-\gamma\Wcal)^{-1}$ the {\it resolvent channel}.\\
    i) The Neumann series expansion of the resolvent channel %
    shows it is the expectation of random unitary channels $\Wcal^k$ where $k$ follows the geometric distribution $p_k=(1-\gamma)\gamma^k$, $k\in \N$. \\
    ii) For $\gamma=0$, $(1-\gamma)(\id-\gamma\Wcal)^{-1}|_{\gamma=0}=\id$ is the identity channel, whereas $\lim_{\gamma\to 1}(1-\gamma)(\id-\gamma\Wcal)^{-1}={\mathcal P}_1$, the spectral projector of $\Wcal$ associated with the eigenvalue 1. \\
    iii) Corollary \ref{cor:tracechannel} also holds for $\gamma>1$, as can be seen from the same arguments applied to the identity $(1-\gamma)(\id-\gamma\Wcal)^{-1}=\frac{\gamma-1}{\gamma}\Wcal^{-1} (\id-\gamma^{-1} \Wcal^{-1})^{-1}$.
\end{remarks}

\subsection{Invariant States of the Simple RQC}

We consider here $0\leq \alpha<1$ only, since for $\alpha=1$, the RQC reduces to the well known unitary conjugation $\Wcal$. 
The invariant states under the simple RQC \eqref{eq:phi0} are determined by the equation
\begin{align}\label{eq:statstat0}
    \Phi(\varrho)=\alpha W \varrho W^* +(1-\alpha)\sum_{i=1}^N \Tr (P_i \varrho)\pi_i=\varrho,
\end{align}
where $\varrho\in \D(\HH)$. This equation is equivalent to
\begin{align}\label{eq:eveq0}
    \varrho=\sum_{j=1}^N \Tr (P_j \varrho) (1-\alpha)(\id-\alpha\Wcal)^{-1}(\pi_j).
\end{align}
 Introducing the populations $\tilde\varrho^i=\Tr(P_i\varrho)\ge 0$ which satisfy $\sum_i \tilde\varrho^i=1$, we deduce from \eqref{eq:eveq0}
\begin{align}
    \Tr(P_i\varrho)=\sum_{j=1}^N \Tr (P_i (1-\alpha)(\id-\alpha\Wcal)^{-1}(\pi_j))\Tr(P_j\varrho), \ \ \forall \ 1\leq i \leq N.
\end{align}
This calls for the following:
\begin{definition}
    For $0\leq \alpha<1$, the coupling matrix $T\in M_N(\R^+)$ of the RQC is the column stochastic matrix given by
    \begin{align}\label{eq:coupling0}
        T_{ij}=\Tr (P_i (1-\alpha)(\id-\alpha\Wcal)^{-1}(\pi_j)), \ \ 1\leq i,j\leq N.
    \end{align}
\end{definition}
\noindent Thanks to Corollary \ref{cor:tracechannel}, $(1-\alpha)(\id-\alpha\Wcal)^{-1}(\pi_j)$ is a state since $\pi_j$ is  a state, so that $T_{ij}\geq 0$ and as $\sum_iP_i=\Id$, $\sum_{i} T_{ij}=\Tr (1-\alpha)(\id-\alpha\Wcal)^{-1}(\pi_j)=1$, as announced.\\

Hence, the population vector of any invariant state of $\Phi$ is an invariant probability vector of the stochastic coupling matrix $T$
\begin{align}
    \tilde\varrho = T \tilde \varrho, \ \ \text{where} \ \ \tilde\varrho\ =\begin{pmatrix}
        \tilde\varrho^1  \cdots  \tilde\varrho^{N}
    \end{pmatrix}^{\mathsf T}\in (\R^+)^{N}.
\end{align}
Conversely, \eqref{eq:eveq0} shows that any probability vector invariant under the coupling matrix $T$ gives rise to a state in $\D(\HH)$ invariant under $\Phi$. By the Perron-Frobenius theorem, $T$ always admits such an invariant vector. 

To assess the number of independent invariant operators under $\Phi$, we need to consider $\ker (\Phi -\id)$ and seek solutions to \eqref{eq:statstat0} in $\B(\HH)$.
\begin{remark}\label{rem:Cval}
    In case $\dim \ker (\Phi -\id)>1$, $\tilde\varrho^i$ and $\Tr (\varrho)$ may take arbitrary complex values and the notions of trace normalization and populations do not necessarily apply. To ease the exposition, we will abuse terminology and continue to refer to $\tilde\varrho^i$ as populations.
\end{remark}
These arguments establish the main result about simple RQC:

\begin{proposition}
    \label{prop:stoch0}
    For $0\leq \alpha<1$, and $T\in M_{N}(\R^+)$ the stochastic coupling matrix \eqref{eq:coupling0} considered as an operator on $\C^{N}$, it holds
    \begin{align}\label{eq:dimkerphi0}
        \dim \ker (\Phi -\id)=\dim \ker (T -\Id)\geq 1.
    \end{align}
    More precisely, the map $\tilde X=(\tilde X^1,\dots, \tilde X^N)^{\mathsf T}\mapsto X$, where
    \begin{equation}
X=\sum_{i=1}^{N}\tilde X^i(1-\alpha) (\id-\alpha\Wcal)^{-1}(\pi_i),
\label{eq:rhoinv0}
\end{equation}
    is a linear isomorphism from $\ker (T-\Id)$ onto $\ker (\Phi-\id)$, with inverse $X\mapsto(\Tr(P_iX))_{i=1}^N$.
\end{proposition}

As a consequence, 
the determination of the invariant states of the simple RQC is reduced to the determination of the invariant probability vectors of the stochastic coupling matrix, a lower dimensionality problem. Note that $\dim \B(\HH)=(\dim \HH)^2$ while $\dim \C^{N}=N$, where $N\leq \dim \HH$ for projectors, see Remark \ref{rem:povm}.

\medskip
We get a simple criterion ensuring uniqueness of the invariant state of the simple RQC, thanks to the well known property: if $T>0$ entrywise, i.e. $T_{ij}>0$ for all $1\leq i, j\leq N$, then $\dim\ker(T-\Id)=1$ and $T$ admits a unique invariant probability vector, whose entries are strictly positive; see  e.g. \cite{N}. As a consequence:
\begin{lemma}\label{lem:faithful0}
   Let $0\leq\alpha<1$ and assume that all reset states $\pi_i$, $1\leq i\leq N$, are faithful, i.e.\ $\pi_i>0$. Then the simple RQC $\Phi$ given by \eqref{eq:phi0} admits a unique invariant state, given by \eqref{eq:rhoinv0}, which is faithful.
\end{lemma}
\begin{proof}
    By Corollary \ref{cor:tracechannel}, the faithfulness of the reset states $\pi_i$ and since $P_i\neq0$ for all $i$, we have $T>0$ entrywise. Let $\tilde \varrho=(\tilde\varrho^1,\dots, \tilde\varrho^N)^{\mathsf T}$ be the unique invariant vector of $T$ with strictly positive entries. Then \eqref{eq:rhoinv0} is invariant and faithful.
\end{proof}
\begin{remarks}\label{rem:ebc}
    i) Actually, faithful reset states imply the stronger property that $\Phi$ is positivity improving, which further implies $\Sp (\Phi)\cap \T \setminus \{1\}=\emptyset$, see Definitions \ref{def:posPhiprop} and the short account of Evans-Høegh-Krohn's theory below. Thus we further have $\lim_{n\to \infty} \Phi^n(\cdot) =\Tr(\cdot)\varrho$, exponentially fast, where $\varrho$ is given by \eqref{eq:rhoinv0}, and the limiting operator $X\mapsto\Tr(X)\varrho$ is the EBC \eqref{eq:ebc} with $N=1$ and $\pi_1=\varrho$\\
    ii) The case $\alpha=0$ is the pure reset $\Phi|_{\alpha=0}=\Mcal$. As observed in \cite{AKL+},  
    its spectrum can be read off almost completely from the stochastic coupling matrix
    $T_0=(\Tr(P_i\pi_j))_{ij}$: one has
    \[
    \Sp(\Phi|_{\alpha=0})\setminus\{0\}=\Sp(T_0)\setminus\{0\},
    \]
    with the same Jordan blocks for the nonzero eigenvalues.

    \noindent iii) The dynamics of the EBC $\Phi|_{\alpha=0}$ is driven by the stochastic matrix $T_0$ in the sense
    $$(\Phi|_{\alpha=0})^n(\sigma)=\sum_{i,j=1}^N \pi_i (T_0^{n-1})_{ij}\Tr (P_j\sigma), \ \ \forall n\in \N^*,$$
    as easily verified.

\noindent iv) In the case $N=1$, $\pi_1=\pi$ arbitrary (see Remark~\ref{rk:samepi}), $\Phi(X)=\alpha\Wcal(X)+(1-\alpha)\Tr(X)\pi$.
Hence the traceless operators $V=\{X\in\B(\HH):\Tr X=0\}$ form an invariant subspace on which $\Phi=\alpha\Wcal$.
Since $\B(\HH)=\C\varrho\oplus V$, where $\varrho$ is the invariant state \eqref{eq:rhoinv0}, we get
\[
\Sp(\Phi)=\{1\}\cup\Sp(\alpha\Wcal|_V),\qquad \Phi^n(X)=\Tr(X)\varrho+\alpha^n\Wcal^n\big(X-\Tr(X)\varrho\big).
\]
If $e^{i\theta_r}$ are the eigenvalues of $W$, with multiplicities $g_r$, then $\Sp(\alpha\Wcal|_V)$ consists of the
$\alpha e^{i(\theta_r-\theta_s)}$, $\theta_r\neq\theta_s$, and of $\alpha$ with multiplicity $\sum_rg_r^2-1$, which is
nonzero if and only if $\dim\HH\ge2$.
\end{remarks}
When the reset states $\pi_i$ are not all faithful, the spectral picture of $\Phi$ is much richer. For the structured Reset Quantum Channels that we study in this paper, the reset states are never faithful. Moreover, they lend themselves to richer interpretations as well.

\section{Two-Subspace RQCs}\label{sec:twosubs}

In this section, we define and discuss Reset Quantum Channels $\Phi:\B(\HH)\to \B(\HH)$ when the Hilbert space $\HH$ is partitioned into two orthogonal subspaces. The extension to the case where $\HH$ is structured into $r > 2$ mutually orthogonal subspaces is dealt with in Section \ref{sec:rreservoir}.
\medskip

Let $\HH$ be a finite-dimensional Hilbert space with an orthogonal decomposition $\HH=\Hs\oplus\Ho$, and $P\in \B(\HH)$ and $Q\in \B(\HH)$ be the corresponding complementary orthogonal projectors such that $\Hs=\ran (P)$ and $\Ho=\ran (Q)$. We will sometimes write $d=\dim \HH$, $d_P=\dim \Hs$ and $d_Q=\dim \Ho$. For an operator $B\in \B(\HH)$ such that $B=PBP$, we will identify the restriction $B|_{\Hs}$ with $PBP$, and write $B\in\B(\Hs)$. The same identifications hold if  $B=QBQ$. 
For a block-diagonal operator $B=B_P\oplus B_Q$ with respect to $\HH=\Hs\oplus\Ho$ (with $B_P=PBP$, $B_Q=QBQ$), we will also write $B=B_P+B_Q$, depending on the context.
Similarly, for states that are reduced by $\Ho$, respectively $\Hs$, we write $\D(\Ho)$, respectively  $\D(\Hs)$.
\medskip

The definition of the Reset Quantum Channel (RQC) $\Phi$ requires the following ingredients:\medskip

\noindent{\bf Data:}
We assume the existence of \\
i)  a unitary block-diagonal operator $W=W_P\oplus W_Q$, where the restrictions $W_P:=W|_{\Hs}$ and  $W_Q:=W|_{\Ho}$ are unitary.\\
ii) two families of nonzero mutually orthogonal projectors $\{P_i\}_{i=1}^{N_P}$, $\{Q_i\}_{i=1}^{N_Q}$, $N_P, N_Q\geq 1$, such that
\[
P=\sum_{i=1}^{N_P}P_i,\qquad Q=\sum_{i=1}^{N_Q}Q_i,\qquad P+Q=\Id,
\]
with $P_iP_j=\delta_{ij}P_i$ and $Q_iQ_j=\delta_{ij}Q_i$.\\
iii) two families of states $\{\pi_i\}_{i=1}^{N_Q}\subset\D(\Hs)$ and $\{\tau_i\}_{i=1}^{N_P}\subset\D(\Ho)$. 
\\
iv) two weights $\alpha, \beta\in [0,1]$. \\ 

We will denote this data by $\{W, \{P_i\}, \{Q_i\}, \{\pi_i\}, \{\tau_i\}, \alpha, \beta\}$.

\begin{definition}\label{def:rqc2}
The Two-subspace Reset Quantum Channel $\Phi\in \B(\B(\HH))$ associated
with the data above is given for any $X\in \B(\HH)$ by
\begin{equation}
\Phi(X):=\alpha\,W_P PX P W_P^*
+\beta\,W_Q QX Q W_Q^*
+(1-\alpha)\sum_{i=1}^{N_P}\Tr(P_iX)\,\tau_i
+(1-\beta)\sum_{i=1}^{N_Q}\Tr(Q_iX)\,\pi_i .
\label{eq:phi}
\end{equation}
\end{definition}

Under the action of $\Phi$, the evolution of a state 
is partly unitarily implemented by $W_P$ or $W_Q$ within the corresponding subspaces, with weights $\alpha$ in $\Hs$ and $\beta$ in $\Ho$. 
The reset mechanism couples these subspaces with weights $1-\alpha$, respectively $1-\beta$, via the EBCs $\sum_{i=1}^{N_P}\Tr(P_i\cdot)\,\tau_i$, respectively $\sum_{i=1}^{N_Q}\Tr(Q_i \cdot)\,\pi_i$, which transfer the populations within the subspaces $\{P_i\HH\}_{i=1}^{N_P}$, respectively $\{Q_i\HH\}_{i=1}^{N_Q}$, to the prescribed states $\{\tau_i\}_{i=1}^{N_P}\subset\D(\Ho)$, respectively $\{\pi_i\}_{i=1}^{N_Q}\subset\D(\Hs)$.

\medskip

Remark \ref{rem:povm} applies here as well: the projectors can be replaced by nonzero operators $P_i\geq 0$ and $Q_i\geq 0$ such that $\sum_iP_i=P$ and $\sum_iQ_i=Q$. Indeed, this implies $0\leq P_i \leq P$, from which we deduce that for any $\phi_Q\in \Ho$, $\|P_i^{1/2}\phi_Q\|=0$, so that $P_i=P P_i P$, and  similarly $Q_i=Q Q_i Q$.

\begin{proposition}
\label{prop:cptp} For all choices of data $\{W, \{P_i\}, \{Q_i\}, \{\pi_i\}, \{\tau_i\}, \alpha, \beta\}$, the RQC 
$\Phi\in \B(\B(\HH))$ defined by \eqref{eq:phi} is completely positive and trace-preserving (CPTP).
\end{proposition}

\begin{proof}
$\Phi$ is a non negative linear combination of completely positive maps: the first two terms are Kraus maps
$X\mapsto KX K^*$ with $K_P=\sqrt\alpha\,W_PP$ and $K_Q=\sqrt\beta\,W_QQ$, and the last two terms are EBCs, hence $\Phi$ is completely positive.
For trace preservation, using
cyclicity of the trace, unitarity of $W_P$, $W_Q$, $\Tr(\tau_i)=\Tr(\pi_i)=1$, and  $\sum_iP_i=P$, $\sum_iQ_i=Q$, we get for any $X\in \B(\HH)$,
\begin{align*}
\Tr\Phi(X)&=\alpha\Tr(PX)+\beta\Tr(QX)+(1-\alpha)\Tr(PX)+(1-\beta)\Tr(QX)\\
&=\Tr\big((P+Q)X\big)=\Tr(X).\qedhere
\end{align*}
\end{proof}

\begin{remarks}
\label{rk:nonunital}
i) For $\alpha=\beta=1$, 
$\Phi(X)=%
(W_P\oplus W_Q)(PXP+QXQ)(W_P\oplus W_Q)^*$ is not quite a unitary conjugation.\\
For $\alpha=\beta=0$, 
$\Phi(X)=\sum_{i=1}^{N_P}\Tr(P_iX)\,\tau_i+\sum_{i=1}^{N_Q}\Tr(Q_iX)\,\pi_i$
is an EBC %
\cite{H0, HSR, AKL+, WYP+}.\\
ii) $\Phi$ is generally not unital:
$\Phi(\Id)=\alpha P+\beta Q+(1-\alpha)\sum_i\Tr(P_i)\tau_i+(1-\beta)\sum_i\Tr(Q_i)\pi_i$, which equals
$\Id=P+Q$ if and only if $(1-\alpha)\sum_i\Tr(P_i)\tau_i=(1-\beta)Q$ and
$(1-\beta)\sum_i\Tr(Q_i)\pi_i=(1-\alpha)P$. These identities hold for $\alpha=\beta=1$, and are non-generic otherwise. \\
iii) The adjoint of $\Phi$ with respect to the Hilbert--Schmidt inner product
is given by
\begin{equation}
\Phi^*(X)=\alpha\,PW_P^*XW_PP+\beta\,QW_Q^*XW_QQ
+(1-\alpha)\sum_{i=1}^{N_P}P_i\Tr(\tau_iX)+(1-\beta)\sum_{i=1}^{N_Q}Q_i\Tr(\pi_iX).
\label{eq:phistar}
\end{equation}
One
checks explicitly that $\Phi^*(\Id)=\alpha P+\beta Q+(1-\alpha)\sum_iP_i+(1-\beta)\sum_iQ_i=P+Q=\Id$.\\
iv) If $\Delta\in \B(\B(\HH))$ denotes the CPTP projector acting as $\Delta(B)=PBP+QBQ$, we have
\begin{align}\label{eq:phidiag}
    \Phi=\Phi\circ \Delta=\Delta \circ \Phi,
\end{align}
so that any operator that is off-diagonal with respect to $\HH=\Hs\oplus \Ho$ belongs to the kernel of $\Phi$. In other words, $\Phi$ is reduced by the projector $\Delta$ and vanishes on $\ran(\id-\Delta)$. This projector will also play a role in Section \ref{sec:lindblad}.
\end{remarks}

The traces onto $\Hs$ and $\Ho$ of the iterates of $\Phi$ on $X$ evolve according to a $2\times 2$ Markov process:

\begin{lemma}\label{lem:Sigma}
For all data and all $\alpha,\beta\in[0,1]$, and for every $X\in\B(\HH)$,
\[
\begin{pmatrix}\Tr(P\Phi(X))\\ \Tr(Q\Phi(X))\end{pmatrix}
=\Sigma\begin{pmatrix}\Tr(PX)\\ \Tr(QX)\end{pmatrix},\qquad
\Sigma=\begin{pmatrix}\alpha&1-\beta\\ 1-\alpha&\beta\end{pmatrix}.
\]
Moreover, $\Sp(\Sigma)=\{1,\alpha+\beta-1\}\subseteq\Sp(\Phi)$.
\end{lemma}
\begin{remark}\label{rem:35}
If $0<\alpha+\beta<2$, then $|\alpha+\beta-1|<1$, and for every $X\in\B(\HH)$,
\[
\lim_{n\to\infty}\Tr(P\Phi^n(X))=\frac{1-\beta}{2-\alpha-\beta}\,\Tr(X),\qquad
\lim_{n\to\infty}\Tr(Q\Phi^n(X))=\frac{1-\alpha}{2-\alpha-\beta}\,\Tr(X),
\]
exponentially fast, with rate $|\alpha+\beta-1|$.
\end{remark}
\begin{proof}
By \eqref{eq:phistar}, $\Phi^*(P)=\alpha P+(1-\beta)Q$ and $\Phi^*(Q)=(1-\alpha)P+\beta Q$. The vector identity
follows from $\Tr(P\Phi(X))=\Tr(\Phi^*(P)X)$ and $\Tr(Q\Phi(X))=\Tr(\Phi^*(Q)X)$.

\smallskip
\noindent The matrix has spectrum $\Sp(\Sigma)=\{1,\alpha+\beta-1\}$.
Let $\lambda\in\Sp(\Sigma)$ and $y\in \R^2$ with $\Sigma^*y=\lambda y$. Let $L=y_1P+y_2Q\neq0$ so that, by linearity,
\[
\Phi^*(L)=\big(\alpha y_1+(1-\alpha)y_2\big)P+\big((1-\beta)y_1+\beta y_2\big)Q
=(\Sigma^*y)_1\,P+(\Sigma^*y)_2\,Q=\lambda L .
\]
Thus $\lambda\in\Sp(\Phi^*)=\Sp(\Phi)$.
\end{proof}

\medskip
In the next section, we will make use of the properties of conjugation operators defined on the subspaces $\Hs$ and $\Ho$. We make explicit the notation and conventions used:

\smallskip
\noindent For $X\in \B(\Hs)$, we write $\Wcal_P(X)=W_PXW_P^*$ for the conjugation superoperator on $\Hs$, and $\Id_P$ for the identity on $\Hs$. The identity as a superoperator on $\B(\Hs)$ is denoted as $\id_P$. Then $\B(\Hs)$ equipped with the Hilbert-Schmidt inner product is a Hilbert space. Similar expressions hold on $\Ho$, by changing $P$ for $Q$.
In particular, we get with  $W=W_P\oplus W_Q$, $\Phi=\Delta\circ \Wcal \circ \Delta$ for $\alpha=\beta=1$, see Remarks \ref{rk:nonunital} i) and iv).

\smallskip
Lemma \ref{lem:resolvent} and Corollary \ref{cor:tracechannel} remain in force when considered on the Hilbert spaces $(\B(\Hs),\HSip{\cdot}{\cdot})$ or $(\B(\Ho), \HSip{\cdot}{\cdot})$, with extra indices in the operators to indicate which spaces are considered. For example, we will write $(1-\gamma)(\id_P-\gamma\Wcal_P)^{-1}$ for $\gamma\in[0,1[$, the quantum channel on $\B(\Hs)$ that preserves faithfulness on $\D(\Hs)$.

The next  sections are devoted to the analysis of the spectrum and invariant eigenoperators of $\Phi$. The invariant states which contain the physical information on the large time times behavior of the system at hand, and the rest of the spectrum provides information on the approach to the asymptotic state at large times.

\section{Reduction to Coupling Matrices}
\label{sec:reduction}

Throughout this section, $0\leq\alpha,\beta<1$. The cases where $\alpha$ or $\beta$ equals one are different and will be dealt with in Section~\ref{sec:limits}. We reduce the eigenvalue equation $\Phi(X)=\lambda X$ to an eigenvalue problem
for a matrix of size $N_P$, built from the populations. We then apply this reduction to three values of $\lambda$ that
are special for all data: $\lambda=1$, which gives the invariant states, $\lambda=\alpha+\beta-1$, which gives a robust
eigenvalue together with an explicit eigenoperator, and $\lambda=0$, which gives part of the kernel. We finally discuss the
uniqueness of the invariant state.

\subsection{Reduction to the Populations}
\label{sec:lambdared}

For $X\in\B(\HH)$, we write $\tilde X_P=(\Tr(P_iX))_{i=1}^{N_P}\in\C^{N_P}$ and $\tilde X_Q=(\Tr(Q_jX))_{j=1}^{N_Q}\in\C^{N_Q}$,
and, by a slight abuse of language, as in Remark \ref{rem:Cval}, call them the population vectors of $X$, although they are populations in the proper
sense, that is nonnegative with total sum $1$, only when $X$ is a state, see Remark~\ref{rem:Cval}. For an invariant
state $\varrho$, they are denoted $\tilde\varrho_P$ and $\tilde\varrho_Q$.

\begin{definition}\label{def:coupling}
Let $0\leq\alpha,\beta<1$ and $\lambda\in\C\setminus\big(\alpha\Sp(\Wcal_P)\cup\beta\Sp(\Wcal_Q)\big)$. The coupling matrices
$T_P(\lambda)\in M_{N_P,N_Q}(\C)$ and $T_Q(\lambda)\in M_{N_Q,N_P}(\C)$ are given by
\begin{equation}\label{eq:coupling}
\begin{aligned}
T_P(\lambda)_{ij}&=\Tr\!\big[P_i(1-\alpha)(\lambda\id_P-\alpha\Wcal_P)^{-1}(\pi_j)\big],\\
T_Q(\lambda)_{kl}&=\Tr\!\big[Q_k(1-\beta)(\lambda\id_Q-\beta\Wcal_Q)^{-1}(\tau_l)\big],
\end{aligned}
\end{equation}
where $1\leq i\leq N_P$, $1\leq j\leq N_Q$ and $1\leq k\leq N_Q$, $1\leq l\leq N_P$. We set $T_{PQ}(\lambda)=T_P(\lambda)T_Q(\lambda)$ and $T_{QP}(\lambda)=T_Q(\lambda)T_P(\lambda)$. Since $1\notin\alpha\Sp(\Wcal_P)\cup\beta\Sp(\Wcal_Q)$, we write $T_P=T_P(1)$, $T_Q=T_Q(1)$, $T_{PQ}=T_{PQ}(1)$ and $T_{QP}=T_{QP}(1)$.
\end{definition}
\noindent By Corollary \ref{cor:tracechannel}, the entries of $T_P$ and $T_Q$ are nonnegative. Moreover, $P=\sum_i P_i$ and
$Q=\sum_i Q_i$ yield $\sum_i (T_P)_{ij}=\sum_k(T_Q)_{kl}=1$, so $T_P$ and $T_Q$ are column stochastic. More generally, by
the trace identity \eqref{eq:traceid0}, the columns of $T_P(\lambda)$ and $T_Q(\lambda)$ sum to
$\frac{1-\alpha}{\lambda-\alpha}$ and $\frac{1-\beta}{\lambda-\beta}$, respectively.

\begin{theorem}\label{prop:lambdared}
Let $\lambda\in\C\setminus\big(\alpha\Sp(\Wcal_P)\cup\beta\Sp(\Wcal_Q)\big)$. For
$\tilde X_P(\lambda)\in\ker(T_{PQ}(\lambda)-\Id)$, set $\tilde X_Q(\lambda)=\frac{1-\alpha}{1-\beta}T_Q(\lambda)\tilde X_P(\lambda)$ and
\begin{equation}
X_\lambda=(1-\beta)\sum_{j=1}^{N_Q}\tilde X_Q^j(\lambda)\,(\lambda\id_P-\alpha\Wcal_P)^{-1}(\pi_j)
+(1-\alpha)\sum_{i=1}^{N_P}\tilde X_P^i(\lambda)\,(\lambda\id_Q-\beta\Wcal_Q)^{-1}(\tau_i).
\label{eq:rholam}
\end{equation}
Then $\tilde X_P(\lambda)\mapsto X_\lambda$ is a linear isomorphism from $\ker(T_{PQ}(\lambda)-\Id)$ onto the block-diagonal
part $\ker(\Phi-\lambda)\cap\ran\Delta$, with inverse $X\mapsto(\Tr(P_iX))_{i=1}^{N_P}$. Moreover
$\ker(\Phi-\lambda)\subseteq\ran\Delta$ if $\lambda\neq0$.
\end{theorem}

\begin{remark}\label{rk:redkernel}
For $\lambda=0$, which requires $0<\alpha,\beta<1$, Theorem~\ref{prop:lambdared} only describes the block-diagonal part of $\ker\Phi$. By \eqref{eq:phidiag}, $\ker\Phi=\ran(\id-\Delta)\oplus\big(\ker\Phi\cap\ran\Delta\big)$, see Section~\ref{sec:kernel}.
\end{remark}

\begin{proof}
Let $X_\lambda\in\ran\Delta$, with population vectors $\tilde X_P(\lambda)=(\Tr(P_iX_\lambda))_{i=1}^{N_P}$ and
$\tilde X_Q(\lambda)=(\Tr(Q_kX_\lambda))_{k=1}^{N_Q}$. By \eqref{eq:phidiag}, $\Phi(X_\lambda)$ is block-diagonal, as
$X_\lambda$ is. Hence, by \eqref{eq:phi}, $\Phi(X_\lambda)=\lambda X_\lambda$ is equivalent to the two block equations
\begin{equation}
(\lambda\id_P-\alpha\Wcal_P)(PX_\lambda P)=(1-\beta)\sum_{j=1}^{N_Q}\tilde X_Q^j(\lambda)\,\pi_j,\qquad
(\lambda\id_Q-\beta\Wcal_Q)(QX_\lambda Q)=(1-\alpha)\sum_{i=1}^{N_P}\tilde X_P^i(\lambda)\,\tau_i .
\label{eq:blocklam}
\end{equation}
Inverting the operators in \eqref{eq:blocklam}, this is equivalent to
\begin{equation}
PX_\lambda P=(1-\beta)\sum_{j=1}^{N_Q}\tilde X_Q^j(\lambda)\,(\lambda\id_P-\alpha\Wcal_P)^{-1}(\pi_j),\qquad
QX_\lambda Q=(1-\alpha)\sum_{i=1}^{N_P}\tilde X_P^i(\lambda)\,(\lambda\id_Q-\beta\Wcal_Q)^{-1}(\tau_i).
\label{eq:blocksinv}
\end{equation}
Tracing the first block against $P_i$ and the second against $Q_k$ gives
\[
\tilde X_P(\lambda)=\frac{1-\beta}{1-\alpha}\,T_P(\lambda)\tilde X_Q(\lambda),\qquad
\tilde X_Q(\lambda)=\frac{1-\alpha}{1-\beta}\,T_Q(\lambda)\tilde X_P(\lambda).
\]
Eliminating $\tilde X_Q(\lambda)$ gives $\tilde X_P(\lambda)=T_{PQ}(\lambda)\tilde X_P(\lambda)$, and $X_\lambda=PX_\lambda P+QX_\lambda Q$
is given by \eqref{eq:rholam}.

Conversely, let $\tilde X_P(\lambda)\in\ker(T_{PQ}(\lambda)-\Id)$, and define $X_\lambda$ by \eqref{eq:rholam}. Its blocks are
given by \eqref{eq:blocksinv}, and tracing them as above gives
\[
\Tr(P_iX_\lambda)=\frac{1-\beta}{1-\alpha}\big(T_P(\lambda)\tilde X_Q(\lambda)\big)_i=\big(T_{PQ}(\lambda)\tilde X_P(\lambda)\big)_i
=\tilde X_P^i(\lambda),\qquad
\Tr(Q_kX_\lambda)=\tilde X_Q^k(\lambda).
\]
So the populations of $X_\lambda$ are $\tilde X_P(\lambda)$ and $\tilde X_Q(\lambda)$, and \eqref{eq:blocksinv} is equivalent to
\eqref{eq:blocklam}, that is, to $\Phi(X_\lambda)=\lambda X_\lambda$. The two maps are thus inverse to each other.

\smallskip
\noindent Finally, if $\lambda\neq0$ and $\Phi(X_\lambda)=\lambda X_\lambda$, then
$X_\lambda=\lambda^{-1}\Phi(X_\lambda)=\lambda^{-1}\Delta(\Phi(X_\lambda))\in\ran\Delta$ by \eqref{eq:phidiag}.
\end{proof}

\begin{remark} \label{rem:colsum}
By Theorem~\ref{prop:lambdared}, the nonzero eigenvalues of $\Phi$ outside $\alpha\Sp(\Wcal_P)\cup\beta\Sp(\Wcal_Q)$
are exactly the $\lambda$ with $1\in\Sp(T_{PQ}(\lambda))$. This condition depends on the data in general. However, as noted after Definition \ref{def:coupling}, every column of $T_{PQ}(\lambda)$ sums to 
\[
s(\lambda)=\frac{(1-\alpha)(1-\beta)}{(\lambda-\alpha)(\lambda-\beta)}.
\]
Hence $s(\lambda)\in\Sp(T_{PQ}(\lambda))$, and $\ker(T_{PQ}(\lambda)-\Id)\neq\{0\}$ as soon as $s(\lambda)=1$, whatever the data. Since
\[
s(\lambda)=1\iff(\lambda-1)\big(\lambda-(\alpha+\beta-1)\big)=0 ,
\]
Theorem~\ref{prop:lambdared} provides eigenoperators of $\Phi$ for $\lambda=1$ and $\lambda=\alpha+\beta-1$, whenever $T_{PQ}(\lambda)$ is defined, see section \ref{sec:fixed} and \ref{sec:robust}. The value $\lambda=0$ is studied in Section~\ref{sec:kernel}. The other solutions depend on the data and are not explicit in general.
\end{remark}

\subsection{Invariant States}
\label{sec:fixed}

For $\lambda=1$, Theorem~\ref{prop:lambdared} describes the fixed points of $\Phi$. The analysis is thus reduced to that of a classical Markov process, given by the column stochastic matrix $T_{PQ}$. As in the previous sections, we denote by $\varrho$ an invariant state of $\Phi$, that is, a solution of
\begin{align}
\Phi(\varrho)=\varrho, \ \ \varrho\in \D(\HH).
\label{eq:statstat}
\end{align}
By \eqref{eq:phidiag}, any invariant state is block-diagonal, $\varrho=P\varrho P+Q\varrho Q$. Since $0\leq\alpha,\beta<1$, one has
$1\notin\alpha\Sp(\Wcal_P)\cup\beta\Sp(\Wcal_Q)$, and \eqref{eq:statstat} reduces to the block equations \eqref{eq:blocklam}
with $\lambda=1$ and $X_1=\varrho$. We first note that the total populations of the two subspaces are fixed by Remark \ref{rem:35}:
\begin{align}\label{eq:traceval}
\Tr(P\varrho)=\frac{1-\beta}{2-\alpha-\beta},\qquad
\Tr(Q\varrho)=\frac{1-\alpha}{2-\alpha-\beta}.
\end{align}
These populations can take any value in $(0,1)$.

\begin{corollary} \label{prop:stoch}
One has $\dim \ker(\Phi-\id)=\dim \ker(T_{PQ}-\Id)=\dim \ker(T_{QP}-\Id)$. Moreover, $\varrho$ is an invariant state of $\Phi$ if and only if it is given by \eqref{eq:rholam} with $\lambda=1$, that is,
\begin{equation}\label{eq:rhoinv}
\varrho=(1-\beta)\sum_{i=1}^{N_Q}\tilde\varrho_Q^i\,(\id_P-\alpha\Wcal_P)^{-1}(\pi_i)+(1-\alpha)\sum_{i=1}^{N_P}\tilde\varrho_P^i\,(\id_Q-\beta\Wcal_Q)^{-1}(\tau_i),
\end{equation}
where $\tilde\varrho_P\in\ker(T_{PQ}-\Id)\cap(\R^+)^{N_P}$, with $\sum_i\tilde\varrho_P^i=\frac{1-\beta}{2-\alpha-\beta}$, and
$\tilde\varrho_Q=\frac{1-\alpha}{1-\beta}T_Q\tilde\varrho_P$.
\end{corollary}

\begin{proof}
Since $T_P$ and $T_Q$ are column stochastic, so are $T_{PQ}$ and $T_{QP}$, and $1$ is an eigenvalue of both. As
$1\notin\alpha\Sp(\Wcal_P)\cup\beta\Sp(\Wcal_Q)$, Theorem~\ref{prop:lambdared} with $\lambda=1$ gives
$\dim\ker(\Phi-\id)=\dim\ker(T_{PQ}-\Id)$. The second equality follows from Flanders' theorem \cite{F}.

\smallskip
Let $\varrho$ be an invariant state. By Theorem~\ref{prop:lambdared}, it is given by \eqref{eq:rhoinv}, where
$\tilde\varrho_P\in\ker(T_{PQ}-\Id)$ is its population vector. These populations are nonnegative, and their sum
$\Tr(P\varrho)$ is given by \eqref{eq:traceval}.

\smallskip
Conversely, let $\tilde\varrho_P$ be as in the statement, which exists by the Perron--Frobenius theorem. Since $T_Q$ is
column stochastic with nonnegative entries, $\tilde\varrho_Q\in(\R^+)^{N_Q}$ and
$\sum_i\tilde\varrho_Q^i=\frac{1-\alpha}{1-\beta}\sum_i\tilde\varrho_P^i$. By Corollary~\ref{cor:tracechannel}, \eqref{eq:rhoinv}
writes $\varrho$ as a combination of the states $(1-\alpha)(\id_P-\alpha\Wcal_P)^{-1}(\pi_i)$ and
$(1-\beta)(\id_Q-\beta\Wcal_Q)^{-1}(\tau_i)$, with nonnegative weights $\frac{1-\beta}{1-\alpha}\tilde\varrho_Q^i$ and
$\frac{1-\alpha}{1-\beta}\tilde\varrho_P^i$. Their sum is
\[
\frac{1-\beta}{1-\alpha}\sum_i\tilde\varrho_Q^i+\frac{1-\alpha}{1-\beta}\sum_i\tilde\varrho_P^i
=\Big(1+\frac{1-\alpha}{1-\beta}\Big)\sum_i\tilde\varrho_P^i=\frac{2-\alpha-\beta}{1-\beta}\cdot\frac{1-\beta}{2-\alpha-\beta}=1 .
\]
Hence $\varrho$ is a state, and it is invariant by Theorem~\ref{prop:lambdared}.
\end{proof}

\begin{remarks}
i) If $\tilde\varrho_Q\in \ker (T_{QP}-\Id)\cap(\R^+)^{N_Q}$ with $\sum_i\tilde\varrho_Q^i=\frac{1-\alpha}{2-\alpha-\beta}$, Formula \eqref{eq:rhoinv} holds with $\tilde\varrho_P=\frac{1-\beta}{1-\alpha}T_P \tilde\varrho_Q$.\\
ii) The choice of the stochastic matrix to consider may be dictated by its size, since in general $N_P\neq N_Q$, with $N_P\leq \dim \Hs$, $N_Q\leq \dim \Ho$ for projectors, independent of $\dim \HH$.
\end{remarks}

\begin{application}[System-Reservoir]\label{ex:NP1}
Consider the RQC $\Phi$  dubbed system-reservoir in the introduction, with data characterized by $N_P=1$, $N_Q\geq 1$, so that $P_1=P$, $\tau_1=\tau\in \D(\Ho)$. Then, $T_{PQ}=1$, as a one by one matrix, so that $\dim \ker (\Phi-\id)=1$ by Corollary~\ref{prop:stoch} and ${\tilde\varrho_P}=\frac{1-\beta}{2-\alpha-\beta}\in \R^{+*}$. Since $T_Q \in M_{N_Q,1}(\R^+)$, it can be considered as a probability vector with components $T_Q^i=(1-\beta)\Tr\!\big[Q_i(\id_Q-\beta\Wcal_Q)^{-1}(\tau)\big]$, $1\leq i\leq N_Q$. Hence, ${\tilde\varrho_Q}=\frac{1-\alpha}{2-\alpha-\beta}T_Q\in (\R^+)^{N_Q}$. Thus the unique invariant state of $\Phi$ is given by
\begin{align}\label{eq:statenp1}
    \varrho=\frac{1-\beta}{2-\alpha-\beta}(1-\alpha)(\id_P-\alpha\Wcal_P)^{-1}\big(\sum_iT_Q^i\pi_i\big)
+\frac{1-\alpha}{2-\alpha-\beta}(1-\beta)(\id_Q-\beta\Wcal_Q)^{-1}(\tau).
\end{align}
If, in addition, $[W_Q,\tau]=0$, \eqref{eq:statenp1} reduces to \eqref{eq:ressys} discussed in the introduction. 
\end{application}

\subsection{A Robust Eigenvalue}
\label{sec:robust}
We determine here an eigenoperator for the robust eigenvalue $\alpha+\beta-1$ of Lemma \ref{lem:Sigma}:
\begin{lemma}
\label{lem:abeig}
Let $\lambda=\alpha+\beta-1\notin\alpha\Sp(\Wcal_P)\cup\beta\Sp(\Wcal_Q)$. Then
$\ker(T_{PQ}(\lambda)-\Id)\neq\{0\}$, and for every nonzero $\tilde X_P(\lambda)$ in it, $X_\lambda$ given by
\eqref{eq:rholam} is a nonzero eigenoperator of $\Phi$ for the eigenvalue $\lambda$.
\end{lemma}

\begin{proof}
By Remark~\ref{rem:colsum}, $s(\lambda)=1$, so that $1\in\Sp(T_{PQ}(\lambda))$. Theorem~\ref{prop:lambdared} then gives the
eigenoperator $X_\lambda$, which is nonzero since its population vector $\tilde X_P(\lambda)$ is. 
\end{proof}

\begin{remark}
\label{rk:abzero}
The condition $\alpha+\beta-1\notin\alpha\Sp(\Wcal_P)\cup\beta\Sp(\Wcal_Q)$ fails only if $\beta=1-2\alpha$ and
$-1\in\Sp(\Wcal_P)$, or $\alpha=1-2\beta$ and $-1\in\Sp(\Wcal_Q)$. Then $\alpha+\beta-1$ is still an eigenvalue of $\Phi$ by
Lemma~\ref{lem:Sigma}, but there may be no eigenoperator of the form \eqref{eq:rholam}, and $\Phi$ may fail to be
diagonalizable, see Remark~\ref{rk:nondiag}.
\end{remark}

\subsection{The Kernel}
\label{sec:kernel}

By \eqref{eq:phidiag}, $\ker\Phi=\ran(\id-\Delta)\oplus\big(\ker\Phi\cap\ran\Delta\big)$, so that $\dim\ker\Phi\ge2\,d_Pd_Q$. We analyze the block-diagonal part with Theorem~\ref{prop:lambdared} for $\lambda=0$.

\begin{proposition}
\label{prop:kernel}
For $1\leq i\leq N_P$, $1\leq j\leq N_Q$ and $0<\alpha,\beta<1$, set
\[
(T_P^\infty)_{ij}=\Tr\big(P_i\Wcal_P^{-1}(\pi_j)\big),\qquad
(T_Q^\infty)_{ji}=\Tr\big(Q_j\Wcal_Q^{-1}(\tau_i)\big).
\]
There is a nonzero block-diagonal $X\in\ker(\Phi)$ if and only if
$\frac{\alpha\beta}{(1-\alpha)(1-\beta)}\in\Sp(T_P^\infty T_Q^\infty)$, where $T_P^\infty T_Q^\infty\in M_{N_P}(\R^+)$ is
column stochastic.
\end{proposition}

The matrices $T_P^\infty$ and $T_Q^\infty$ are the formal limits of $T_P$ and $T_Q$ in \eqref{eq:coupling} as $\alpha\to\infty$
and $\beta\to\infty$, respectively, which explains the notation; see Remark~\ref{expdyn} iii).

\begin{proof}
Since $\alpha,\beta>0$, $0\notin\alpha\Sp(\Wcal_P)\cup\beta\Sp(\Wcal_Q)$, and Theorem~\ref{prop:lambdared} applies
with $\lambda=0$. As %
\[
T_P(0)=-\frac{1-\alpha}{\alpha}\,T_P^\infty,\qquad T_Q(0)=-\frac{1-\beta}{\beta}\,T_Q^\infty,\qquad
T_{PQ}(0)=\frac{(1-\alpha)(1-\beta)}{\alpha\beta}\,T_P^\infty T_Q^\infty .
\]
Hence a nonzero block-diagonal $X\in\ker\Phi$ exists if and only if $1\in\Sp(T_{PQ}(0))$, i.e.\
$\frac{\alpha\beta}{(1-\alpha)(1-\beta)}\in\Sp(T_P^\infty T_Q^\infty)$. Finally,
$\sigma_j=\Wcal_P^{-1}(\pi_j)=W_P^*\pi_jW_P$ is a state on $\Hs$, so that $(T_P^\infty)_{ij}=\Tr(P_i\sigma_j)\ge0$ and
$\sum_{i=1}^{N_P}(T_P^\infty)_{ij}=\Tr(P\sigma_j)=1$. Hence $T_P^\infty$ is column stochastic, and so is $T_Q^\infty$ by
the same argument. So is their product $T_P^\infty T_Q^\infty$.
\end{proof}

\begin{remarks}\label{rem:ab}
Set $\kappa=\frac{\alpha\beta}{(1-\alpha)(1-\beta)}>0$.

\smallskip
\noindent i) The matrices $T_P^\infty$ and $T_Q^\infty$ are column stochastic. Hence so is their product, whose spectral radius
is therefore equal to one. Moreover, $\kappa>1$ if and only if $\alpha+\beta>1$.
\begin{itemize}[leftmargin=*]
\item If $\alpha+\beta>1$, then $\kappa\notin\Sp(T_P^\infty T_Q^\infty)$, and the kernel of $\Phi$ consists exactly of the
off-diagonal blocks.
\item If $\alpha+\beta=1$, then $\kappa=1\in\Sp(T_P^\infty T_Q^\infty)$, and $\ker\Phi$ always contains a nonzero block-diagonal element given by Lemma~\ref{lem:abeig}.
\item If $\alpha+\beta<1$, then $0<\kappa<1$, and $\ker\Phi$ may contain more elements, depending on
$T_P^\infty T_Q^\infty$.
\end{itemize}

\smallskip
\noindent ii) Since $\kappa\neq0$, and $\Sp(T_P^\infty T_Q^\infty)\setminus\{0\}=\Sp(T_Q^\infty T_P^\infty)\setminus\{0\}$ by
Flanders' theorem \cite{F}, the condition of Proposition~\ref{prop:kernel} is equivalent to $\kappa\in\Sp(T_Q^\infty T_P^\infty)$.
Here $T_Q^\infty T_P^\infty\in M_{N_Q}(\R^+)$ is also column stochastic.

\end{remarks}
\medskip

\subsection{Uniqueness}
\label{sec:unique}

We say that the reset states are faithful if $\pi_i>0$ on $\Hs$ for all $1\leq i\leq N_Q$ and $\tau_i>0$ on $\Ho$
for all $1\leq i\leq N_P$.
Uniqueness of the invariant state is equivalent to $1$ being a simple eigenvalue of the stochastic matrix $T_{PQ}$ or of $T_{QP}$. In the following we focus on $T_{PQ}$, with the understanding that all properties can be adapted to $T_{QP}$, mutatis mutandis. \medskip

Uniqueness of the invariant state holds in particular if the stochastic matrix $T_{PQ}$ is irreducible, in which case it has an eigenvector with strictly positive components. Moreover, if $T_{PQ}$  is irreducible and aperiodic, then the simple eigenvalue one is the only modulus one eigenvalue of $T_{PQ}$. Furthermore, if $T_{PQ}>0$ entrywise, then $T_{PQ}$ is irreducible and aperiodic. See e.g. \cite{N} for more details.  

\medskip
\noindent Let us provide conditions ensuring uniqueness of the invariant state. Recall
\begin{align}\label{eq:matelTPQ}
(T_{PQ})_{ij}&=(T_PT_Q)_{ij} \\ \nonumber
&=(1-\alpha)(1-\beta)\sum_{k=1}^{N_Q}\Tr\big[P_i(\id_P-\alpha\Wcal_P)^{-1}(\pi_k)\big]\,
\Tr\big[Q_k(\id_Q-\beta\Wcal_Q)^{-1}(\tau_j)\big],
\end{align}
By Lemma \ref{lem:resolvent}, the resolvents expand as
$(\id_P-\alpha\Wcal_P)^{-1}=\sum_{n\in\N}\alpha^n\Wcal_P^n$ and $(\id_Q-\beta\Wcal_Q)^{-1}=\sum_{m\in\N}\beta^m\Wcal_Q^m$.
Hence
\begin{equation}\label{eq:TPQseries}
(T_{PQ})_{ij}=(1-\alpha)(1-\beta)\sum_{k=1}^{N_Q}\sum_{n,m\in\N}\alpha^n\beta^m\,\Tr[P_i\Wcal_P^n(\pi_k)]\,\Tr[Q_k\Wcal_Q^m(\tau_j)].
\end{equation}
The next lemma says that $T_{PQ}$ is entrywise strictly positive exactly when $W$ has sufficient mixing.
\begin{lemma}
\label{lem:mixing}
$T_{PQ}>0$ entrywise if and only if for all $1\leq i,j\leq N_P$, there exist $1\leq k\leq N_Q$ and $n,m\in\N$ such that
$$\alpha^n\beta^m\,\Tr[P_i\Wcal_P^n(\pi_k)]\,\Tr[Q_k\Wcal_Q^m(\tau_j)]>0.$$
\end{lemma}
\begin{proof}
The maps $\Wcal_P^n$ and $\Wcal_Q^m$ send states to states, so all the terms of \eqref{eq:TPQseries} are nonnegative, and
$(1-\alpha)(1-\beta)>0$. Hence $(T_{PQ})_{ij}>0$ if and only if one of these terms is positive.
\end{proof}
When $0<\alpha,\beta<1$, the factor $\alpha^n\beta^m$ is positive and can be dropped. When $\alpha=0$, only $n=0$ is
allowed, and similarly for $\beta=0$.

\medskip
We also provide a simpler more explicit criterion similar to that used in Lemma \ref{lem:faithful0}:
\begin{lemma}
\label{lem:faithful}
If all $\pi_i$, $1\leq i\leq N_Q$, are faithful on $\Hs$, then $T_{PQ}>0$ entrywise.
\end{lemma}
\begin{proof}
Let $1\leq i,j\leq N_P$. Since $T_Q$ is column stochastic, there is $k$ with $(T_Q)_{kj}>0$, so that
$\beta^m\Tr[Q_k\Wcal_Q^m(\tau_j)]>0$ for some $m$. Since $\pi_k>0$ on $\Hs$ and $P_i\neq0$, the term $n=0$ gives
$\Tr[P_i\pi_k]>0$. Lemma~\ref{lem:mixing} gives $T_{PQ}>0$ entrywise.
\end{proof}

\begin{remarks}
    i) By symmetry, the same conclusion about uniqueness of the invariant state holds if all reset states $\tau_i$, $1\leq i\leq N_P$, are faithful.

\noindent
ii) Note that $T_{PQ}>0$ entrywise does not say anything about the faithfulness of the unique invariant state $\varrho$ itself, as Application \ref{ex:NP1} shows. Consider $W=\Id$ so that the restriction of the invariant state \eqref{eq:statenp1} to $\Ho$ is proportional to $\tau\in \D(\Ho)$, which may have a non trivial kernel while $T_{PQ}=1$. 

\noindent
iii) Uniqueness of the invariant state holds generically, even when no reset state is faithful. More precisely, let
$0<\alpha<1$, fix all the data except $W_P$, and let $W_P$ be Haar distributed on $\Ucal(\Hs)$. Then the invariant state
is unique for almost every $W_P$.

\smallskip\noindent
Indeed, for $1\leq i\leq N_P$ and $1\leq k\leq N_Q$, set $f_{ik}(W)=\Tr[P_iW\pi_kW^*]$. This is a real polynomial in the
real and imaginary parts of the entries of $W$, and it does not vanish identically. To see this, let $\psi$ be a unit
eigenvector of $\pi_k$ with eigenvalue $p>0$, and $\phi\in\ran P_i$ a unit vector. Completing $\psi$ and $\phi$ into
orthonormal bases of $\Hs$ yields a unitary $W_0$ with $W_0\psi=\phi$, and 
\[
f_{ik}(W_0)\geq\langle\phi,W_0\pi_kW_0^*\phi\rangle=\langle\psi,\pi_k\psi\rangle=p>0 .
\]
If $P_i$ is only nonnegative, see Remark \ref{rem:povm}, we take for $\phi$ a unit eigenvector of $P_i$ with eigenvalue $e>0$, and the lower bound becomes $e\,p>0$.
By \cite[Prop.~3]{LSD+}, the zero set of $f_{ik}$ therefore has vanishing Haar measure. Outside the union on $i,k$ of these zero sets, the term $n=1$ in the expansion of $T_P$ gives
$(T_P)_{ik}\geq(1-\alpha)\alpha f_{ik}(W_P)>0$ for all $i,k$. Since $T_Q$ is column stochastic, Lemma~\ref{lem:mixing}
applies with $n=1$, and $T_{PQ}>0$ entrywise.

\smallskip\noindent
If $\alpha=0<\beta$, the same argument with $W_Q$ random gives $T_{QP}>0$. For $\alpha=\beta=0$, the matrices $T_P$ and
$T_Q$ do not depend on the unitaries, and uniqueness may fail for all of them. This is the case, for instance, when $N_P=N_Q\geq2$ and the
data split into blocks $P_k\HH$, $Q_k\HH$, with $\pi_k\in\D(P_k\HH)$ and $\tau_k\in\D(Q_k\HH)$.

\end{remarks}

\section{Peripheral Spectrum}

\noindent Stronger spectral results on $\Phi$ follow from the irreducibility theory of Schwarz maps, as developed by Evans and H\o egh-Krohn in \cite{EH-K}, see also \cite{Schrader}, which we briefly recall here in a  form suitable for our purpose.\medskip

\begin{definition} \label{def:posPhiprop} Let $\Psi$ be a positive linear map on $\B(\HH)$.\\
i) $\Psi$ is a Schwarz map if
\begin{align}
  \Psi(\Id)=\Id  \ \ \text{and} \ \ \Psi(X^*X)\ge\Psi(X)^*\Psi(X), \ \forall X\in \B(\HH).
\end{align}
ii) $\Psi$ is irreducible if for any nonzero $X, Y\geq 0$ such that $\Tr (XY)=0$, there exists $k\in\N^*$ such
that $\Tr (X \Psi^k(Y))>0$.\\
iii) $\Psi$ is positivity improving if $\Psi(X)>0$ for all nonzero $X\geq 0$.\\
iv) $\Psi$ is primitive if $\Psi^k$ is positivity improving for some $k\in\N^*$.
\end{definition}

We have that $\Psi$ is irreducible if and only if $\Psi^*$ is.
If $\Psi$ is positivity improving, so are all its powers $\Psi^m$, $m\in\N^*$, which are therefore irreducible.
\medskip

Let $\Psi$ be a quantum channel on $\B(\HH)$. Its dual $\Psi^*$ is unital and completely positive, hence $2$-positive, so that $\Psi^*$ is a Schwarz map, and $\Sp(\Psi)=\Sp(\Psi^*)$. If $\Psi^*$ is irreducible, the Evans-H\o egh-Krohn theorems for irreducible Schwarz maps state, among other things, that $1$ is a simple eigenvalue of $\Psi$ \cite[Thm.~2.3]{EH-K}, and that the peripheral spectrum of $\Psi$ is a finite subgroup $\Gamma\subset\T$ consisting of simple eigenvalues. If moreover all the powers of $\Psi^*$ are irreducible, then $\Gamma=\{1\}$ \cite[Thm.~4.4]{EH-K}. For a quantum channel $\Psi$, primitivity is equivalent to $1$ being a simple eigenvalue, with {eigenvector given by a faithful state}, and the
only eigenvalue on the unit circle, and this is also equivalent to the existence of the limit $\Psi^n$ as $n\to \infty$, e.g.\ \cite[Thm.~6.7]{Wolf} . 

Statements of this sort are sometimes called quantum Perron-Frobenius theorems. They allow us to prove 
\begin{theorem}
\label{thm:unique}
Let $0\leq\alpha,\beta<1$ with $\alpha+\beta>0$. If all reset states are faithful, then the invariant state $\varrho$ is unique and faithful. Moreover, 
the peripheral spectrum is trivial, $\Sp(\Phi)\cap\T=\{1\}$, and
$\Phi^n(\sigma)\to\varrho$ as $n\to\infty$ for every initial state $\sigma\in \D(\HH)$, exponentially fast.
\end{theorem}

The proof relies on the following proposition.

\begin{proposition}
\label{prop:primitive}
Let $0\leq\alpha,\beta<1$ with $\alpha+\beta>0$. If all reset states are faithful, then $\Phi^2$ is positivity
improving.
\end{proposition}

\begin{proof}
Let $\sigma\ge0$, $\sigma\neq0$. We show that $\Phi^2(\sigma)>0$, using \eqref{eq:phi}. Since $\sum_iP_i+\sum_jQ_j=\Id$, at least one of the numbers
$\Tr(P_i\sigma)$, $\Tr(Q_j\sigma)$ is strictly positive. Suppose some $\Tr(P_{i_0}\sigma)>0$ (the other
case is symmetric). Then the reset part contains $(1-\alpha)\Tr(P_{i_0}\sigma)\,\tau_{i_0}$, and since
$\tau_{i_0}$ is faithful on $\Ho$,
\[
Q\,\Phi(\sigma)\,Q\ \ge\ (1-\alpha)\Tr(P_{i_0}\sigma)\,\tau_{i_0}\ >\ 0\quad\text{on }\Ho ,
\]
so $\Phi(\sigma)$ is faithful on $\Ho$. Applying $\Phi$ once more, every $\Tr(Q_j\Phi(\sigma))>0$, so the
reset part of $\Phi^2(\sigma)$ contains $(1-\beta)\Tr(Q_j\Phi(\sigma))\,\pi_j$ for every $j$, with each
$\pi_j$ faithful on $\Hs$; hence
\[
P\,\Phi^2(\sigma)\,P\ \ge\ (1-\beta)\sum_j\Tr\big(Q_j\Phi(\sigma)\big)\,\pi_j\ >\ 0\quad\text{on }\Hs .
\]
Moreover $\Phi^2(\sigma)$ is faithful on $\Ho$. If $\beta>0$, this holds since $\beta\,W_Q\big(Q\Phi(\sigma)Q\big)W_Q^*$ is a
conjugation of the strictly positive $Q\Phi(\sigma)Q$ by the unitary $W_Q$. If $\beta=0$, then $\alpha>0$ and $P\Phi(\sigma)P\geq\alpha\,W_PP\sigma PW_P^*\neq0$, so that $\Tr(P_i\Phi(\sigma))>0$ for some $i$, and $Q\Phi^2(\sigma)Q\geq(1-\alpha)\Tr(P_i\Phi(\sigma))\,\tau_i>0$. Being faithful on both $\Hs$
and $\Ho$, and block-diagonal, $\Phi^2(\sigma)$ is strictly positive on $\HH$. Thus $\Phi^2$ is positivity
improving.
\end{proof}

The Theorem then follows by the facts recalled above.

\begin{remarks}
\label{rk:decreasing}
i) The convergence rate is governed by the spectral gap. Set $r_2=\max\{|\lambda|:\lambda\in\Sp(\Phi)
\setminus\{1\}\}<1$ and let $k$ be the size of the largest Jordan block of $\Phi$ at an eigenvalue of modulus $r_2$.
If $r_2>0$, the spectral convergence estimate \cite[Thm.~8.23]{Wolf} gives $C>0$ such that, for every state $\sigma$ and every $n\ge1$, $\trnorm{\Phi^n(\sigma)-\varrho}\le C\,n^{\,k-1}\,r_2^{\,n}$.
Hence, for every $\varepsilon>0$ there is $C_\varepsilon>0$ with $\trnorm{\Phi^n(\sigma)-\varrho}\le C_\varepsilon\,e^{-n(|\ln r_2|-\varepsilon)}$, so that the convergence is exponential at any rate smaller than $|\ln r_2|$. If $r_2=0$, then $\Phi^n(\sigma)=\varrho$ for all $n\geq k$.\\
ii) If $d_P\geq2$, then $\Phi$ is not positivity improving.
Consider a pure state $\sigma=|\varphi_P\rangle\langle\varphi_P|$ supported in $\Hs$. Then
\[
\Phi(\sigma)=\alpha\,|W_P\varphi_P\rangle\langle W_P\varphi_P|
+(1-\alpha)\sum_{i=1}^{N_P}\Tr(P_i\sigma)\,\tau_i ,
\]
where the second term is supported on $\Ho$. Since $d_P\geq2$, there is a nonzero $x\in\Hs$ with
$x\perp W_P\varphi_P$, and $\langle x,\Phi(\sigma)x\rangle=0$, so $\Phi(\sigma)$ is not faithful. This fact does not
depend on the reset states. The same holds if $d_Q\geq2$, with $P$ and $Q$ exchanged.\\
iii) For $\alpha=\beta=0$, $\alpha+\beta-1=-1$ is an eigenvalue of $\Phi$ by Lemma~\ref{lem:Sigma}, so $\Phi$ is not primitive, even when all reset states are faithful.
\end{remarks}

\section{The Rest of the Spectrum}
\label{sec:spectrum}

Having settled the peripheral spectrum, we now look at the eigenvalues lying inside the open unit
disk $\mathbb D$. We compute the full spectrum explicitly when $N_P=1$, and, in terms of the coupling matrix, when the reset states commute with the unitaries. We then perturb the commutative case.

\begin{figure}[H]%
\centering
\includegraphics[width=\textwidth]{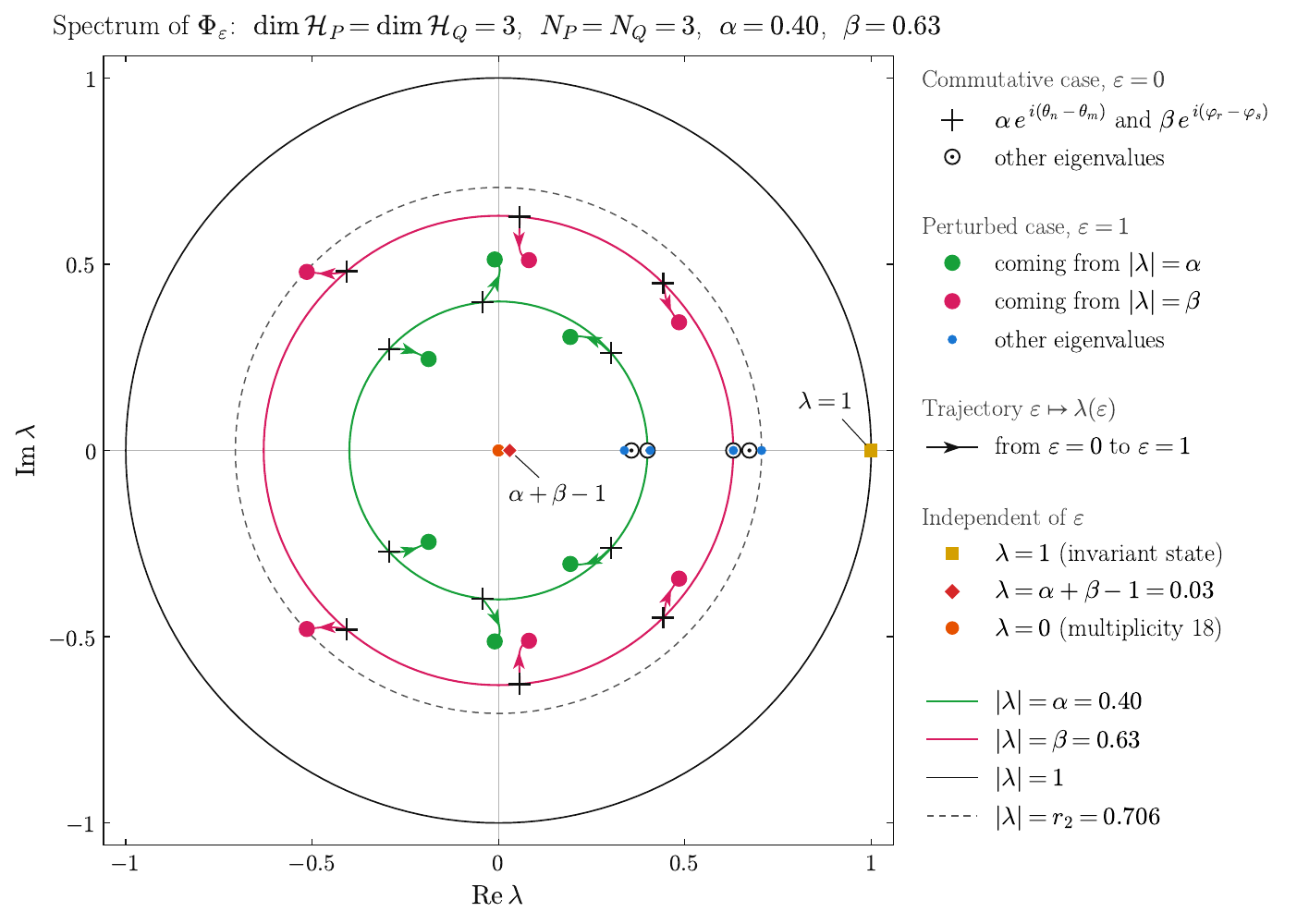}
\caption{Spectrum of $\Phi_\varepsilon$ for $\dim\Hs=\dim\Ho=3$, with $\alpha=0.40$ and $\beta=0.63$. Dots are
the simple eigenvalues, the orange dot is the degenerate eigenvalue $0$, the red diamond is $\alpha+\beta-1$, the yellow square is $\lambda=1$, and the black and dashed circles are $|\lambda|=1$ and $|\lambda|=r_2$.
Black crosses mark the eigenvalues $\alpha\,e^{i(\theta_n-\theta_m)}$ and $\beta\,e^{i(\varphi_r-\varphi_s)}$, of the rescaled conjugations $\alpha\Wcal_P, \beta \Wcal_Q$, where
$\{e^{i\theta_n}\}$ and $\{e^{i\varphi_r}\}$ are the eigenvalues of $W_P$ and $W_Q$; they lie on the green
circle $|\lambda|=\alpha$ and on the magenta circle $|\lambda|=\beta$.
Each curve is the trajectory of an eigenvalue from $\varepsilon=0$ to $\varepsilon=1$. Circled dots and blue dots are the other eigenvalues at $\varepsilon=0$ and $\varepsilon=1$.}
\label{fig:spectre}
\end{figure}

\medskip 
\noindent To illustrate this section, Figure~\ref{fig:spectre} shows the spectrum of a family of RQC depending on a parameter $\varepsilon\in [0,1]$, $\Phi_\varepsilon$ for two subspaces of dimension $3$, with
$\varepsilon-$ dependent reset states. For $\varepsilon=0$, these states commute with $W_P$ and $W_Q$ and the eigenvalues of the conjugations $\alpha\Wcal_P$ and $\beta\Wcal_Q$ on their subspaces belong to $\Sp (\Phi_0)$, see Section \ref{sec:commutative}, whereas this is no longer the case for $\varepsilon>0$, Section~\ref{sec:perturbation}. For $\varepsilon=1$, $r_2$ is as in Remark~\ref{rk:decreasing}~i), and $1-r_2$ is the spectral gap. Note the constant eigenvalues $1$ and $\alpha+\beta-1$. 

\subsection{%
System-Reservoir RQC}\label{sec:np1} 
For $N_P=1$, write $\tau=\tau_1$. Then $\Phi$ is the system-reservoir RQC of Application~\ref{ex:NP1}, that is
\[
\Phi(X)=\alpha\,W_P PX P W_P^*+\beta\,W_Q QX Q W_Q^*+(1-\alpha)\Tr(PX)\,\tau+(1-\beta)\sum_{j=1}^{N_Q}\Tr(Q_jX)\,\pi_j ,
\]
as in \eqref{phisysres} but without the condition $[W_Q,\tau]=0$, and its unique invariant state is \eqref{eq:statenp1}. This also covers the case where all the $\tau_i$ coincide, since then
$\sum_i\Tr(P_iX)\tau_i=\Tr(PX)\tau$. In this case, the spectrum of $\Phi$ is explicit without any commutativity assumption.

\begin{lemma}\label{lem:singlereset}
Let $N_P=1$ and $V_P=\{X\in\B(\Hs):\Tr X=0\}$. Then $V_P$ is invariant under $\Phi$, and
$\Phi|_{V_P}=\alpha\Wcal_P|_{V_P}$. 
\end{lemma}

\begin{proof}
For $X\in V_P$, one has $X=PXP$, so $QXQ=0$, $\Tr(Q_jX)=0$ and $\Tr(PX)=\Tr X=0$. Hence \eqref{eq:phi} gives
$\Phi(X)=\alpha\Wcal_P(X)\in V_P$.
\end{proof}

Let $e^{i\theta_n}$, $1\le n\le d_P$, and $e^{i\varphi_r}$, $1\le r\le d_Q$, be the eigenvalues of $W_P$ and $W_Q$, repeated according to their multiplicities $g_i$ and $h_j$, with phases $\theta_n,\varphi_r\in[0,2\pi)$, and set $m_P=\sum_ig_i^2$, $m_Q=\sum_jh_j^2$. Spectra are understood as multisets: each eigenvalue is repeated according to its algebraic multiplicity, $\{a\}^k$ denotes $a$ repeated $k$ times.

\begin{remark}\label{rk:specWP}
One has $\Sp(\Wcal_P)=\{e^{i(\theta_n-\theta_m)}\}_{1\le n,m\le d_P}$, where $1$ has multiplicity $m_P$. Since $\Wcal_P$ is unitary, $P$ is an eigenoperator
for $1$, and its orthogonal complement in $\B(\Hs)$ is $V_P$, we get
$\Sp(\alpha\Wcal_P|_{V_P})=\{\alpha e^{i(\theta_n-\theta_m)}\}_{\theta_n\neq\theta_m}\cup\{\alpha\}^{m_P-1}$, where $m_P-1\geq1$ if and only if $d_P\ge2$. In the same way, with $V_Q=\{X\in\B(\Ho):\Tr X=0\}$,
$\Sp(\beta\Wcal_Q|_{V_Q})=\{\beta e^{i(\varphi_r-\varphi_s)}\}_{\varphi_r\neq\varphi_s}\cup\{\beta\}^{m_Q-1}$.
\end{remark}

\begin{proposition}\label{lem:singleblock}
Let $N_P=1$. Then
\[
\Sp(\Phi)=\{0\}^{2d_Pd_Q}\cup\{\alpha e^{i(\theta_n-\theta_m)}\}_{\theta_n\neq\theta_m}\cup\{\beta e^{i(\varphi_r-\varphi_s)}\}_{\varphi_r\neq\varphi_s}
\cup\{\alpha\}^{m_P-1}\cup\{\beta\}^{m_Q-1}\cup\{1,\ \alpha+\beta-1\}.
\]
\end{proposition}

\begin{proof}
Let $F=\mathrm{span}(P/d_P,Q/d_Q)$. Every $X\in\B(\HH)$ reads
\[
X=\big(X-\Delta(X)\big)+\Big(PXP-\Tr(PX)\tfrac{P}{d_P}\Big)+\Big(QXQ-\Tr(QX)\tfrac{Q}{d_Q}\Big)+\Tr(PX)\tfrac{P}{d_P}+\Tr(QX)\tfrac{Q}{d_Q},
\]
and the dimensions $2d_Pd_Q$, $d_P^2-1$, $d_Q^2-1$ and $2$ add up to $(d_P+d_Q)^2$. Hence
\[
\B(\HH)=\ran(\id-\Delta)\oplus V_P\oplus V_Q\oplus F .
\]
The first three spaces lie in the kernel of $\Pi(X)=\big(\Tr(PX),\Tr(QX)\big)^{\mathsf T}$, so $\Pi(X)$ gives the
coordinates of the component of $X$ on $F$, in the basis $(P/d_P,Q/d_Q)$. We now compute $\Phi$ on each space.
It vanishes on $\ran(\id-\Delta)$ by \eqref{eq:phidiag}, and $\Phi=\alpha\Wcal_P$ on $V_P$ by
Lemma~\ref{lem:singlereset}. For $X\in V_Q$, one has $X=QXQ$, so $PXP=0$ and $\Tr(PX)=0$, and \eqref{eq:phi} gives
\[
\Phi(X)=\beta\,\Wcal_Q(X)+(1-\beta)\sum_{j=1}^{N_Q}\Tr(Q_jX)\,\pi_j ,
\]
where $\Wcal_Q(X)\in V_Q$, and the second term lies in $\B(\Hs)$ with trace $(1-\beta)\Tr(QX)=0$, hence in $V_P$.
Finally, $\Pi\circ\Phi=\Sigma\circ\Pi$ by Lemma~\ref{lem:Sigma}, so the block of $\Phi$ on $F$ is $\Sigma$. Thus, in the
decomposition $\B(\HH)=\ran(\id-\Delta)\oplus V_P\oplus V_Q\oplus F$,
\[
\Phi=\begin{pmatrix}0&0&0&0\\ 0&\alpha\Wcal_P|_{V_P}&\ast&\ast\\ 0&0&\beta\Wcal_Q|_{V_Q}&\ast\\ 0&0&0&\Sigma\end{pmatrix},
\]
which is block upper triangular. Hence
\[
\Sp(\Phi)=\{0\}^{2d_Pd_Q}\cup\Sp(\alpha\Wcal_P|_{V_P})\cup\Sp(\beta\Wcal_Q|_{V_Q})\cup\Sp(\Sigma),
\]
with $\Sp(\Sigma)=\{1,\alpha+\beta-1\}$ by Lemma~\ref{lem:Sigma}, and Remark~\ref{rk:specWP} gives the result. 
\end{proof}

\begin{remark} i) The case $N_Q=1$
is symmetric.\\
ii) 
Let $0\leq\alpha,\beta<1$, $\lambda=\alpha+\beta-1$, and assume $\lambda\notin\alpha\Sp(\Wcal_P)\cup\beta\Sp(\Wcal_Q)$. For $N_P=1$,
$T_{PQ}(\lambda)$ is the $1\times1$ matrix $s(\lambda)=1$, and the eigenoperator of Lemma~\ref{lem:abeig} reads
\[
\sum_{j=1}^{N_Q}T_Q(\lambda)_{j1}\,(\lambda\id_P-\alpha\Wcal_P)^{-1}(\pi_j)+(\lambda\id_Q-\beta\Wcal_Q)^{-1}(\tau).
\]
Furthermore, if $N_Q=1$ write $\pi=\pi_1$; then $T_Q(\lambda)_{11}=-\frac{1-\beta}{1-\alpha}$, so it is proportional to
$(1-\beta)(\lambda\id_P-\alpha\Wcal_P)^{-1}(\pi)-(1-\alpha)(\lambda\id_Q-\beta\Wcal_Q)^{-1}(\tau)$.
\end{remark}

\begin{remark}\label{rk:nondiag}
This example illustrates Remark~\ref{rk:abzero}. Let $N_P=N_Q=1$, $\alpha\in (0,1/2]\setminus\{1/3\}$ and $\beta=1-2\alpha$, so that $\alpha+\beta-1=-\alpha$. Assume 
$E=\ker(\Wcal_P+\id_P)\neq\{0\}$, and  $\pi$ is not orthogonal to $E$. We show that the
eigenoperators of $\Phi$ for $-\alpha$ are exactly the nonzero elements of $E$.

Every $Y\in E$ satisfies $\Tr Y=-\Tr Y$, so $E\subseteq V_P$, and $\Phi(Y)=\alpha\Wcal_P(Y)=-\alpha Y$ by Lemma~\ref{lem:singlereset}.
Conversely, let $X\neq0$ with $\Phi(X)=-\alpha X$. Since $\Phi$ is trace-preserving, $\Tr X=\Tr\Phi(X)=-\alpha\Tr X$,
so $\Tr X=0$. By \eqref{eq:phi}, the block on $\Hs$ of this equation reads
\[
(-\alpha\id_P-\alpha\Wcal_P)(PXP)=(1-\beta)\Tr(QX)\,\pi .
\]
Take $Y\in E$ with $\HSip{Y}{\pi}\neq0$. Since $\Wcal_P$ is unitary, $\Wcal_P^*(Y)=-Y$, so the adjoint
of $-\alpha\id_P-\alpha\Wcal_P$ annihilates $Y$. Taking the inner product with $Y$ gives
\[
0=\HSip{Y}{(-\alpha\id_P-\alpha\Wcal_P)(PXP)}=(1-\beta)\Tr(QX)\,\HSip{Y}{\pi},
\]
hence $\Tr(QX)=0$, and $\Tr(PX)=0$ since $\Tr X=0$. As $X=-\alpha^{-1}\Phi(X)$ is block-diagonal, the two blocks of \eqref{eq:phi} give $\alpha\Wcal_P(PXP)=-\alpha PXP$ and $\beta\Wcal_Q(QXQ)=-\alpha QXQ$. Since
$-\alpha\notin\beta\Sp(\Wcal_Q)$, $QXQ=0$, and $X=PXP\in E$. 

Thus the geometric multiplicity of $-\alpha$ is $\dim E$, while its algebraic multiplicity is $\dim E+1$ by
Proposition \ref{lem:singleblock}: $\dim E$ from $\Sp(\alpha\Wcal_P|_{V_P})$, plus one from $\alpha+\beta-1$. Hence $\Phi$ is not
diagonalizable.
\end{remark}

\subsection{Spectral Properties in the Commutative Case}
\label{sec:commutative}

When the reset states commute with the unitary part of the channel, the full spectrum can be computed in terms of the coupling matrix. As in the case $N_P=1$, the reason is that $\Phi$ is block triangular, now in the matrix basis associated with the eigenvectors of $W_P$ and $W_Q$.
 
\begin{proposition}
\label{prop:commut}
Let $0\leq\alpha,\beta<1$ and assume $[\tau_i,W_Q]=[\pi_j,W_P]=0$ for all $i,j$. With the notation of Section~\ref{sec:np1}, let $z$ be the algebraic multiplicity of $0$ in $\Sp(T_{PQ})$. Then
\[
\Sp(\Phi)=\{0\}^{2d_Pd_Q}\cup\{\alpha e^{i(\theta_n-\theta_m)}\}_{\theta_n\neq\theta_m}\cup\{\beta e^{i(\varphi_r-\varphi_s)}\}_{\varphi_r\neq\varphi_s}
\cup\{\alpha\}^{m_P-N_P+z}\cup\{\beta\}^{m_Q-N_P+z}\cup\Lambda,
\]
where $\Lambda$ consists of the two roots of $(\lambda-\alpha)(\lambda-\beta)=(1-\alpha)(1-\beta)\mu$, for each eigenvalue $\mu\neq0$ of $T_{PQ}$.
\end{proposition}

\begin{proof}
Let $W_P\, p_n=e^{i\theta_n}p_n$ and $W_Q\, q_r=e^{i\varphi_r}q_r$ so that $\{p_n\}$ and $\{q_r\}$ are orthonormal eigenbases. For all $n,m$,
$\Wcal_P\big(|p_n\rangle\langle p_m|\big)=e^{i(\theta_n-\theta_m)}|p_n\rangle\langle p_m|$. Hence the commutant of $W_P$ in $\B(\Hs)$ is
\[
\Ccal_P=\{X\in\B(\Hs):[X,W_P]=0\}=\Span\{|p_n\rangle\langle p_m|:\theta_n=\theta_m\},
\]
of dimension $m_P$, on which $\Wcal_P=\id_{\Ccal_P}$. Its orthogonal complement $\Ocal_P=\Span\{|p_n\rangle\langle p_m|:\theta_n\neq\theta_m\}$ in $\B(\Hs)$ is invariant under $\Wcal_P$, which is diagonal there with eigenvalues $e^{i(\theta_n-\theta_m)}$, $\theta_n\neq\theta_m$. We define $\Ccal_Q$ and $\Ocal_Q$ in the same way from $W_Q$. With $\Ccal=\Ccal_P\oplus\Ccal_Q$,
\[
\B(\HH)=\Ccal\oplus\Ocal_P\oplus\Ocal_Q\oplus\ran(\id-\Delta).
\]
By \eqref{eq:phidiag}, $\Phi=0$ on $\ran(\id-\Delta)$, of dimension $2d_Pd_Q$. For $X\in\Ocal_P$, \eqref{eq:phi} gives
$\Phi(X)=\alpha\Wcal_P(X)+(1-\alpha)\sum_i\Tr(P_iX)\,\tau_i$, where $\alpha\Wcal_P(X)\in\Ocal_P$, and $\tau_i\in\Ccal_Q$ since
$[\tau_i,W_Q]=0$. Similarly, $\Phi(X)-\beta\Wcal_Q(X)\in\Ccal_P$ for $X\in\Ocal_Q$. Finally, for $A\in\Ccal_P$ and $B\in\Ccal_Q$,
\[
\Phi(A+B)=\Big(\alpha A+(1-\beta)\sum_j\Tr(Q_jB)\,\pi_j\Big)+\Big(\beta B+(1-\alpha)\sum_i\Tr(P_iA)\,\tau_i\Big)\in\Ccal_P\oplus\Ccal_Q .
\]
Hence, with $\Psi=\Phi|_{\Ccal}$, $\Phi$ is block upper triangular in this decomposition:
\[
\Phi=\begin{pmatrix}
\Psi&\ast&\ast&0\\
0&\alpha\Wcal_P|_{\Ocal_P}&0&0\\
0&0&\beta\Wcal_Q|_{\Ocal_Q}&0\\
0&0&0&0
\end{pmatrix},
\]
where the $\ast$ are the reset terms found above. Its characteristic polynomial is therefore
\[
\chi_\Phi(\lambda)=\chi_\Psi(\lambda)\prod_{\theta_n\neq\theta_m}\big(\lambda-\alpha e^{i(\theta_n-\theta_m)}\big)\prod_{\varphi_r\neq\varphi_s}\big(\lambda-\beta e^{i(\varphi_r-\varphi_s)}\big)\,\lambda^{2d_Pd_Q}.
\]

\smallskip
It remains to compute $\chi_\Psi$. Consider the linear maps
\[
a:\left\{\begin{array}{ccl}
\Ccal_P&\to&\C^{N_P}\\
A&\mapsto&\big(\Tr(P_iA)\big)_{i=1}^{N_P}
\end{array}\right.
\qquad
p:\left\{\begin{array}{ccl}
\C^{N_Q}&\to&\Ccal_P\\
y&\mapsto&\sum_{j=1}^{N_Q}y_j\,\pi_j
\end{array}\right.
\]
\[
b:\left\{\begin{array}{ccl}
\Ccal_Q&\to&\C^{N_Q}\\
B&\mapsto&\big(\Tr(Q_jB)\big)_{j=1}^{N_Q}
\end{array}\right.
\qquad
t:\left\{\begin{array}{ccl}
\C^{N_P}&\to&\Ccal_Q\\
x&\mapsto&\sum_{i=1}^{N_P}x_i\,\tau_i
\end{array}\right.
\]
Since $\Wcal_P(\pi_j)=\pi_j$ and $\Wcal_Q(\tau_i)=\tau_i$, Definition~\ref{def:coupling} gives $(T_P)_{ij}=\Tr(P_i\pi_j)$ and
$(T_Q)_{ji}=\Tr(Q_j\tau_i)$, that is, $T_P$ and $T_Q$ are the matrices of $a\circ p$ and $b\circ t$ in the canonical bases. In the decomposition $\Ccal=\Ccal_P\oplus\Ccal_Q$,
\[
\lambda\,\id_{\Ccal}-\Psi=\begin{pmatrix}(\lambda-\alpha)\,\id_{\Ccal_P}&-(1-\beta)\,p\circ b\\ -(1-\alpha)\,t\circ a&(\lambda-\beta)\,\id_{\Ccal_Q}\end{pmatrix}.
\]
The upper left block is invertible, with determinant $(\lambda-\alpha)^{m_P}$, and its Schur complement is
\[
(\lambda-\beta)\,\id_{\Ccal_Q}-\frac{(1-\alpha)(1-\beta)}{\lambda-\alpha}\,t\circ a\circ p\circ b=(\lambda-\beta)\big(\id_{\Ccal_Q}-s(\lambda)\,t\circ a\circ p\circ b\big).
\]
Hence, by the Schur complement formula \cite{HoJ},
\[
\chi_\Psi(\lambda)=(\lambda-\alpha)^{m_P}(\lambda-\beta)^{m_Q}\det\big(\id_{\Ccal_Q}-s(\lambda)\,t\circ a\circ p\circ b\big).
\]
Let $U=s(\lambda)\,t:\C^{N_P}\to\Ccal_Q$ and $V=a\circ p\circ b:\Ccal_Q\to\C^{N_P}$. By Flanders' theorem \cite{F}, $U\circ V$ and $V\circ U$ have
the same nonzero eigenvalues with the same algebraic multiplicities, so $\det(\id_{\Ccal_Q}-U\circ V)=\det(\Id-V\circ U)$. Since the matrix of $V\circ U=s(\lambda)\,(a\circ p)\circ(b\circ t)$ in the canonical basis is $s(\lambda)\,T_PT_Q=s(\lambda)\,T_{PQ}$, this gives
\[
\chi_\Psi(\lambda)=(\lambda-\alpha)^{m_P}(\lambda-\beta)^{m_Q}\det\big(\Id-s(\lambda)\,T_{PQ}\big).
\]
This identity holds for $\lambda\notin\{\alpha,\beta\}$. Since $\chi_\Psi$ is a polynomial and the right-hand side is a rational
function of $\lambda$, the two coincide as rational functions, so the multiplicities of $\alpha$ and $\beta$ can be read
on the right-hand side after simplification.
Let $\mu_1,\dots,\mu_{N_P}$ be the eigenvalues of $T_{PQ}$. The $z$ zero eigenvalues contribute a factor $1$ to $\det(\Id-s(\lambda)T_{PQ})=\prod_k(1-s(\lambda)\mu_k)$, and each
$\mu_k\neq0$ contributes
\[
1-s(\lambda)\mu_k=\frac{(\lambda-\alpha)(\lambda-\beta)-(1-\alpha)(1-\beta)\mu_k}{(\lambda-\alpha)(\lambda-\beta)} .
\]
Hence
\[
\chi_\Psi(\lambda)=(\lambda-\alpha)^{m_P-N_P+z}(\lambda-\beta)^{m_Q-N_P+z}\prod_{\mu_k\neq0}\Big((\lambda-\alpha)(\lambda-\beta)-(1-\alpha)(1-\beta)\mu_k\Big),
\]
where the exponents are nonnegative since $N_P-z\le\mathrm{rk}(T_{PQ})\le\min(m_P,m_Q)$, as $T_{PQ}$ is the matrix of $a\circ p\circ b\circ t$, which factorizes through $\Ccal_P$ and $\Ccal_Q$. Together with the expression of $\chi_\Phi$, this gives the result.
\end{proof}

\begin{remarks} \
i) Since $T_{PQ}$ is column stochastic, $1\in\Sp(T_{PQ})$, and for $\mu=1$ the two roots in $\Lambda$ are $1$ and $\alpha+\beta-1$, see Remark \ref{rem:colsum}.\\
ii) In the commutative case, $T_{PQ}(\lambda)=s(\lambda)T_{PQ}$, so the condition $1\in\Sp(T_{PQ}(\lambda))$ of Theorem \ref{prop:lambdared} reads $\frac{(\lambda-\alpha)(\lambda-\beta)}{(1-\alpha)(1-\beta)}\in\Sp(T_{PQ})$, which gives $\Lambda$. Proposition~\ref{prop:commut} adds the algebraic multiplicities, and the eigenvalues lying in $\alpha\Sp(\Wcal_P)\cup\beta\Sp(\Wcal_Q)$, where the $\lambda$-reduction does not apply.\\
iii) For $N_P=N_Q=1$, $T_{PQ}=(1)$ and $z=0$: one recovers Proposition \ref{lem:singleblock} above in the commutative case. For $d_P=d_Q=1$, moreover $m_P=m_Q=1$, so $\Sp(\Phi)=\{1,\alpha+\beta-1,0,0\}$, and, when $\alpha,\beta>0$, neither $\alpha$ nor $\beta$ is an eigenvalue.
\end{remarks}

\subsection{Perturbation of the Commutative Case}
\label{sec:perturbation}

In the commutative case of Section~\ref{sec:commutative}, most eigenvalues of $\Phi$ are exactly the scaled
conjugation eigenvalues $\alpha\,e^{i(\theta_n-\theta_m)}$ and $\beta\,e^{i(\varphi_r-\varphi_s)}$, so they
sit exactly on the circles of radius $\alpha$ and $\beta$. We now break the commutativity by tilting one reset state in a
direction that does not commute with the corresponding unitary, and we compute their displacement at
first order of perturbation theory. We focus on the modulus-$\alpha$ eigenvalues, coming from $W_P$; the analysis of the
modulus-$\beta$ case is similar.%
 
\medskip
\noindent Throughout, $\Phi_0$ denotes the unperturbed commutative channel of Section~\ref{sec:commutative},
that is the channel \eqref{eq:phi} whose reset states all commute with their unitaries,
$[\pi_j,W_P]=[\tau_i,W_Q]=0$ for all $i,j$. We perturb a single reset state with support in $\Hs$: we fix a state
$\omega\in\D(\Hs)$ that does not commute with $W_P$, and replace
\[
\pi_1\ \longmapsto\ \pi_1^{(\varepsilon)}=\pi_1+\varepsilon\,\Delta\pi_1,\qquad
\Delta\pi_1=\omega-\pi_1,\quad[\omega,W_P]\neq0 ,
\]
for $\varepsilon\in[0,1]$, all the other reset states being left unchanged, hence still commuting with
their unitaries. Note that $\Delta\pi_1$ is a difference of two states, so $\Tr(\Delta\pi_1)=0$; and
$\pi_1^{(\varepsilon)}=(1-\varepsilon)\pi_1+\varepsilon\omega$ is a convex combination of two states, hence
a state, so the perturbed channel $\Phi_\varepsilon$, obtained from $\Phi_0$ by this replacement, is again a quantum channel. This is a rank-one perturbation of $\Phi_0$, as the only change in \eqref{eq:phi} is the replacement of
$\pi_1$ by $\pi_1+\varepsilon\Delta\pi_1$ in the contribution $(1-\beta)\Tr(Q_1X)\pi_1$, so that
\begin{equation}
\Phi_\varepsilon=\Phi_0+\varepsilon\,\Theta,\qquad
\Theta=(1-\beta)\,\dket{\Delta\pi_1}\dbra{Q_1}.
\label{eq:rankone}
\end{equation}
Here $\Theta(X)=(1-\beta)\Tr(Q_1X)\,\Delta\pi_1$ is a rank-one operator on $\B(\HH)$, and since
$\Tr(\Delta\pi_1)=0$ it is trace-annihilating, consistently with $\Phi_\varepsilon$ being trace-preserving.
 
\medskip
\noindent We use the eigenbasis of $W_P$, $W_P|p_n\rangle=e^{i\theta_n}|p_n\rangle$. For a simple eigenvalue
$\lambda_0$ of $\Phi_0$, first-order perturbation theory \cite[Ch.~2, \S2]{K} gives
\begin{equation}
\lambda(\varepsilon)=\lambda_0+\varepsilon\,\frac{\Tr\big(L^*\Theta(R)\big)}{\Tr\big(L^*R\big)}+O(\varepsilon^2),
\label{eq:drift}
\end{equation}
with $R,L$ right and left eigenoperators at $\lambda_0$. 

\medskip
\noindent We consider $\lambda_0=\alpha\,e^{i(\theta_n-\theta_m)}$,
$\theta_n\neq\theta_m$, $\alpha>0$, further assuming that $\lambda_0$ is a simple eigenvalue of $\Phi_0$.

\subsubsection{Left and Right Eigenoperators}
 
Since the reset states of $\Phi_0$ commute with the unitaries, $(1-\alpha)(\id_P-\alpha\Wcal_P)^{-1}(\pi_j)=\pi_j$
and $(1-\beta)(\id_Q-\beta\Wcal_Q)^{-1}(\tau_l)=\tau_l$. Hence the coupling matrices of Definition~\ref{def:coupling}
reduce to the overlaps
\[
(T_P)_{kj}=\Tr(P_k\pi_j),\qquad (T_Q)_{jl}=\Tr(Q_j\tau_l).
\]

\medskip
We record the conditions under which the formulas below are well defined. As in Remark~\ref{rem:colsum}, set
\begin{align}\label{eq:a}
s(\lambda_0)=\frac{(1-\alpha)(1-\beta)}{(\lambda_0-\alpha)(\lambda_0-\beta)}, \ \text{where } \ \lambda_0-\alpha\neq0\ \ \text{and} \  \lambda_0-\beta\neq 0.
\end{align}

\noindent We note that $\det\big(\Id-s(\lambda_0)\,T_P\,T_Q\big)\neq0$, i.e.\ $1$ is not an eigenvalue of $s(\lambda_0)\,T_P\,T_Q$.
If this is not the case, then $(\lambda_0-\alpha)(\lambda_0-\beta)=(1-\alpha)(1-\beta)\mu$ for some eigenvalue $\mu\neq0$ of $T_PT_Q$, so that $\lambda_0\in\Lambda$ and $\lambda_0$ has multiplicity at least two by Proposition~\ref{prop:commut}, a contradiction.

\begin{remark}
Since $\lambda_0\in\alpha\Sp(\Wcal_P)$, Theorem \ref{prop:lambdared} does not apply. Indeed, the eigenoperators
\eqref{eq:rholam} it provides are combinations of resolvents applied to the reset states, so they cannot have components along 
$|p_n\rangle\langle p_m|$, which lies in the kernel of $\lambda_0\id_P-\alpha\Wcal_P$. This is why $R$ is built
directly in Lemma~\ref{lem:LR}.
\end{remark}

\begin{lemma}
\label{lem:LR}
For $\lambda_0=\alpha\,e^{i(\theta_n-\theta_m)}$, $\theta_n\neq\theta_m$, $\alpha>0$, a simple eigenvalue of $\Phi_0$,
$L=|p_n\rangle\langle p_m|$ is a left eigenoperator of $\Phi_0$ associated with $\lambda_0$, and there exist $C\in\C^{N_Q}$ and $K\in\C^{N_P}$ such that
\[
R=|p_n\rangle\langle p_m|+\sum_jC_j\,\pi_j+\sum_lK_l\,\tau_l
\]
is a right eigenoperator of $\Phi_0$ for $\lambda_0$. The vectors $C$ and $K$ are given by \eqref{eq:val} in Appendix~\ref{app:LR}.
\end{lemma}

\begin{proof}
See Appendix \ref{app:LR}.
\end{proof}

\subsubsection{The First-Order Shift}
 
\begin{proposition}
\label{prop:shift} Under the assumptions of Lemma~\ref{lem:LR},
we have $\Tr(L^*R)=1$ and
\begin{equation}
\lambda(\varepsilon)=\lambda_0+\varepsilon\,(1-\beta)(T_Q K)_1\,\langle p_n|\Delta\pi_1
|p_m\rangle+O(\varepsilon^2).
\label{eq:shift}
\end{equation}
\end{proposition}

\begin{proof}
$\Tr(L^*R)=\langle p_n|R|p_m\rangle=1$, since $\sum_jC_j\langle p_n|\pi_j|p_m\rangle=0$ ($[W_P,\pi_j]=0$ and $\theta_n\neq \theta_m$)
and $\sum_lK_l\tau_l$ is supported on $\Ho$. For the numerator, \eqref{eq:rankone} gives $\Theta(R)=(1-\beta)\Tr(Q_1R)\Delta\pi_1$
with $\Tr(Q_1R)=\sum_lK_l\Tr(Q_1\tau_l)=\sum_lK_l(T_Q)_{1l}$, since $|p_n\rangle\langle p_m|$ and the $\pi_j$ are supported on $\Hs$, so that $\Tr(L^*\Theta(R))=(1-\beta)(\sum_lK_l
(T_Q)_{1l})\langle p_n|\Delta\pi_1|p_m\rangle$. Inserting this into \eqref{eq:drift} yields \eqref{eq:shift}.
\end{proof}

\begin{example}\label{ex:shift}
Let $0<\alpha<1$, $0\leq\beta<1$ and $d_P=d_Q=N_P=N_Q=2$, with orthonormal bases $(p_1,p_2)$ of $\Hs$ and $(q_1,q_2)$ of $\Ho$, and take
\[
\begin{aligned}
&W_P=|p_1\rangle\langle p_1|+i\,|p_2\rangle\langle p_2|, &\qquad &W_Q=\Id_Q,\\
&P_1=|u\rangle\langle u|,\quad P_2=\Id_P-P_1,\quad u=\tfrac{1}{\sqrt2}(p_1+p_2), &\qquad &Q_j=|q_j\rangle\langle q_j|,\\
&\pi_1=\pi_2=\tfrac12P, &\qquad &\tau_j=|q_j\rangle\langle q_j|,
\end{aligned}
\]
and perturb $\pi_1$ with $\omega=P_1$. Then $T_Q=\Id$, all entries of $T_P$ equal $\frac12$, and $\lambda_0=-i\alpha$ is a simple eigenvalue of $\Phi_0$ by Proposition~\ref{prop:commut}. A straightforward computation shows that
\[
\lambda(\varepsilon)=-i\alpha-\varepsilon\,\frac{(1-\alpha)(1-\beta)}{4(\beta+i\alpha)}+O(\varepsilon^2),
\]
whose first-order term never vanishes.
\end{example}

\subsubsection{Several Simultaneous Perturbations}

The shift formula extends at once to several perturbations acting at the same time.%
Suppose we perturb several reset states simultaneously,
\[
\pi_i\mapsto\pi_i+\varepsilon\,\Delta\pi_i\quad(i\in I),\qquad
\tau_l\mapsto\tau_l+\varepsilon\,\Delta\tau_l\quad(l\in J),
\]
for some index sets $I,J$. The perturbation of $\Phi$ is the sum of the individual ones,
\[
\Theta=\sum_{i\in I}\Theta^{\pi}_i+\sum_{l\in J}\Theta^{\tau}_l,\qquad
\Theta^{\pi}_i(X)=(1-\beta)\Tr(Q_iX)\,\Delta\pi_i,\quad
\Theta^{\tau}_l(X)=(1-\alpha)\Tr(P_lX)\,\Delta\tau_l .
\]

Since $L$, $R$ and $\Tr(L^*R)=1$ do not involve the perturbation, only the numerator of \eqref{eq:drift} changes.
For $\lambda_0=\alpha\,e^{i(\theta_n-\theta_m)}$, by linearity,
\[
\Tr\big(L^*\Theta(R)\big)=(1-\beta)\sum_{i\in I}\Tr(Q_iR)\,\langle p_n|\Delta\pi_i|p_m\rangle
+(1-\alpha)\sum_{l\in J}\Tr(P_lR)\,\langle p_n|\Delta\tau_l|p_m\rangle .
\]
Since $\Delta\tau_l=Q\,\Delta\tau_l\,Q$ and $Q|p_m\rangle=0$, every term of the second sum vanishes. Exchanging $P$
and $Q$, the perturbations of the $\pi_i$ do not contribute for $\mu_0=\beta\,e^{i(\varphi_r-\varphi_s)}$. With
$\Tr(Q_iR)=(T_QK)_i$, this gives
\begin{equation}
\begin{aligned}
\lambda(\varepsilon)&=\lambda_0+\varepsilon\,(1-\beta)\sum_{i\in I}\big(T_QK\big)_i\,\langle p_n|\Delta\pi_i|p_m\rangle
+O(\varepsilon^2),\\
\mu(\varepsilon)&=\mu_0+\varepsilon\,(1-\alpha)\sum_{l\in J}\big(T_PK'\big)_l\,\langle q_r|\Delta\tau_l|q_s\rangle
+O(\varepsilon^2),
\end{aligned}
\label{eq:multishift}
\end{equation}
where $K'$ is the vector $K$ of Lemma~\ref{lem:LR} with $P$ and $Q$ exchanged, at $\mu_0$, so that the right eigenoperator
at $\mu_0$ reads $|q_r\rangle\langle q_s|+\sum_jK'_j\pi_j+\sum_lC'_l\tau_l$. Thus, at first order, each circle only
feels the reset states supported in the corresponding subspace: %
$\lambda_0$ moves for a generic choice of the $\Delta\pi_i$ as soon as $(T_QK)_i\neq0$ for some $i\in I$, and $\mu_0$ moves for a
generic choice of the $\Delta\tau_l$ as soon as $(T_PK')_l\neq0$ for some $l\in J$.

\section{Limiting Parameter Values}
\label{sec:limits}

For completeness, we consider here the limiting values of $\alpha,\beta$ left out above in the analysis of the invariant states of $\Phi$ given by \eqref{eq:phi}. We keep the notation of Sections~\ref{sec:np1} and~\ref{sec:commutative}. In particular, $\Ccal_P$ and $\Ccal_Q$ denote the commutants of $W_P$ and $W_Q$ in $\B(\Hs)$ and $\B(\Ho)$, of dimensions $m_P$ and $m_Q$, see the proof of Proposition~\ref{prop:commut}.

\medskip
Two kinds of dimensions appear below: $\ker(\Phi-\id)$ is a complex subspace of $\B(\HH)$,
whereas the invariant states form a convex set of self-adjoint operators, whose natural size is the real
dimension of its affine hull \cite[\S2.1]{BV}. The following lemma relates the two.

\begin{lemma}
\label{lem:affdim}
Let $\HH$ be a finite-dimensional Hilbert space and $\Vcal\subseteq\B(\HH)$ a
complex subspace, stable under $X\mapsto X^*$, containing $\Id_{\HH}$, of complex dimension $m$. Then
$\Vcal_h=\{X\in\Vcal:X=X^*\}$ is a real vector space of dimension $m$, and the convex set
$\mathcal S(\Vcal)=\{\varrho\in\Vcal:\varrho\ge0,\ \Tr\varrho=1\}$ has affine hull $\{X\in\Vcal_h:\Tr X=1\}$, of
real dimension $m-1$.
\end{lemma}
 
\begin{proof}
See Appendix \ref{app:affdim}.
\end{proof}

\begin{remarks}
\label{rk:affdim}
i) For $\Vcal=\B(\HH)$, Lemma~\ref{lem:affdim} reduces to the classical fact that the density matrices form a convex
set of real dimension $(\dim\HH)^2-1$, see \cite[\S8.4]{BZ}.\\
ii) The assumption $\Id_{\HH}\in\Vcal$ is only used in the second part of the proof. It provides a positive
definite state $\varrho_0\in\Vcal$, around which one can move in every traceless direction of $\Vcal_h$. The
proof works as well if $\Vcal$ contains some positive definite operator $A$, with $\varrho_0=A/\Tr A$.
\end{remarks}

\begin{proposition}
\label{prop:alpha1}
Let $\alpha=1$ and $0 \le\beta <1$. Then
\[
\ker(\Phi-\id)=\{\varrho_P\oplus0:\ \varrho_P\in\Ccal_P\},
\]
a complex vector space of dimension $m_P$, and the invariant states form the convex set
$\{\varrho_P\oplus0:\ \varrho_P\in\Ccal_P,\ \varrho_P\ge0,\ \Tr\varrho_P=1\}$, whose affine hull has real
dimension $m_P-1$.
\end{proposition}
 
\begin{proof}
When $\alpha=1$ the reset from $\Hs$ to $\Ho$ disappears and
\[
\Phi(\varrho)=W_PP\varrho PW_P^*+\beta\,W_QQ\varrho QW_Q^*+(1-\beta)\sum_i\Tr(Q_i\varrho)\pi_i .
\]
By \eqref{eq:phidiag}, a fixed point is block-diagonal, $\varrho=\varrho_P\oplus\varrho_Q$, and $\Phi(\varrho)=\varrho$ splits into
\[
\varrho_P=W_P\varrho_PW_P^*+(1-\beta)\sum_i\Tr(Q_i\varrho_Q)\pi_i,\qquad \varrho_Q=\beta\,W_Q\varrho_QW_Q^* .
\]
The second equation reads $(\id_Q-\beta\Wcal_Q)(\varrho_Q)=0$. Since $\beta<1$, $\id_Q-\beta\Wcal_Q$ is invertible
by Lemma~\ref{lem:resolvent}, so $\varrho_Q=0$. The first equation then
reduces to $\varrho_P=W_P\varrho_PW_P^*$, i.e.\ $\varrho_P\in\Ccal_P$. Conversely
every $\varrho_P\oplus0$ with $\varrho_P\in\Ccal_P$ is fixed by $\Phi$. This proves the description of
$\ker(\Phi-\id)$, whose dimension is $\dim_\C\Ccal_P=m_P$.

\medskip
The invariant states are the elements $\varrho_P\oplus0$ of $\ker(\Phi-\id)$ with $\varrho_P
\ge0$ and $\Tr\varrho_P=1$, i.e.\ the image of $\mathcal S(\Ccal_P)$ under $\varrho_P\mapsto\varrho_P\oplus0$.
The commutant $\Ccal_P$ is $*$-stable (as $W_P$ is normal, $[X,W_P]=0$ implies $[X^*,W_P]=0$) and contains
$\Id_P$, so Lemma \ref{lem:affdim} applies: the affine hull of
$\mathcal S(\Ccal_P)$ has real dimension $m_P-1$, and so has its image under the affine injection $\varrho_P
\mapsto\varrho_P\oplus0$.
\end{proof}
 
\begin{proposition}
\label{prop:bothone}
Let $\alpha=\beta=1$. Then
\[
\ker(\Phi-\id)=\{\varrho_P\oplus\varrho_Q:\ \varrho_P\in\Ccal_P,\ \varrho_Q\in\Ccal_Q\},
\]
a complex vector space of dimension $m_P+m_Q$, and the invariant states form the convex set
$\{\varrho_P\oplus\varrho_Q:\ \varrho_P\in\Ccal_P,\ \varrho_Q\in\Ccal_Q,\ \varrho_P,\varrho_Q\ge0,\
\Tr\varrho_P+\Tr\varrho_Q=1\}$, whose affine hull has real dimension $m_P+m_Q-1$.
\end{proposition}
 
\begin{proof}
When $\alpha=\beta=1$ both resets disappear and $\Phi(\varrho)=W_PP\varrho PW_P^*+W_QQ\varrho QW_Q^*$.
By \eqref{eq:phidiag}, a fixed point is block-diagonal, $\varrho=\varrho_P\oplus\varrho_Q$, and
$\Phi(\varrho)=\varrho$ splits into the two uncoupled equations
\[
\varrho_P=W_P\varrho_PW_P^*,\qquad \varrho_Q=W_Q\varrho_QW_Q^* ,
\]
i.e.\ $\varrho_P\in\Ccal_P$ and $\varrho_Q\in\Ccal_Q$; conversely every such $\varrho_P\oplus
\varrho_Q$ is fixed by $\Phi$. Hence $\ker(\Phi-\id)=\Ccal_P\oplus\Ccal_Q$, of dimension $m_P+m_Q$.
 
The invariant states are the elements of $\ker(\Phi-\id)$ that are positive with unit trace,
i.e.\ $\mathcal S(\Ccal_P\oplus\Ccal_Q)$. The subspace $\Ccal_P\oplus\Ccal_Q\subseteq\B(\HH)$ is $*$-stable and
contains $\Id_P\oplus\Id_Q=\Id$, so Lemma \ref{lem:affdim} applies with $m=m_P+m_Q$: the affine hull of the
invariant states has real dimension $m_P+m_Q-1$.
\end{proof}

The case $\beta=1$, $0\leq\alpha<1$, is symmetric: $\ker(\Phi-\id)=\{0\oplus\varrho_Q:\ \varrho_Q\in\Ccal_Q\}$, of dimension $m_Q$, and the invariant states are the $0\oplus\varrho_Q$ with $\varrho_Q\in\Ccal_Q$ a state.

\section{Random Unitary Dynamics} \label{sec:randhaar}

So far the unitary dynamics were fixed; we now take one of them random and consider the expectation of the
corresponding random invariant state. In general, computing that invariant state requires the fixed vector of
$T_{PQ}=T_PT_Q$, whose dependence on the unitaries $W_P,W_Q$ is too intricate to get access to the expectation. We therefore work exclusively in the system-reservoir setting, where the invariant state is
explicit and built from the resolvent channels $(1-\gamma)(\id-\gamma\Wcal)^{-1}$. Since the expectation of a
random quantum channel is again a quantum channel (a standard fact; see \cite[Sec.~2]{Schrader}), the
expectation can be carried out on the resolvent directly. \\
We recall the invariant state \eqref{eq:statenp1} in the system-reservoir case, Application \ref{ex:NP1}: with $N_P=1$,
$\tau_1=\tau$, it reads, emphasizing the dependence on the channel in the notation,
\begin{align}
\varrho(\Phi)=\frac{1-\beta}{2-\alpha-\beta}(1-\alpha)(\id_P-\alpha\Wcal_P)^{-1}\Big(\sum_iT_Q^i\pi_i\Big)
+\frac{1-\alpha}{2-\alpha-\beta}(1-\beta)(\id_Q-\beta\Wcal_Q)^{-1}(\tau),
\label{eq:rhorecall}
\end{align}
where $T_Q^i=(1-\beta)\Tr[Q_i(\id_Q-\beta\Wcal_Q)^{-1}(\tau)]$. 
We take the generator $W_P$ of the dynamics of the system random and keep that of the reservoir $W_Q$ fixed, so that only the resolvent on $\Hs$ is affected. More precisely, $W_P=W_P(\omega)$ depends on a random parameter $\omega$ of a probability space $(\Omega,\mathcal F,\Prob)$. We write $\Phi_\omega$ for the corresponding random RQC and $\varrho(\Phi_\omega)$ for its invariant state, given by \eqref{eq:rhorecall} with $W_P(\omega)$ in place of $W_P$; all expectations are taken with respect to $\omega$. This is justified by the fact that we might want the dynamics of the reservoir to commute with $\tau\in \D(\Ho)$, while that of the system is {\it a priori} arbitrary. 
Alongside $\E[\varrho(\Phi_\omega)]$ we also compute the
expectation of the channel $\E[\Phi_\omega]$ and its invariant state $\varrho(\E[\Phi_\omega])$; comparing $\E[\varrho(\Phi_\omega)]$ and $\varrho(\E[\Phi_\omega])$ measures whether averaging and taking the invariant state commute, in which case the simpler channel $\E[\Phi_\omega]$ serves as a
proxy for the random model.

\medskip

We treat two types of randomness: $W$ with fixed deterministic eigenprojectors and independent random eigenvalues, and $W$
Haar distributed.

\subsection{Random Matrix with Deterministic Projectors}
\label{sec:randomdet}

We start with the expectation value of the resolvent in a general abstract framework.
 \medskip

{\bf Hypothesis R1:} The unitary $W=\sum_jR_j\,e^{i\theta_j}\in \B(\HH)$ has fixed orthogonal spectral projectors $R_j$ and random eigenvalues given by i.i.d.\ phases $\theta_j\in [0, 2\pi)$. 

\medskip
We write $\varphi(k)=\E[e^{ik\theta_1}]$ for the common characteristic function of the random phases and define the pinching quantum channel $\Diag_W$ by $\Diag_W(\cdot )=\sum_jR_j\cdot R_j$ on $\B(\HH)$.

\begin{proposition}
\label{prop:avg}
Under {\bf R1}, for every operator $A\in\B(\HH)$ and every $\gamma\in[0,1[$,
\begin{equation}
\E\big[(1-\gamma)(\id-\gamma\Wcal)^{-1}(A)\big]=\sum_jR_jAR_j+\Big((1-\gamma)\sum_{k\ge0}\gamma^k|\varphi(k)|^2\Big)
\sum_{j\neq l}R_jAR_l .
\label{eq:avg}
\end{equation}
For uniformly distributed phases ($\varphi(k)=\delta_{k,0}$) this yields the quantum channel
\begin{equation}
\E\big[(1-\gamma)(\id-\gamma\Wcal)^{-1}(A)\big]=(1-\gamma)\,A+\gamma\,\Diag_W(A).%
\label{eq:avguniform}
\end{equation}
\end{proposition}
 
\begin{proof}
Since $\Wcal(R_jAR_l)=e^{i(\theta_j-\theta_l)}R_jAR_l$,
\[
(1-\gamma)(\id-\gamma\Wcal)^{-1}(A)=(1-\gamma)\sum_{j,l}\frac{R_jAR_l}{1-\gamma\,e^{i(\theta_j-\theta_l)}} .
\]
For $j=l$, %
we have $\sum_jR_jAR_j$. For $j\neq l$, expanding the
geometric series and using independence,
\[
\E\Big[\frac{1}{1-\gamma\,e^{i(\theta_j-\theta_l)}}\Big]
=\sum_{k\ge0}\gamma^k\,\E[e^{ik\theta_j}]\,\E[e^{-ik\theta_l}]=\sum_{k\ge0}\gamma^k|\varphi(k)|^2 ,
\]
which gives \eqref{eq:avg}. For uniform phases only $k=0$ survives, so the off-diagonal coefficient is
$(1-\gamma)$; with $\sum_{j\neq l}R_jAR_l=A-\Diag_W(A)$, \eqref{eq:avg} becomes $(1-\gamma)A+\gamma\Diag_W(A)$.
\end{proof}
\begin{remark}
For uniformly distributed phases, and fixed eigenprojectors, we thus get that the expectation of the resolvent  channel is a convex combination of the identity and the pinching with respect to the eigenprojectors of the unitary $W$.
\end{remark}

Let us apply this to the invariant state \eqref{eq:rhorecall} of the RQC of the system-reservoir setting. We assume the unitary $W_P$ is random, while $W_Q$ is fixed. %
We take the expectation of (the $\Hs$-block of) \eqref{eq:rhorecall}, when $W_P$ satisfies {\bf R1} with uniformly distributed phases. Proposition~\ref{prop:avg} gives, with $\eta=\sum_iT_Q^i \pi_i\in \D(\Hs)$,
\begin{equation}
\E[\varrho(\Phi_\omega)]=\frac{1-\beta}{2-\alpha-\beta}\big[(1-\alpha)\,\eta+\alpha\,\Diag_{W_P}(\eta)\big]
\ + \frac{(1-\alpha)(1-\beta)}{2-\alpha-\beta}\,(\id_Q-\beta\Wcal_Q)^{-1}(\tau).
\label{eq:avgness}
\end{equation}
Note that in case $[W_Q, \tau]=0$, $\eta=\sum_i\Tr(Q_i \tau) \pi_i$ and the $Q$-part of $\E[\varrho(\Phi_\omega)]$ is $\frac{1-\alpha}{2-\alpha-\beta}\tau$.

\medskip
 
We now compute the expectation of the system-reservoir RQC and its invariant state.
 
\begin{proposition}
\label{prop:avgchanphase}
If $W_P$ satisfies {\bf R1} with uniformly distributed phases, the expectation of $\Phi_\omega$ of \eqref{phisysres} is
\begin{equation}
\E[\Phi_\omega](X)=\alpha\,\Diag_{W_P}(PXP)+\beta\,W_QQXQW_Q^*+(1-\alpha)\Tr(PX)\,\tau+(1-\beta)\sum_i\Tr(Q_iX)\,\pi_i.
\label{eq:avgchanphase}
\end{equation}
Furthermore, if $0\leq\alpha,\beta<1$, its invariant state is exactly $\E[\varrho(\Phi_\omega)]$ of \eqref{eq:avgness}: $$\varrho(\E[\Phi_\omega])=\E[\varrho(\Phi_\omega)].$$
\end{proposition}
 
\begin{proof}
Only the first term of $\Phi_\omega$ involves $W_P$. For uniform phases, $\E[e^{i(\theta_j-\theta_l)}]=\delta_{jl}$,
so $$\E[W_P\,PXP\,W_P^*]=\sum_{j,l}\E[e^{i(\theta_j-\theta_l)}]\,R_j(PXP)R_l=\sum_jR_j(PXP)R_j=\Diag_{W_P}(PXP);$$
the other terms are unchanged, giving \eqref{eq:avgchanphase}. 

Since $\E[\Phi_\omega](X)$ is block-diagonal for every $X$, any solution of $\E[\Phi_\omega](X)=X$ reads $X=X_P+X_Q$, with $X_P=PXP$ and $X_Q=QXQ$. 
The projections of the equation on  $\Ho$ and $\Hs$ yield
$$X_Q=\Tr(X_P)(1-\alpha)(\id_Q-\beta\Wcal_Q)^{-1}(\tau), \ \ (\id_P-\alpha \Diag_{W_P})(X_P)=(1-\beta)\sum_{i}\Tr(Q_i X_Q)\pi_i.$$ Inserting  $X_Q$ in the second equation, we have, see below \eqref{eq:rhorecall}, $\Tr(Q_i X_Q)=\Tr(X_P)\frac{1-\alpha}{1-\beta} T_Q^i$, so that $\sum_{i}\Tr(Q_i X_Q)\pi_i=\Tr(X_P)\frac{1-\alpha}{1-\beta}\eta$. Now, the pinching channel $\Diag_{W_P}$ is a projector, so $\id_P-\alpha \Diag_{W_P}$ is invertible with inverse $(\id_P-\alpha \Diag_{W_P})^{-1}=\id_P+\frac{\alpha}{1-\alpha}\Diag_{W_P}$. Altogether we have the unique solution, up to the constant $\Tr(X_P)$
$$ 
X=\Tr(X_P)\big[ (1-\alpha)(\id_P+\frac{\alpha}{1-\alpha}\Diag_{W_P})(\eta)+(1-\alpha)(\id_Q-\beta\Wcal_Q)^{-1}(\tau)\big].
$$
Choosing $\Tr(X_P)=\frac{1-\beta}{2-\alpha-\beta}$, see \eqref{eq:traceval}, we get that $\E[\varrho(\Phi_\omega)]$ is the unique invariant state of $\E[\Phi_\omega]$. 
\end{proof}

\subsection{Haar Random Unitaries}
\label{sec:haar}

We now consider the expectation value of the resolvent when $W$ is Haar distributed, in the same abstract framework.
 \medskip

{\bf Hypothesis R2:} The unitary $W\in \B(\HH)$ is Haar distributed on the unitary group $\Ucal_d(\HH)$, where $d=\dim\HH$.

\medskip
For $n\ge0$, let,
\begin{equation}
\Ecal_n(A)=\E\big[\Wcal^n(A)\big]=\E\big[W^n A\,(W^*)^n\big].
\label{eq:Ecalm}
\end{equation}
 
\begin{proposition}
\label{prop:haaravg}
Under {\bf R2}, for every $A\in\B(\HH)$ and every $n>0$,
\begin{equation}
\Ecal_n(A)=\frac{\Tr A}{d}\,\Id+m(n)\Big(A-\frac{\Tr A}{d}\,\Id\Big),
\qquad m(n)=\frac{\min(n,d)-1}{d^2-1}.
\label{eq:mm}
\end{equation}
In particular $m(1)=0$ and $m(n)=\tfrac{1}{d+1}$ for $n\ge d$.
\end{proposition}
We extend the formula to $n=0$ by setting $m(0)=1$.\\

To prove this result on the expectation of the random dynamics implemented by a Haar distributed unitary, we use the following form of Schur's lemma, see \cite{Meckes}, Lemma 1.10, adapted to our setting.
 
\begin{lemma}[Schur]
\label{lem:schur}
Let $\HH$ be a finite-dimensional complex Hilbert space and $\mathcal G\subseteq\B(\HH)$ act
irreducibly on $\HH$, that is, $\mathcal G$ leaves no subspace other than $\{0\}$ and $\HH$ invariant. If
$\varphi\in\B(\HH)$ commutes with every element of $\mathcal G$, then $\varphi=\lambda\,\Id$ for some
scalar $\lambda\in\C$.
\end{lemma}
 
\begin{proof}[Proof of Proposition~\ref{prop:haaravg}]
The Haar measure on $\Ucal_d(\HH)$ is bi-invariant: for
fixed unitaries $U,V$, the matrix $UWV$ has the same law as $W$ \cite{Meckes}. %
Fix a unitary $V$ and compute, inserting
$VV^*=V^*V=\Id$ between the powers, 
\begin{align*}
\Ecal_n(VAV^*)&=\E\big[W^n\,VAV^*\,(W^*)^n\big]
=\E\big[V\,(V^*W V)^n\,A\,(V^*W^*V)^n\,V^*\big]\\
&=V\,\E\big[(V^*WV)^n\,A\,\big((V^*WV)^*\big)^n\big]\,V^* .
\end{align*}
Since $V^*WV$ has the same law as $W$, the expectation on the right equals
$\E[W^nA(W^*)^n]=\Ecal_n(A)$. Hence
\[
\Ecal_n(VAV^*)=V\,\Ecal_n(A)\,V^*\qquad\text{for any unitary }V,
\]
that is, $\Ecal_n$ commutes with every conjugation $\Wcal_V:A\mapsto VAV^*$.

Under the conjugation $\Wcal_V$ by  $V\in \Ucal_d(\HH)$, the space
$\B(\HH)$ splits into two invariant subspaces,
\[
\B(\HH)=\C\,\Id\ \oplus\ \B_0,\qquad \B_0=\{A:\Tr A=0\},
\]
and both are irreducible \cite[\S15.3, \S15.5]{FultonHarris} under $\{\Wcal_V\}_{V\in\Ucal_d(\HH)}$. Because $\Ecal_n$
commutes with every $\Wcal_V$, it preserves each subspace, and its restriction to each commutes
with the irreducible family $\{\Wcal_V\}$ acting there. By Schur's Lemma~\ref{lem:schur}, $\Ecal_n$ acts as a scalar on
each. On $\C\,\Id$ the scalar is $1$, since $\Ecal_n(\Id)=\E[W^n\Id(W^*)^n]=\Id$; call $m(n)$ the scalar
on $\B_0$. Writing $A=\tfrac{\Tr A}{d}\Id+\big(A-\tfrac{\Tr A}{d}\Id\big)$, with the first term in
$\C\,\Id$ and the second in $\B_0$, gives the first part of \eqref{eq:mm}.

We compute the scalar $m(n)$ by taking the trace of $\Ecal_n$, viewed as a
linear map on the $d^2$-dimensional Hilbert space $\B(\HH)$ with Hilbert-Schmidt inner product. From the block form \eqref{eq:mm}, $\Ecal_n$ equals
$1$ on the $1$-dimensional space $\C\,\Id$ and $m(n)$ on the $(d^2-1)$-dimensional space $\B_0$, so
\[
\Tr\Ecal_n=1\cdot1+m(n)\,(d^2-1).
\]
On the other hand, for a fixed matrix $M$ the conjugation $A\mapsto MAM^*$ has trace $|\Tr M|^2$; indeed,
fixing an orthonormal basis $\{e_i\}_{i=1}^d$ of $\HH$, the operators $\{|e_i\rangle\langle e_j|\}_{i,j}$ form an
orthonormal basis of $\B(\HH)$, and
$\Tr(A\mapsto MAM^*)=\sum_{i,j}\langle e_i,Me_i\rangle\overline{\langle e_j,Me_j\rangle}=|\Tr M|^2$. Taking
$M=W^n$, we get the expectation, 
\[
\Tr\Ecal_n=\E\,|\Tr(W^n)|^2 .
\]
By the exact moment formula on $\Ucal_d(\HH)$, which goes back to Diaconis--Shahshahani \cite{DS} and is stated in
\cite[Prop.~3.11]{Meckes}, $\E\,|\Tr(W^n)|^2=\min(n,d)$ for every $n\ge1$. Equating the two
expressions for $\Tr\Ecal_n$ gives $1+m(n)(d^2-1)=\min(n,d)$, i.e.\ \eqref{eq:mm}.
\end{proof}
 
We now compute the expectation of the resolvent channel.

\begin{proposition}
\label{prop:f}
Under {\bf R2}, for every $A\in\B(\HH)$ and every $\gamma\in[0,1[$, for $W$ unitary and Haar distributed,
\begin{equation}
\E\big[(1-\gamma)(\id-\gamma\Wcal)^{-1}(A)\big]=\frac{\Tr A}{d}\,\Id+f(\gamma,d)\,\Big(A-\frac{\Tr A}{d}\,\Id\Big),
\label{eq:fdef}
\end{equation}
where
\begin{equation}
f(\gamma,d)=1-\gamma+\frac{\gamma^2\,(1-\gamma^{\,d-1})}{(1-\gamma)\,(d^2-1)} .
\label{eq:fclosed}
\end{equation}
The function $\gamma\mapsto f(\gamma,d)$ is strictly decreasing and convex on $[0,1[$, with $f(0,d)=1$,
$f(\gamma,d)\to\tfrac1{d+1}$ as $\gamma\to1$, and $f(\gamma,d)-(1-\gamma)=O(1/d^2)$ as $d\to\infty$.
\end{proposition}
\begin{remark}
For Haar distributed unitaries, the expectation of the resolvent channel is the depolarizing channel with parameter $f(\gamma,d)$. 
\end{remark}

\begin{proof}
The scalar part is invariant since $\Wcal(\Id)=\Id$ gives $(1-\gamma)(\id-\gamma\Wcal)^{-1}(\Id)=\Id$.
For the traceless part $A_0=A-\tfrac{\Tr A}{d}\Id$,
Proposition~\ref{prop:haaravg} gives $\Ecal_n(A_0)=m(n)\,A_0$ for all $n\ge0$. Since $\gamma<1$ and
$\|\Wcal^n(A_0)\|=\|A_0\|$, the Neumann series converges uniformly in $W$, and taking expectation term by term yields
\[
\E\big[(1-\gamma)(\id-\gamma\Wcal)^{-1}(A_0)\big]=\Big((1-\gamma)\sum_{n\ge0}\gamma^n\,m(n)\Big)A_0 ,
\]
which, combined with the scalar part above, gives \eqref{eq:fdef} with $f(\gamma,d)=(1-\gamma)\sum_{n\ge0}\gamma^nm(n)$. Inserting
$m(n)=\frac{\min(n,d)-1}{d^2-1}$ (with $m(0)=1$) and summing the resulting elementary geometric series gives
the closed form \eqref{eq:fclosed}. 
For the stated properties, we use the equivalent finite form of \eqref{eq:fclosed}, obtained from
$\frac{\gamma^2(1-\gamma^{d-1})}{1-\gamma}=\sum_{k=2}^d\gamma^k$:
\begin{equation}
f(\gamma,d)=1-\gamma+\frac1{d^2-1}\sum_{k=2}^d\gamma^k .
\label{eq:ffinite}
\end{equation}
Differentiating \eqref{eq:ffinite} twice,
\[
f'(\gamma,d)=-1+\frac1{d^2-1}\sum_{k=2}^d k\,\gamma^{k-1},\qquad
f''(\gamma,d)=\frac1{d^2-1}\sum_{k=2}^d k(k-1)\,\gamma^{k-2}\ge0,
\]
so $f$ is convex. On $[0,1[$, bounding $\sum_{k=2}^dk\gamma^{k-1}<\sum_{k=2}^dk=\frac{(d-1)(d+2)}{2}$ gives
\[
f'(\gamma,d)<-1+\frac{d+2}{2(d+1)}=-\frac{d}{2(d+1)}<0,
\]
so $f$ is strictly decreasing. 
Finally $f(0,d)=1$; as $\gamma\to1$ the sum in \eqref{eq:ffinite} tends to
$\frac{d-1}{d^2-1}=\frac1{d+1}$. Hence, for fixed $\gamma$,  $f(\gamma,d)-(1-\gamma)=O(1/d^2)$ as
$d\to\infty$.
\end{proof}

Taking the expectation of the invariant state when $W_P$ satisfies {\bf R2}, by applying Proposition~\ref{prop:f} with $\gamma=\alpha$, $d=d_P$ and $A=\eta=\sum_iT_Q^i \pi_i\in \D(\Hs)$, we get
\begin{equation}
\E[\varrho(\Phi_\omega)]=\frac{1-\beta}{2-\alpha-\beta}\Big[\frac{\Id_P}{d_P}+f(\alpha,d_P)\,\Big(\eta-\frac{\Id_P}{d_P}\Big)\Big]
 + \frac{(1-\alpha)(1-\beta)}{2-\alpha-\beta}\,(\id_Q-\beta\Wcal_Q)^{-1}(\tau).
\label{eq:avgnessHaar}
\end{equation}
Hence, taking the expectation of the invariant state over Haar distributed dynamics in $\Hs$ yields a reservoir dependent depolarized state.

\begin{proposition}
\label{prop:avgchanHaar}
If $W_P$ satisfies {\bf R2}, the expectation of $\Phi_\omega$ of \eqref{phisysres} is
\begin{equation}
\E[\Phi_\omega](X)=\alpha\,\Tr(PX)\,\frac{\Id_P}{d_P}+\beta\,W_QQXQW_Q^*+(1-\alpha)\Tr(PX)\,\tau+(1-\beta)\sum_i
\Tr(Q_iX)\,\pi_i ,
\label{eq:avgchanHaar}
\end{equation}
and its unique invariant state reads %
\begin{equation}
\varrho(\E[\Phi_\omega])=\frac{1-\beta}{2-\alpha-\beta}\Big[\frac{\Id_P}{d_P}+(1-\alpha)\,\Big(\eta-\frac{\Id_P}{d_P}\Big)\Big]+ \frac{(1-\alpha)(1-\beta)}{2-\alpha-\beta}\,(\id_Q-\beta\Wcal_Q)^{-1}(\tau).
\label{eq:nessHaar}
\end{equation}
\end{proposition}

\begin{proof}
Taking the expectation of \eqref{phisysres} over $W_P$, where only the first term depends on $W_P$, we have thanks to %
Proposition~\ref{prop:haaravg} at $n=1$ %
\[
\E[\alpha W_P(PXP)W_P^*]=\alpha\,\frac{\Tr(PX)\,\Id_P}{d_P}.
\]
The other terms are unchanged, giving \eqref{eq:avgchanHaar}. The invariant state is computed as in the proof of Proposition \ref{prop:avgchanphase}. Indeed, both $\Diag_{W_P}(\cdot)$ and $\Tr(\cdot P)\Id_P/d_P$ are projectors, so that the same algebraic considerations yield uniqueness of the invariant state and the form of its $\Hs$-block which reads here 
$$\varrho(\E[\Phi_\omega])_P=\frac{1-\beta}{2-\alpha-\beta}\Big[(1-\alpha)\,\eta+\alpha\,\Tr(\eta P)\frac{\Id_P}{d_P}\Big]=\frac{1-\beta}{2-\alpha-\beta}\Big[\frac{\Id_P}{d_P}+(1-\alpha)\,\Big(\eta-\frac{\Id_P}{d_P}\Big)\Big].$$

This yields \eqref{eq:nessHaar}.
\end{proof}
The two states $\E[\varrho(\Phi_\omega)]$ and $\varrho(\E[\Phi_\omega])$ differ only on the $\Hs$-block, through the coefficient of $\eta-\frac{\Id_P}{d_P}$. Therefore we immediately get from Proposition \ref{prop:f}
\begin{corollary}
\label{cor:approx} Under the assumptions of Proposition~\ref{prop:avgchanHaar},
\begin{align}
\big\|\E[\varrho(\Phi_\omega)]-\varrho(\E[\Phi_\omega])\big\|_1
&=\frac{1-\beta}{2-\alpha-\beta}\,\frac{\alpha^2(1-\alpha^{\,d_P-1})}{(1-\alpha)(d_P^2-1)}\,\|\eta-\Id_P/d_P\|_1
=O\!\left(\frac1{d_P^2}\right),\ \ d_P\to\infty.
\label{eq:diff}
\end{align}
\end{corollary}
Therefore, in the limit of large dimensions of $\Hs$, the expectation of the invariant state over Haar distributed random dynamics in $\Hs$, coincides with the invariant state of the expected channel, which replaces a unitary dynamics by a reset to the maximally mixed state.

\section{Generalization to \texorpdfstring{$r$}{r} Reservoirs}
\label{sec:rreservoir}

We consider here the generalization of the two-subspace model of the previous sections to a Hilbert space structured into the direct sum of $r\ge2$ subspaces. All the structural results carry over, and several of them are in fact clearer in the general setting, where the role of each ingredient is isolated.

\subsection{The Channel and its Invariant State}
\label{sec:rsetting}

\noindent{\bf Data.}
Let $\HH=\bigoplus_{k=1}^r\HH_k$ be an orthogonal decomposition into $r$ subspaces, with $d_k=\dim\HH_k$. We assume the existence
of:\\
i) a unitary block-diagonal operator $W=\bigoplus_{k=1}^r W_k$, each $W_k:=W|_{\HH_k}$ unitary on
$\HH_k$;\\
ii) for each $1\le k\le r$, a family of nonzero mutually orthogonal projectors $\{P^{(k)}_i\}_{i=1}^{N_k}$,
$N_k\ge1$, with
\[
P^{(k)}=\sum_{i=1}^{N_k}P^{(k)}_i,\qquad \sum_{k=1}^r P^{(k)}=\Id,
\]
where $P^{(k)}$ is the orthogonal projection onto $\HH_k$, $P^{(k)}_iP^{(\ell)}_j=\delta_{ij}\delta_{k\ell}P^{(k)}_i$; %
iii) for every $1\leq \ell\neq  k\leq r$, and each $1\le i\le
N_\ell$, a state $\pi^{(k \ell)}_i\in\D(\HH_k)$;\\
iv) for each $1\le k\le r$, a weight $\alpha_k\in[0,1]$; \\
v) a column stochastic matrix $\Gamma=(\gamma_{k\ell})_{k,\ell=1}^r$ with $\gamma_{kk}=0$.%

In the data above, the indices $\ell$ and $k$ of the state $\pi^{(k \ell)}_i$ label the source and target subspaces, respectively, while the matrix elements $\gamma_{k\ell}$ denote the routing probability from subspace $\ell$
to subspace $k$.
 
The data is gathered in the notation
$\big\{\{W_k\},\{P^{(k)}_i\},\{\pi^{(k\ell)}_i\},\{\alpha_k\},\{\gamma_{k\ell}\}\big\}$.
\begin{definition} The $r$-subspace RQC associated with the data above acts on $X\in \B(\HH)$ as
    \begin{equation}
\Phi(X)=\sum_{\ell=1}^r\alpha_\ell\,W_\ell P^{(\ell)}X P^{(\ell)}W_\ell^*
+\sum_{\ell=1}^r(1-\alpha_\ell)\sum_{k\neq\ell}\gamma_{k\ell}\sum_{i=1}^{N_\ell}
\Tr\big(P^{(\ell)}_iX\big)\,\pi^{(k\ell)}_i .
\label{eq:phir}
\end{equation}
\end{definition}

\noindent The dynamics is coherently implemented within the subspaces $\HH_k$ by
the unitaries $W_k$. For $r=2$ we recover \eqref{eq:phi}, with $\HH_1=\Hs$, $\HH_2=\Ho$, $W_1=W_P$, $W_2=W_Q$,
$\alpha_1=\alpha$, $\alpha_2=\beta$, the projectors $P^{(1)}_i=P_i$ ($N_1=N_P$) and $P^{(2)}_j=Q_j$ ($N_2=N_Q$),
the only possible weights $\gamma_{21}=\gamma_{12}=1$, and the reset states
\[
\pi^{(21)}_i=\tau_i\in\D(\Ho),\quad 1\le i\le N_P,\qquad\qquad
\pi^{(12)}_j=\pi_j\in\D(\Hs),\quad 1\le j\le N_Q .
\]
As in Remark \ref{rem:povm}, the projectors $P^{(k)}_i$ can be replaced by nonzero operators $P^{(k)}_i\geq 0$ such that $\sum_{i=1}^{N_k}P^{(k)}_i=P^{(k)}$ for all $1\leq k\leq r$.

\begin{proposition}
\label{prop:rcptp} For any choice of data $\big\{\{W_k\},\{P^{(k)}_i\},\{\pi^{(k\ell)}_i\},\{\alpha_k\},\{\gamma_{k\ell}\}\big\}$, the RQC 
$\Phi$ in \eqref{eq:phir} is a quantum channel.
\end{proposition}
 
\begin{proof}
Complete positivity follows as in the two-subspace case (Proposition~\ref{prop:cptp}), %
while trace preservation uses $\Tr(\pi^{(k\ell)}_i)=1$, $\sum_i\Tr(P^{(\ell)}_iX)=\Tr(P^{(\ell)}X)$
and $\sum_{k\neq\ell}\gamma_{k\ell}=1$:
\[
\Tr\Phi(X)=\sum_\ell\alpha_\ell\Tr(P^{(\ell)}X)+\sum_\ell(1-\alpha_\ell)\Tr(P^{(\ell)}X)
=\sum_\ell\Tr(P^{(\ell)}X)=\Tr(X).\qedhere
\]
\end{proof}

We now determine the fixed points of $\Phi$ for $0\le\alpha_k<1$, following Section~\ref{sec:fixed}. As for $r=2$,
every fixed point $X$ is block-diagonal, since no term of \eqref{eq:phir} produces off-diagonal blocks. We write
$X=\bigoplus_kX_k$ with $X_k=P^{(k)}XP^{(k)}$, $\Wcal_k(X)=W_kXW_k^*$ for the conjugation on $\B(\HH_k)$, $\id_k$ for
the identity superoperator on $\B(\HH_k)$, and $X^{(k)}_i=\Tr(P^{(k)}_iX)$ for the populations of block $k$. The
equation $\Phi(X)=X$ restricted to $\HH_k$ reads
\[
(\id_k-\alpha_k\Wcal_k)(X_k)=\sum_{\substack{\ell=1\\ \ell\neq k}}^r(1-\alpha_\ell)\,\gamma_{k\ell}\sum_{j=1}^{N_\ell}
X^{(\ell)}_j\,\pi^{(k\ell)}_j .
\]
Since $\alpha_k<1$, $\id_k-\alpha_k\Wcal_k$ is invertible, and
\begin{equation}
X_k=\sum_{\substack{\ell=1\\ \ell\neq k}}^r(1-\alpha_\ell)\,\gamma_{k\ell}\sum_{j=1}^{N_\ell}X^{(\ell)}_j\,
(\id_k-\alpha_k\Wcal_k)^{-1}(\pi^{(k\ell)}_j).
\label{eq:rblockrho}
\end{equation}
As in Definition~\ref{def:coupling}, we introduce coupling matrices.

\begin{definition}\label{def:couplingr}
For $0\le\alpha_k<1$, $1\le k\le r$, and for $1\le k\neq\ell\le r$, the coupling matrices
$T^{(k\ell)}\in M_{N_k,N_\ell}(\R^+)$ are given by
\begin{align}\label{eq:couplingr}
T^{(k\ell)}_{ij}=\Tr\big[P^{(k)}_i\,(1-\alpha_k)(\id_k-\alpha_k\Wcal_k)^{-1}(\pi^{(k\ell)}_j)\big],\ \ 1\leq i\leq N_k,\ 1\leq j\leq N_\ell.
\end{align}
\end{definition}
As for $r=2$, they are column stochastic, since $(1-\alpha_k)(\id_k-\alpha_k\Wcal_k)^{-1}(\pi^{(k\ell)}_j)$ is a state on
$\HH_k$ (Corollary~\ref{cor:tracechannel}) and $\sum_{i=1}^{N_k}P^{(k)}_i=P^{(k)}$.

\medskip
Let $N=\sum_kN_k$, and for each $k$ let $X^{(k)}=(X^{(k)}_i)_{1\le i\le N_k}\in\C^{N_k}$ be the population
vector of the block $k$; we write $\tilde X=(X^{(1)},\dots,X^{(r)})^{\mathsf T}\in\C^N$. Tracing
\eqref{eq:rblockrho} against $P^{(k)}_i$ gives
\[
X^{(k)}_i=\sum_{\ell\neq k}\frac{1-\alpha_\ell}{1-\alpha_k}\,\gamma_{k\ell}\sum_{j=1}^{N_\ell}
T^{(k\ell)}_{ij}\,X^{(\ell)}_j ,\qquad 1\le k\le r,\ 1\le i\le N_k ,
\]
that is, the finite linear system $\tilde X=M\tilde X$, where $M$ is the $N\times N$ block matrix
\begin{equation}
M=\begin{pmatrix}
0 & M^{(12)} & \cdots & M^{(1r)}\\
M^{(21)} & 0 & \cdots & M^{(2r)}\\
\vdots & \vdots & \ddots & \vdots\\
M^{(r1)} & M^{(r2)} & \cdots & 0
\end{pmatrix},
\qquad
M^{(k\ell)}_{ij}=\frac{1-\alpha_\ell}{1-\alpha_k}\,\gamma_{k\ell}\,T^{(k\ell)}_{ij},
\label{eq:MT}
\end{equation}
the block $M^{(k\ell)}$ being of size $N_k\times N_\ell$, with $1\le i\le N_k$, $1\le j\le N_\ell$.

\medskip
While it is not stochastic in general, the matrix $M$ is similar to a stochastic matrix:

\begin{lemma}
\label{lem:rcolsum}
Let $D$ be the positive diagonal matrix $D=\operatorname{diag}\big((1-\alpha_k)\,\Id_{N_k}\big)_{1\le
k\le r}$. Then the matrix $\widetilde M=DMD^{-1}$ is column stochastic, so
$1\in\Sp(\widetilde M)=\Sp(M)$, and $M$ admits a nonnegative eigenvector $\tilde \varrho\in (\R^+)^N$.
\end{lemma}
 
\begin{proof}
Since $D$ acts blockwise, $\widetilde M^{(kk)}=0$ and, for $k\neq\ell$,
$\widetilde M^{(k\ell)}=(1-\alpha_k)\,M^{(k\ell)}\,(1-\alpha_\ell)^{-1}=\gamma_{k\ell}\,T^{(k\ell)}$, which has nonnegative
entries. By column stochasticity of $T^{(k\ell)}$ and of $\Gamma$, the column of $\widetilde M$ indexed by $(\ell,j)$
sums to
\[
\sum_{\substack{k=1\\ k\neq\ell}}^r\gamma_{k\ell}\sum_{i=1}^{N_k}T^{(k\ell)}_{ij}=\sum_{\substack{k=1\\ k\neq\ell}}^r\gamma_{k\ell}=1 .
\]
By Perron--Frobenius, $\widetilde M$ has a nonnegative fixed vector $Y\in(\R^+)^N$, and $\tilde X=D^{-1}Y$ is a
nonnegative fixed vector of $M$.
\end{proof}

The invariant states of $\Phi$ are thus obtained from the nonnegative fixed vectors of $M$.

\begin{theorem}
\label{thm:rrhoinfty}
(i) Let $\tilde\varrho=(\tilde\varrho^{(k)}_i)\in(\R^+)^N$ solve $M\tilde\varrho=\tilde\varrho$ with
$\sum_{k=1}^r\sum_{i=1}^{N_k}\tilde\varrho^{(k)}_i=1$. Then
\begin{equation}
\varrho=\sum_{k=1}^r\sum_{\substack{\ell=1\\ \ell\neq k}}^r(1-\alpha_\ell)\,\gamma_{k\ell}\sum_{j=1}^{N_\ell}
\tilde\varrho^{(\ell)}_j\,(\id_k-\alpha_k\Wcal_k)^{-1}(\pi^{(k\ell)}_j)
\label{eq:rrhoinfty}
\end{equation}
is an invariant state of the RQC $\Phi$ \eqref{eq:phir}.
(ii) For every invariant state $\varrho$, the block populations $t_k=\Tr(P^{(k)}\varrho)$
satisfy $s=\Gamma s$, where $s_k=(1-\alpha_k)\,t_k$. If $1$ is a simple eigenvalue of $\Gamma$, then $t_k\propto s_k/(1-\alpha_k)$, with $s$ the Perron vector of $\Gamma$ and $\sum_{k=1}^rt_k=1$.
\end{theorem}
 
\begin{proof}
(i) Positivity follows from the fact that $(1-\alpha_k)(\id_k-\alpha_k\Wcal_k)^{-1}$ is a quantum channel
(Corollary \ref{cor:tracechannel}) and from the nonnegativity of the weights. By definition of $T^{(k\ell)}$ and
\eqref{eq:MT}, the populations of $\varrho$ are
\begin{equation}
\label{eq:rpop}
\Tr\big(P^{(k)}_i\varrho\big)=\sum_{\substack{\ell=1\\ \ell\neq k}}^r\frac{1-\alpha_\ell}{1-\alpha_k}\,\gamma_{k\ell}\sum_{j=1}^{N_\ell}
T^{(k\ell)}_{ij}\,\tilde\varrho^{(\ell)}_j=(M\tilde\varrho)_{(k,i)}=\tilde\varrho^{(k)}_i ,
\end{equation}
the last equality being $M\tilde\varrho=\tilde\varrho$. Summing over $i$ and $k$, and using
$\sum_{i=1}^{N_k}P^{(k)}_i=P^{(k)}$, $\sum_{k=1}^rP^{(k)}=\Id$, we get
\[
\Tr(\varrho)=\sum_{k=1}^r\sum_{i=1}^{N_k}\Tr\big(P^{(k)}_i\varrho\big)
=\sum_{k=1}^r\sum_{i=1}^{N_k}(M\tilde\varrho)_{(k,i)}=\sum_{k=1}^r\sum_{i=1}^{N_k}\tilde\varrho^{(k)}_i=1 .
\]
Invariance is then checked block by block as in
the case $r=2$: the component of $\Phi(\varrho)$ on $\HH_k$ is $\alpha_k\Wcal_k(\varrho_k)+
(\id_k-\alpha_k\Wcal_k)(\varrho_k)=\varrho_k$, where $\varrho_k=P^{(k)}\varrho P^{(k)}$.
 
\smallskip
(ii) Any invariant state is block-diagonal and satisfies \eqref{eq:rblockrho}. Tracing it against
$P^{(k)}=\sum_{i=1}^{N_k}P^{(k)}_i$ and using $(1-\alpha_k)\Tr\big[(\id_k-\alpha_k\Wcal_k)^{-1}(X)\big]=\Tr(X)$ and
$\Tr(\pi^{(k\ell)}_j)=1$ gives $(1-\alpha_k)t_k=\sum_{\ell=1,\,\ell\neq k}^r(1-\alpha_\ell)\gamma_{k\ell}t_\ell$, i.e.\
$s=\Gamma s$. If $1$ is a simple eigenvalue of $\Gamma$, $s$ is a multiple of the Perron vector, fixed by
$\sum_{k=1}^rt_k=1$.
\end{proof}

\begin{lemma}
\label{lem:rkernel}
For $\tilde v=(\tilde v^{(k)}_i)\in\C^N$, let $\Psi(\tilde v)\in\B(\HH)$ be defined by the right-hand side of
\eqref{eq:rrhoinfty} with $\tilde\varrho$ replaced by $\tilde v$. Then $\Psi$ is an isomorphism from
$\ker(M-\Id)$ onto $\ker(\Phi-\id)$, whose inverse maps $X$ to its population vector
$\tilde X=(\Tr(P^{(k)}_iX))_{k,i}$. In particular
\[
\dim\ker(\Phi-\id)=\dim\ker(M-\Id)=\dim\ker(\widetilde M-\Id).
\]
\end{lemma}
 
\begin{proof}
Let $\tilde v\in\ker(M-\Id)$. Apart from positivity and normalization, the proof of Theorem
\ref{thm:rrhoinfty}(i) only uses $M\tilde\varrho=\tilde\varrho$, so it applies to $\tilde v$: $\Psi(\tilde v)$ is a
fixed point of $\Phi$, and by \eqref{eq:rpop} its population vector is $\tilde v$.
 
Conversely, let $X\in\ker(\Phi-\id)$. As for invariant states, $X$ is block-diagonal and its blocks satisfy
\eqref{eq:rblockrho} with the populations $\tilde X$ of $X$, which reads $X=\Psi(\tilde X)$. Taking populations,
\eqref{eq:rpop} gives $\tilde X=M\tilde X$.
 
Hence $\Psi:\ker(M-\Id)\to\ker(\Phi-\id)$ and $X\mapsto\tilde X$ are inverse to each other. Finally
$\widetilde M=DMD^{-1}$ gives $\ker(M-\Id)=D^{-1}\ker(\widetilde M-\Id)$.
\end{proof}

In particular, if $1$ is a simple eigenvalue of $\widetilde M$, then $\dim\ker(\Phi-\id)=1$, and the invariant
state of Theorem~\ref{thm:rrhoinfty} is unique.

\begin{remarks}
\label{rk:relimination}
\begin{itemize}[leftmargin=*]
\item[i)] For $r=2$ one can eliminate one subspace and reduce to a single matrix $T_{PQ}$, as in
Section~\ref{sec:fixed}. For $r\ge3$ no such global elimination is available in general, so one keeps the full $N\times N$
matrix $M$; the stochastic structure is recovered through the diagonal similarity $\widetilde M=DMD^{-1}$ of
Lemma~\ref{lem:rcolsum}.
\item[ii)] The proof of (ii) uses neither $M$ nor the explicit formula \eqref{eq:rrhoinfty}:
only the trace-preserving property of the normalized resolvent and the normalization of the reset states enter.
Hence the block populations of any invariant state depend only on the weights $\alpha_k$ and the routing
$\Gamma$, and not on the unitaries $W_k$, the reset states $\pi^{(k\ell)}_j$, or the sub-projectors $P^{(k)}_i$.
\item[iii)] When $1$ is not a simple eigenvalue of $\Gamma$ (that is, when the routing graph has several closed classes, see \cite{N}), $s$ is no longer uniquely determined by $\Gamma$,
and the block populations depend on the invariant state considered.
\item[iv)] For $r=2$, $\Gamma=\left(\begin{smallmatrix}0&1\\1&0\end{smallmatrix}\right)$, $s=(\tfrac12,\tfrac12)$, and
(ii) gives back \eqref{eq:traceval}.
\end{itemize}
\end{remarks}

\subsection{Robust Part of the Spectrum}
\label{sec:msec}

The invariant state was reconstructed from the block populations $\varrho^{(k)}_i$, which solve the reduced system
$\tilde\varrho=M\tilde\varrho$. We now use the subspace populations $t_k=\Tr(P^{(k)}\varrho)$ to extract part of the
spectrum of $\Phi$. As for $r=2$ (Lemma~\ref{lem:Sigma}), the projectors $P^{(1)},\dots,P^{(r)}$ span a subspace of
$\B(\HH)$ that is invariant under $\Phi^*$, on which $\Phi^*$ is the $r\times r$ matrix $\Sigma^{*}$, where
\[
\Sigma_{k\ell}=\alpha_k\,\delta_{k\ell}+(1-\alpha_\ell)\,\gamma_{k\ell}\,(1-\delta_{k\ell}).
\]
Its eigenvalues are eigenvalues of $\Phi$, for every choice of parameters.

\begin{theorem}\label{thm:inclusion}
For every $1\le k\le r$,
\[
\Phi^*\big(P^{(k)}\big)=\alpha_k\,P^{(k)}+\sum_{\substack{\ell=1\\ \ell\neq k}}^r(1-\alpha_\ell)\,\gamma_{k\ell}\,P^{(\ell)} .
\]
Hence $\Tr\big(P^{(k)}\Phi(X)\big)=\sum_{\ell=1}^r\Sigma_{k\ell}\Tr\big(P^{(\ell)}X\big)$ for all $X\in\B(\HH)$, the
matrix $\Sigma$ is column stochastic, and $\Sp(\Sigma)\subseteq\Sp(\Phi)$.
\end{theorem}
 
\begin{proof}
As in \eqref{eq:phistar}, the adjoint of \eqref{eq:phir} is
\begin{align*}
\Phi^*(X)&=\sum_{\ell=1}^r\alpha_\ell\,P^{(\ell)}W_\ell^*XW_\ell P^{(\ell)}
+\sum_{\ell=1}^r(1-\alpha_\ell)\sum_{\substack{k'=1\\ k'\neq\ell}}^r\gamma_{k'\ell}\sum_{i=1}^{N_\ell}P^{(\ell)}_i\,\Tr\big(\pi^{(k'\ell)}_iX\big).
\end{align*}
For $X=P^{(k)}$, the first sum reduces to $\alpha_kP^{(k)}$. In the second one,
$\Tr(\pi^{(k'\ell)}_iP^{(k)})=\delta_{k'k}$ and $\sum_{i=1}^{N_\ell}P^{(\ell)}_i=P^{(\ell)}$. This gives the
expression of $\Phi^*(P^{(k)})$. The identity on the populations
follows from $\Tr(P^{(k)}\Phi(X))=\Tr(\Phi^*(P^{(k)})X)$. Finally, the entries of $\Sigma$ are nonnegative, and for every $\ell$,
\[
\sum_{k=1}^r\Sigma_{k\ell}=\alpha_\ell+(1-\alpha_\ell)\sum_{\substack{k=1\\ k\neq\ell}}^r\gamma_{k\ell}=\alpha_\ell+(1-\alpha_\ell)=1,
\]
since the columns of $\Gamma$ sum to $1$. Hence $\Sigma$ is column stochastic.
 
\smallskip
Let $\lambda\in\Sp(\Sigma)$ and $y\neq0$ with $\Sigma^{\mathsf T}y=\lambda y$, and set $L=\sum_{k=1}^ry_kP^{(k)}$. The
$P^{(k)}$ are nonzero and supported on orthogonal blocks, hence linearly independent, so $L\neq0$. By linearity and
the expression of $\Phi^*(P^{(k)})=\sum_{\ell=1}^r\Sigma_{k\ell}P^{(\ell)}$,
\[
\Phi^*(L)=\sum_{k=1}^ry_k\sum_{\ell=1}^r\Sigma_{k\ell}\,P^{(\ell)}=\sum_{\ell=1}^r(\Sigma^{*}y)_\ell\,P^{(\ell)}=\lambda L .
\]
Hence $\Sp(\Sigma)\subseteq\Sp(\Phi^*)={\Sp(\Phi)}$. %
\end{proof}

\subsection{Single-Reset Subspaces}
\label{sec:singlereset}

So far, the reset states $\pi^{(kk_0)}_i$ leaving $\HH_{k_0}$ may depend on the block $P^{(k_0)}_i$ the population
leaves from. We now consider the case where they do not.

\begin{definition}\label{def:singlereset}
A subspace $\HH_{k_0}$ is \emph{single-reset} if, for every $k\neq k_0$, $\pi^{(kk_0)}_i=\pi^{(kk_0)}$ for all
$1\le i\le N_{k_0}$.
\end{definition}
According to Remark \ref{rk:samepi}, this amounts to choosing $N_{k_0}=1$. This is the setup of the system-reservoir framework which 
we will further discuss in Section \ref{sec:rsysres}.

\begin{proposition}\label{thm:singlerespec}
Assume $\HH_{k_0}$ is single-reset, and let $V_{k_0}=\{X\in\B(\HH_{k_0}):\Tr X=0\}$. Then $V_{k_0}$ is invariant under
$\Phi$, and $\Phi|_{V_{k_0}}=\alpha_{k_0}\Wcal_{k_0}|_{V_{k_0}}$. In particular,
$\Sp(\alpha_{k_0}\Wcal_{k_0}|_{V_{k_0}})\subseteq\Sp(\Phi)$.
\end{proposition}

\begin{proof}
For $X\in V_{k_0}$, only the unitary term of $\HH_{k_0}$ and the resets leaving $\HH_{k_0}$ survive in
\eqref{eq:phir}. Since $\HH_{k_0}$ is single-reset and $\sum_{i=1}^{N_{k_0}}P^{(k_0)}_i=P^{(k_0)}$, the latter equal
\[
(1-\alpha_{k_0})\sum_{\substack{k=1\\ k\neq k_0}}^r\gamma_{kk_0}\Tr\big(P^{(k_0)}X\big)\,\pi^{(kk_0)}=0,
\]
as $\Tr(P^{(k_0)}X)=\Tr X=0$.
\end{proof}

\begin{remarks}
\begin{itemize}[leftmargin=*]
\item[i)] By the argument of Remark~\ref{rk:samepi}, $\HH_{k_0}$ is single-reset if and only if $\Phi$ coincides with
the RQC obtained by replacing the family $\{P^{(k_0)}_i\}$ by the single projector $P^{(k_0)}$, i.e.\ with $N_{k_0}=1$.
\item[ii)] As in Remark~\ref{rk:specWP}, $\Sp(\alpha_{k_0}\Wcal_{k_0}|_{V_{k_0}})$ consists of the
$\alpha_{k_0}e^{i(\theta_n-\theta_m)}$, $n\neq m$, where $e^{i\theta_n}$ are the eigenvalues of $W_{k_0}$ repeated
according to their multiplicities $g_i$, and of $\alpha_{k_0}$ with multiplicity $d_{k_0}-1$. In total, the eigenvalue
$\alpha_{k_0}$ has multiplicity $\sum_ig_i^2-1$, which is nonzero if and only if $d_{k_0}\ge2$.
\item[iii)] Proposition~\ref{thm:singlerespec} holds whatever the data on the other subspaces. If all subspaces but possibly one, say $\HH_{k_*}$, are single-reset, the proof of Proposition~\ref{lem:singleblock} extends verbatim, with $\Pi(X)=(\Tr(P^{(k)}X))_{k=1}^r$ and Theorem~\ref{thm:inclusion} in place of Lemma~\ref{lem:Sigma}, and gives the whole spectrum. Thus we get the generalization of the results of Section~\ref{sec:np1}: with the multiset convention of Section~\ref{sec:np1},
\[
\Sp(\Phi)=\{0\}^{d^2-\sum_{k=1}^rd_k^2}\cup\Sp(\Sigma)\cup\bigcup_{k=1}^r\Sp\big(\alpha_k\Wcal_k|_{V_k}\big),
\]
where $d=\dim\HH$ and $d_k=\dim\HH_k$. Indeed, for $X\in V_{k_*}$, the reset part of $\Phi(X)$ has traceless components on each $\HH_k$, $k\neq k_*$, so $\Phi|_{\ker\Pi}$ is block upper triangular. This fails in general with two subspaces that are not single-reset.
\end{itemize}
\end{remarks}

\subsection{System-Reservoirs Setup} \label{sec:rsysres}
We consider here the setup describing a small system interacting with several $r-1\geq 2$ reservoirs that are not directly coupled with one another. With the small system subspace $\HH_1$ and the reservoir subspaces $\HH_k$, $2\leq k\leq r$, the coupling structure is encoded in the stochastic routing matrix 
\begin{align}
    \Gamma=\begin{pmatrix}
        0 & 1& 1 & \dots & 1 \\
        \gamma_{21} & 0 & 0 & \dots &0 \\
        \gamma_{31} &  0 & 0 & \dots &0 \\
        \vdots & \vdots &\vdots &\ddots &\vdots\\
        \gamma_{r1} & 0 & 0 & \dots &0
    \end{pmatrix}
\ \ \text{s.t.} \ \ \sum_{k=2}^{r}\gamma_{k1}=1. 
\end{align} 
Consequently, in the eigenvalue equation $\tilde X=M\tilde X$, $M$ takes the form, see \eqref{eq:MT},
\begin{equation}
M=\begin{pmatrix}
0 & M^{(12)} & \cdots & M^{(1r)}\\
M^{(21)} & 0 & \cdots & 0\\
\vdots & \vdots & \ddots & \vdots\\
M^{(r1)} & 0 & \cdots & 0
\end{pmatrix},
\end{equation}
with $M^{(1k)}_{ij}=\frac{1-\alpha_k}{1-\alpha_1}\,T^{(1k)}_{ij},$ and $M^{(k1)}_{ij}=\frac{1-\alpha_1}{1-\alpha_k}\,\gamma_{k1}T^{(k1)}_{ij}$. This structure yields the equivalent relations between the block components of $\tilde X$: 
\begin{align}
    X^{(1)}&=\sum_{k=2}^{r}M^{(1k)}M^{(k1)}X^{(1)}, \ \  %
    X^{(k)}=M^{(k1)}X^{(1)}.
\end{align}
The second key ingredient of the system-reservoirs setup is the existence of a unique reset state from the system to each reservoir, i.e. $N_1=1$. Hence, the stochastic coupling matrices satisfy
\begin{align}
T^{(k1)}&= (t_1(k), t_2(k), \cdots, t_{N_k}(k))^T\in M_{N_k, 1}(\R^+), \ \  \text{with} \sum_{i=1}^{N_k}t_i(k)=1,\nonumber \\
T^{(1k)}&= (1, 1, \dots, 1)\in M_{1,N_k}(\R^+).
\end{align} Furthermore $X^{(1)}\in \C\setminus\{0\}$ while $X^{(k)}\in \C^{N_k}$, $2\leq k\leq r$. Therefore 
\begin{align}
\sum_{k=2}^{r}M^{(1k)}M^{(k1)}=\sum_{k=2}^{r}\gamma_{k1}T^{(1k)}T^{(k1)}=\sum_{k=2}^{r}\gamma_{k1}=1, 
\end{align} so that $\dim(\ker(M-\Id))=1$ with
\begin{align}
    \tilde X&\in \C \Big(
        \frac{1}{1-\alpha_1}, \frac{\gamma_{21}}{1-\alpha_2}\,T^{(21)}, \frac{\gamma_{31}}{1-\alpha_3}\,T^{(31)}, \dots, \frac{\gamma_{r1}}{1-\alpha_r}\,T^{(r1)}
    \Big)^{T}.
\end{align}
Let us adapt the notations of the projectors and reset states to the system-reservoirs case, to make them closer to the two-subspace case:
\begin{align}\label{eq:notsysres}
P^{(1)}&:=P=\Id_{\HH_1}, \ \ P^{(k)}:=Q^{(k)}, \ \ \text{and} \ \ P^{(k)}_i:=Q^{(k)}_i, \nonumber\\ 
    \pi^{(k1)}&:=\tau^{(k)}\in \D(\HH_k), \ \ \text{and} \ \ \pi^{(1k)}_i:=\pi^{(k)}_i\in \D(\HH_1), \ \ 2\leq k \leq r, \ \ 1\leq i\leq N_{k}.
\end{align}
The RQC then reads
\begin{align}\label{eq:physysress}
\Phi(X)&=\alpha_1 W_1 P XPW_1^*+\sum_{k=2}^r\alpha_k\,W_k Q^{(k)}X Q^{(k)}W_k^*
\nonumber\\
&+\sum_{k=2}^r(1-\alpha_1)\gamma_{k1}\Tr(PX)\tau^{(k)}+\sum_{k=2}^r(1-\alpha_k)\sum_{i=1}^{N_k}
\Tr\big(Q^{(k)}_iX\big)\,\pi^{(k)}_i.
\end{align}
\begin{theorem}\label{thm:sys-ress}
    Assume $0\leq \alpha_k<1$ for all $1\leq k\leq r$. Then the unique invariant state of the RQC \eqref{eq:physysress} with \eqref{eq:notsysres} reads 
    \begin{align}
\varrho&=c\Big(\sum_{\substack{\ell=2}}^r\frac{\gamma_{\ell 1}}{1-\alpha_1}\,\sum_{j=1}^{N_\ell}
t_j(\ell)\,(1-\alpha_1)(\id_1-\alpha_1\Wcal_1)^{-1}(\pi^{(\ell)}_j)\nonumber\\
&+\sum_{k=2}^r\,\frac{\gamma_{k1}}{1-\alpha_k}
\,(1-\alpha_k)(\id_k-\alpha_k\Wcal_k)^{-1}(\tau^{(k)})\Big),
\end{align}
where $c=\big[\frac{1}{1-\alpha_1}+\sum_{k=2}^{r}\frac{\gamma_{k1}}{1-\alpha_k}\big]^{-1}$ and $t_j(\ell)=\Tr\big[Q^{(\ell)}_j\,(1-\alpha_\ell)(\id_\ell-\alpha_\ell\Wcal_\ell)^{-1}(\tau^{(\ell)})\big]$.
\end{theorem}
The restriction of the invariant state $\varrho$ in each subspace $\HH_k$ of the reservoir is given by the population $\Tr(Q^{(k)}\varrho)=c\frac{\gamma_{k1}}{1-\alpha_k}$ times the state $(1-\alpha_k)(\id_k-\alpha_k\Wcal_k)^{-1}(\tau^{(k)})$, which coincides with \eqref{eq:statenp1} in case $r=2$. Whenever the state $\tau^{(k)}$ is invariant under the dynamics in $\HH_k$, i.e.  $[W_k,\tau^{(k)}]=0$, this restriction reads $\Tr(Q^{(k)}\varrho)\;\tau^{(k)}$, in line with \eqref{eq:ressys} for $r=2$.\\
 The more interesting restriction of the invariant state to the system subspace $\HH_1$ reads 
 \begin{align}\label{eq:thermstate1}
      c\sum_{\substack{\ell=2}}^r\frac{\gamma_{\ell 1}}{1-\alpha_1}\,\sum_{j=1}^{N_\ell}
t_j(\ell)\,(1-\alpha_1)(\id_1-\alpha_1\Wcal_1)^{-1}(\pi^{(\ell)}_j), 
 \end{align}
 with population $\Tr(P\varrho) = c\sum_{\substack{\ell=2}}^r\frac{\gamma_{\ell 1}}{1-\alpha_1}=\frac{c}{1-\alpha_1}$. The label $2\leq \ell \leq r$ indicates the contribution of the reservoir $\ell$, a state in $\HH_1$ which can be written as
 \begin{align}\label{eq:contl}
     (1-\alpha_1)(\id_1-\alpha_1\Wcal_1)^{-1}\Big(\sum_{j=1}^{N_\ell} t_j(\ell)\,\pi^{(\ell)}_j\Big), \ \ \text{where} \ \ 
     \sum_{j=1}^{N_\ell} t_j(\ell)\,\pi^{(\ell)}_j \in \D(\HH_1).
 \end{align}
 The channel $(1-\alpha_1)(\id_1-\alpha_1\Wcal_1)^{-1}$ describes the effect of the dynamics in the system on these states. %
 Assuming the invariance property in all reservoirs $[W_\ell,\tau^{(\ell)}]=0$ for simplicity, we get $t_j(\ell)=\Tr(Q_j^{(\ell)}\tau^{(\ell)})$ for the $\tau^{(\ell)}$-dependent weight of the contributed state $\pi^{(\ell)}_j$ associated with $Q_j^{(\ell)}$. \\ 
 We further specialize to reset states given by Gibbs states $\tau^{(\ell)}=\sum_{j=1}^{N_\ell}\frac{e^{-\theta_\ell E_j^{(\ell)}}}{Z_\ell} Q_j^{(\ell)}$ in all reservoirs, $\theta_\ell>0$  as in Section \ref{sec:mainresults}, and assume all reservoirs are identical, i.e. $N_\ell=N_0$, $E_j^{(\ell)}=E_j$, in increasing order, $Q_j^{(\ell)}=Q_j$, with $\Tr(Q_j)=1$ for simplicity. The only reservoir dependent data are the (inverse) temperatures $\theta_\ell$, and the reset states $\pi_j^{(\ell)}$. The state contributed to \eqref{eq:contl} by the $\ell$-th reservoir reads
 \begin{align}
     \sum_{j=1}^{N_\ell} t_j(\ell)\,\pi^{(\ell)}_j=\sum_{j=1}^{N_0} \frac{e^{-\theta_\ell E_j}}{Z_\ell}\,\pi^{(\ell)}_j.
 \end{align}
 This setup describes the interaction of a small system with $r-1$ identical thermal reservoirs labeled by $\ell$, at different temperatures, all of which inserts the populations within their subspaces $\{Q_j\HH_\ell\}_{j=1}^{N_0}$ to the states $\pi_j^{(\ell)}$ within the system's subspace according to Boltzmann weights. \\ As a simple illustration assume $r=3$, i.e. two reservoirs and consider the system subspace $\HH_1$ to be a chain spanned by orthonormal vectors $\{e_j\}_{j=1}^{N_0}$, with indices $j$ considered as discrete positions. The reservoir $\ell=2$ inserts populations from the left, $\pi_j^{(2)}=|e_j\rangle\langle e_j|$ while the second reservoir inserts them from the right, $\pi_j^{(3)}=|e_{N_0+1-j}\rangle\langle e_{N_0+1-j}|$. This choice of indices makes $\Phi$ transfer the populations in the lower energy levels of the reservoir to the extremity of the chain closer to the reservoir, while the populations in the higher energy levels are transferred farther from the reservoir. Focusing on the effects of the temperatures, we choose $\gamma_{\ell 1}=1/2$, $\alpha_k=\alpha$, so that the invariant state \eqref{eq:thermstate1} reads, up to the normalization constant,
 \begin{align}\label{eq:chain}
     \sum_{j=1}^{N_0}(1-\alpha)(\id_1-\alpha\Wcal_1)^{-1}\Big[\Big(\frac{e^{-\theta_2 E_j}}{Z_2}+\frac{e^{-\theta_3 E_{N_0+1-j}}}{Z_3}\Big)|e_j\rangle\langle e_j|\Big].
 \end{align}
 We do not dwell on this formula and limit ourselves to a few remarks.
 \begin{remarks}
     i) In case $[W_1, |e_j\rangle\langle e_j|]=0$, the profile of the populations on the sites is given by the numerical factor in the argument of the resolvent. It features higher values at the boundaries of the chain than in the bulk of it, and the end of the chain closest to the coldest reservoir has higher populations than the one closest to the hottest reservoir. For a non trivial dynamics $W_1$, this picture is blurred by the action of the channel  $(1-\alpha)(\id_1-\alpha\Wcal_1)^{-1}$. \\
     ii) Considering a Haar distributed random unitary, and taking the expectation values of the populations on the sites we get thanks to Proposition \ref{prop:f} $$\E \langle e_k|(1-\alpha)(\id_1-\alpha\Wcal_1)^{-1}(|e_j\rangle\langle e_j|) e_k\rangle=\frac{1}{N_0}(1-f(\alpha,N_0))+f(\alpha, N_0)\delta_{kj}.$$ 
     Introducing this expression in \eqref{eq:chain}, the $\delta_{jk}$ term reproduces the profile of Remark i), weighted by $f(\alpha,N_0)$, while the first term adds a uniform background $(1-f(\alpha,N_0))/N_0$ which flattens the distribution.   \\
     iii) For a coherent dynamics $W_1$ which acts like a forward, respectively backwards, shift on sites with odd, respectively even, indices, plus suitable boundary conditions at the extremities of the chain, one can check that the asymptotic populations on the sites display the same general features as in Remark i), supplemented by oscillations.  
\end{remarks}

\section{The Continuous-Time Limit}
\label{sec:continuous}
So far the dynamics is discrete. To obtain a continuous-time description, we let the reset weights
$1-\alpha$ and $1-\beta$ vanish with a time step $h$, at fixed rates per unit of time, and assume the unitaries are generated by Hamiltonian operators over the time step $h$. We then consider the QRC iterated $n$ times, with $n=t/h\to \infty$, $h\to 0$, keeping $t>0$ fixed, to get a semigroup $(e^{t\Lind})_{t\geq 0}$ when restricted to the right set of operators, that we analyze.
For simplicity, we only spell out the details of the two-subspace model of Section \ref{sec:twosubs}.

{\subsection{The Lindblad Generator}}
\label{sec:lindblad}

With $t=nh$, we
assume
\begin{equation}
\alpha(h)=1-\gamma_Ph+O(h^2),\quad \beta(h)=1-\gamma_Qh+O(h^2),\quad
W_\#(h)=e^{-iH_\#h},\quad \#\in\{P,Q\},
\label{eq:scaling}
\end{equation}
with rates $\gamma_P,\gamma_Q>0$ and a Hamiltonian $H=H_P\oplus H_Q$. We make explicit the dependence in $h$ of the corresponding  RQC \eqref{eq:phi}, denoted by $\Phi_h$. To leading order, resets occur at rate $\gamma_\#$, and
between resets the system evolves under $H_\#$.

We now let $h\to0$ with $t=nh$ fixed and show that the discrete iteration $\Phi_h^{\,n}$ becomes a
continuous-time evolution $e^{t\Lind}$. Since all spaces are finite-dimensional, %
a direct telescoping argument gives the convergence, with an explicit rate. \medskip

The limit has to be taken on block-diagonal operators. Indeed, while $e^{t\Lind}|_{t=0}=\id$, for $h=0$ the channel is not the identity, 
but the projection $\Phi_0=\Delta$ onto block-diagonal operators of Remarks~\ref{rk:nonunital}~iv).
By \eqref{eq:phidiag}, $\Phi_h^{\,n}$ vanishes on the cross-blocks for every $n\ge1$, and so does its limit. Since each $e^{t\mathcal{L}'}$, with $\mathcal{L}'\in\B(\B(\HH))$, is invertible, the limit cannot be of this form on $\B(\HH)$. We work instead on $\Acal:=\Delta (\B(\HH))$, the block-diagonal subalgebra
\[
\Acal=P\B(\HH)P \oplus Q\B(\HH)Q =\B(\Hs) \oplus \B(\Ho),
\]
with $\Phi_0|_\Acal=\id_\Acal$, where the continuous-time limit lives. We write
$\Phi_h|_\Acal$ for the restriction of $\Phi_h$ to $\Acal$. This restriction is again a quantum
channel on $\Acal$, for three reasons. First, $\Phi_h$ sends $\Acal$ into $\Acal$. 
Second, it preserves the trace, like $\Phi_h$. Third, it is completely positive: indeed $\Phi_h$
is completely positive on all of $\B(\HH)$, and restricting a completely positive map to a stable
subalgebra keeps it completely positive, see Remark \ref{rk:nonunital} iv) .

\begin{theorem}[Continuous-time limit]
\label{thm:ctlimit}
Under the scaling \eqref{eq:scaling}, and for $t\ge0$ fixed with $nh=t$, the restricted iteration
$\big(\Phi_h|_{\mathcal A}\big)^{n}$ converges in norm $\|\cdot\|_{1\to 1}$, as $h\to0$, to $e^{t\Lind}$ on
$\mathcal A$, where
\begin{equation}
\Lind(X)=-i[H,X]
+\gamma_P\Big(\sum_i\Tr(P_iX)\,\tau_i-X_P\Big)
+\gamma_Q\Big(\sum_j\Tr(Q_jX)\,\pi_j-X_Q\Big),\qquad X\in\mathcal A,
\label{eq:lindblad}
\end{equation}
with $X_P=PX P$, $X_Q=QX Q$.  The semigroup $(e^{t\Lind})_{t\geq 0}$ defines a quantum dynamical semigroup on $\mathcal A$, but not on $\B(\HH)$.

\medskip
\noindent Moreover, for  $n$ large enough, uniformly for $t$ in a bounded interval, we have
\begin{equation}
\big\|\big(\Phi_h|_{\mathcal A}\big)^{n}-e^{t\Lind}\big\|_{1\to1}\le\frac{Kt^2}{n},
\label{eq:rate}
\end{equation}
where $K$ is independent of $t$. 
\end{theorem}
\noindent The generator \eqref{eq:lindblad} is itself of reset type: the coherent part $-i[H,\cdot]$ is
corrected by terms that, at rates $\gamma_P$ and $\gamma_Q$, remove the population of each block and re-prepare it in the
reset states $\tau_i,\pi_j$. One thus gets, in the continuous-time limit, a Lindbladian similar to the reset
form studied in \cite{HJ}.

\medskip
\noindent Although Theorem~\ref{thm:ctlimit} only involves $\mathcal A$, formula \eqref{eq:lindblad} makes sense for every $X\in\B(\HH)$, with $X_P=PXP$ and $X_Q=QXQ$, and we keep the notation $\Lind$ for this extension. Before proving the theorem, we make precise in which sense $\Lind$ is a Lindblad generator: it coincides on $\mathcal A$ with a generator of GKSL form on $\B(\HH)$, from which
it differs on the off-diagonal blocks.

\medskip
We write $\Delta(X)=PXP+QXQ$ for the projection onto $\mathcal A$ of Remarks~\ref{rk:nonunital}~iv), so that $\mathcal A^\perp=\ran(\id-\Delta)=P\B(\HH)Q\oplus Q\B(\HH)P$ is the space of off-diagonal operators, and we set $\bar\gamma=\frac12(\gamma_P+\gamma_Q)$ and $\mathcal U_t(X)=e^{-itH}Xe^{itH}$.

\smallskip
\noindent Let $(e_{ib})$ and $(f_{jd})$ be orthonormal bases of $\ran P_i$ and $\ran Q_j$, and write the reset states in their eigenbases as $\tau_i=\sum_at_{ia}|u_{ia}\rangle\langle u_{ia}|$ and $\pi_j=\sum_cp_{jc}|v_{jc}\rangle\langle v_{jc}|$. Set 

\[\Gamma_{iab}=\sqrt{\gamma_P\,t_{ia}}\;|u_{ia}\rangle\langle e_{ib}|, \ \text{and}\  \Lambda_{jcd}=\sqrt{\gamma_Q\,p_{jc}}\;|v_{jc}\rangle\langle f_{jd}|.\]
We compute %
\[
\sum_{a,b}\Gamma_{iab}X\Gamma_{iab}^*=\gamma_P\Tr(P_iX)\,\tau_i,\qquad
\sum_{c,d}\Lambda_{jcd}X\Lambda_{jcd}^*=\gamma_Q\Tr(Q_jX)\,\pi_j,
\]
and
\[
\sum_{i,a,b}\Gamma_{iab}^*\Gamma_{iab}=\gamma_PP,\qquad \sum_{j,c,d}\Lambda_{jcd}^*\Lambda_{jcd}=\gamma_QQ .
\]
Since $\Gamma_{iab}=Q\Gamma_{iab}P$ and $\Lambda_{jcd}=P\Lambda_{jcd}Q$, these jump operators lie in $\mathcal A^\perp$: they transfer the population from one block to the other, and they do not belong to $\mathcal A$.\\

If the $P_i$ and $Q_j$ are nonnegative operators only, see Remark \ref{rem:povm}, we take for $(e_{ib})_b$ and $(f_{jd})_d$ orthonormal eigenbases, $P_ie_{ib}=\epsilon_{ib}e_{ib}$ and $Q_jf_{jd}=\eta_{jd}f_{jd}$, and we multiply $\Gamma_{iab}$ by $\sqrt{\epsilon_{ib}}$ and $\Lambda_{jcd}$ by $\sqrt{\eta_{jd}}$. The two identities above are unchanged.

\smallskip
\noindent We define on $\B(\HH)$
\[
\begin{aligned}
\widetilde\Lind(X)&=-i[H,X]+\sum_{i,a,b}\Big(\Gamma_{iab}X\Gamma_{iab}^*-\tfrac12\big\{\Gamma_{iab}^*\Gamma_{iab},X\big\}\Big)+\sum_{j,c,d}\Big(\Lambda_{jcd}X\Lambda_{jcd}^*-\tfrac12\big\{\Lambda_{jcd}^*\Lambda_{jcd},X\big\}\Big)\\
&=-i[H,X]+\gamma_P\sum_i\Tr(P_iX)\,\tau_i+\gamma_Q\sum_j\Tr(Q_jX)\,\pi_j-\tfrac12\{\gamma_P P+\gamma_Q Q,X\}.
\end{aligned}
\]

\noindent While $\Lind$ does not have the GKSL form, $\widetilde\Lind$ does, so that
$(e^{t\widetilde\Lind})_{t\geq0}$ is a quantum dynamical semigroup on $\B(\HH)$
\cite{CE, Lindblad}.
However, as the next lemma shows, $(e^{t\widetilde\Lind})_{t\geq0}$ and $(e^{t\Lind})_{t\geq0}$ agree on $\Acal$ as quantum dynamical semigroups, while $(e^{t\Lind})_{t\geq0}$
does not preserve positivity on $B(\HH)$ in general.
\begin{lemma}
\label{lem:Ltilde} With the notation above, we have
\begin{itemize}
\item[i)] $\widetilde\Lind=\Lind-\bar\gamma\,(\id-\Delta)$, and %
\[\widetilde\Lind|_{\mathcal A}=\Lind|_{\mathcal A}, \ \ \widetilde\Lind|_{{\mathcal A}^\perp}=-i[H, \cdot]-\bar\gamma \id_{{\mathcal A}^\perp}\ \ \text{and} \ \ \Lind|_{{\mathcal A}^\perp}=-i[H, \cdot].\]
\item[ii)] $[\Lind,\Delta]=[\widetilde\Lind,\Delta]=0$. 
\item[iii)] For all $t\ge0$, 
\begin{equation}
e^{t\widetilde\Lind}=e^{t\Lind}\circ\Delta+e^{-\bar\gamma t}\mathcal U_t \circ(\id-\Delta),\qquad
e^{t\Lind}=e^{t\widetilde\Lind}\circ\Delta+\mathcal U_t\circ(\id-\Delta).
\label{eq:expLtilde}
\end{equation}
In particular, $e^{t\Lind}|_{\mathcal A}=e^{t\widetilde\Lind}|_{\mathcal A}$.
\end{itemize}
\end{lemma}

\begin{proof}
i) Since $\{P,X\}=2PXP+PXQ+QXP$ and $\{Q,X\}=2QXQ+QXP+PXQ$,
\[
\tfrac12\{\gamma_P P+\gamma_Q Q,X\}=\gamma_PPXP+\gamma_QQXQ+\bar\gamma\,(PXQ+QXP),
\]
which gives $\widetilde\Lind=\Lind-\bar\gamma(\id-\Delta)$. As $\id-\Delta$ vanishes on $\mathcal A$, both operators coincide there. The computation above also yields the expressions of restrictions of $\Lind$ and $\widetilde\Lind$ on ${\mathcal A}^\perp$.\\
ii) Each term of $\Lind$ commutes with $\Delta$: the commutator with the block-diagonal $H$, the reset terms, since $\Tr(P_i\Delta(X))=\Tr(P_iX)$, $\Tr(Q_j\Delta(X))=\Tr(Q_jX)$ and $\tau_i,\pi_j\in\mathcal A$, and the maps $X\mapsto PXP$, $X\mapsto QXQ$. Hence $[\Lind,\Delta]=0$, and $[\widetilde\Lind,\Delta]=0$ by i).%
\\
iii) For $X\in\Acal^\perp$, %
$\Lind(X)=-i[H,X]$, by i).
Since $\Acal^\perp$ is invariant under $\Lind$ by ii), $e^{t\Lind}\circ(\id-\Delta)=\Ucal_t\circ(\id-\Delta)$. By ii), $\Lind$ and $\id-\Delta$ commute, so 
\[
e^{t\tilde\Lind}=e^{t\Lind}\circ\big(\Delta+e^{-t\bar\gamma}(\id-\Delta)\big)=e^{t\Lind}\circ\Delta+e^{-t\bar\gamma}\Ucal_t\circ(\id-\Delta),
\]
which is the first identity. The second identity is proven similarly.
\end{proof}

In words, $e^{t\Lind}$ and $e^{t\tilde\Lind}$ coincide on $\Acal$, and on $\Acal^\perp$, $e^{t\Lind}$ acts as the unitary evolution $\Ucal_t$, whereas $e^{t\tilde\Lind}$ additionally damps the off-diagonal blocks at rate $\bar\gamma$. Moreover, as shown in the example below, $e^{t\Lind}$ on $\B(\HH)$ is not positive, let alone completely positive, in general. 
\begin{example}
\label{rk:whyblockdiag}
Consider $(e^{t\Lind})_{t\geq 0}$ in the simplest setup
$\dim\Hs=\dim\Ho=1$, hence $N_P=N_Q=1$, and take $H=0$. Then the reset states are $\pi=|p\rangle\langle p|$ and
$\tau=|q\rangle\langle q|$, and \eqref{eq:lindblad} reduces to
\[
\Lind(X)=\gamma_P\big(\Tr(PX)\,\tau-X_P\big)+\gamma_Q\big(\Tr(QX)\,\pi-X_Q\big).
\]
Writing a state in the ordered basis $\{p,q\}$ as 
$X=\left(\begin{smallmatrix} \tilde x_P & c\\ \bar c & \tilde x_Q\end{smallmatrix}\right)$ with $\tilde x_P+\tilde x_Q=1$, we
read off the action of $\Lind$ on each entry. The population $\tilde x_P=\Tr(PX)$ evolves as
$\dot {\tilde{x}}_P=\Tr(P\Lind(X))$. Since $\Tr(P\tau)=0$, $\Tr(PX_P)=\tilde x_P$, $\Tr(QX)=\tilde x_Q$, $\Tr(PX_Q)=0$ and
$\Tr(P\pi)=1$, this gives 
\[
\dot {\tilde{x}}_P =%
\gamma_Q-(\gamma_P+\gamma_Q)\,\tilde x_P,
\]
using $\tilde x_Q=1-\tilde x_P$. Solving this differential equation, $\tilde x_P(t)$, $\tilde x_Q(t)$ converge as $t\to\infty$ to the
stationary values, independent of the initial conditions,
\[
\tilde x_P^\infty=\frac{\gamma_Q}{\gamma_P+\gamma_Q},\qquad \tilde x_Q^\infty=\frac{\gamma_P}{\gamma_P+\gamma_Q}.
\]
The off-diagonal element $c$ obeys $\dot c=\langle p|\Lind(X)q\rangle$. All terms $\tau,\pi,X_P,X_Q$ are
diagonal, so they do not contribute to the off-diagonal entries. %
Hence $\dot c=0$ and $c(t)=c(0)$ $\forall t\geq 0$.%

\noindent Take the pure initial state
\[
X_0=\frac12\begin{pmatrix}1&1\\1&1\end{pmatrix}=\tfrac12\big(|p\rangle+|q\rangle\big)
\big(\langle p|+\langle q|\big)\ \ge0 ,
\]
so that $\tilde x_P(0)=\tilde x_Q(0)=\tfrac12$ and $c(0)=\tfrac12$. With $2\bar\gamma=\gamma_P+\gamma_Q$ and $\delta=\tfrac12-\tilde x_P^\infty$,
the evolved state keeps the off-diagonal element frozen %
while the populations converge,
\[
X(t)=\begin{pmatrix} \tilde x_P(t) & \tfrac12\\[2pt] \tfrac12 & \tilde x_Q(t)\end{pmatrix},
\qquad \tilde x_P(t)=\tilde x_P^\infty+\delta\,e^{-2\bar \gamma t},\quad \tilde x_Q(t)=1-\tilde x_P(t).
\]
Its determinant is
\[
\det X(t)=\tilde x_P(t)\,\tilde x_Q(t)-\tfrac14
=\big(\tilde x_P^\infty+\delta e^{-2\bar\gamma t}\big)\big(\tilde x_Q^\infty-\delta e^{-2\bar \gamma t}\big)-\tfrac14
=-\delta^2\big(1-e^{-2\bar\gamma t}\big)^2 ,
\]
which is strictly negative for every $t>0$ as soon as $\delta\neq0$, i.e.\ $\gamma_P\neq\gamma_Q$. So $X(t)$ is no longer a state: $e^{t\Lind}$ does not preserve positivity on $\B(\HH)$, for any $t>0$.
\end{example}

\smallskip
\begin{proof}[Proof of Theorem \ref{thm:ctlimit}]
Take $X\in\mathcal A$, so that $X=X_P\oplus X_Q$ is block-diagonal with
$X_P=PX P$ and $X_Q=QX Q$. Expanding the scaling \eqref{eq:scaling} on such a
$X$, with $W_PX_P W_P^*=X_P-ih[H_P,X_P]+\|X_P\|_1O(h^2)$, similarly for $Q$,
$\alpha=1-\gamma_Ph+O(h^2)$ and $\beta=1-\gamma_Qh+O(h^2)$, we get 
\[
\Phi_h(X)=X-ih[H,X]-h\big(\gamma_PX_P+\gamma_QX_Q\big)
+h\Big(\gamma_P\sum_i\Tr(P_iX)\tau_i+\gamma_Q\sum_j\Tr(Q_jX)\pi_j\Big)+\|X\|_1O(h^2).
\]
The zeroth-order term is $X$ itself, consistent with $\Phi_0|_{\mathcal A}=\id_{\mathcal A}$; the
first-order term defines the generator. Dividing by $h$ and letting $h\to0$,
$h^{-1}(\Phi_h|_{\mathcal A}-\id_{\mathcal A})(X)\to\Lind(X)$ with $\Lind$ as in
\eqref{eq:lindblad}.

\smallskip
Both $\Phi_h|_{\mathcal A}$ and $e^{h\Lind}$ expand as
$\id+h\Lind+O(h^2)$, so their difference, in which the terms $\id_{\mathcal A}$ and $h\Lind$ cancel, is second
order. Precisely, by \eqref{eq:scaling} there exist $h_*>0$ and $C_1\ge0$ such that
\[
\big\|\Phi_h|_{\mathcal A}-(\id_{\mathcal A}+h\Lind)\big\|_{1\to1}\le C_1h^2\qquad(0<h\le h_*),
\]
and, for $0<h\le h_*$, the Taylor remainder of the
exponential gives
\[
\big\|e^{h\Lind}-(\id_{\mathcal A}+h\Lind)\big\|_{1\to1}
\le\sum_{k\ge2}\frac{h^k\|\Lind\|^k}{k!}\le\frac{\|\Lind\|^2}{2}\,e^{h_*\|\Lind\|}\,h^2=:C_2h^2 .
\]
Adding these two bounds,
\[
\big\|\Phi_h|_{\mathcal A}-e^{h\Lind}\big\|_{1\to1}\le Kh^2,\qquad K:=C_1+C_2,
\]
where $K$ depends only on $\|\Lind\|$ and on the fixed data of the model, and not on $t$.

\smallskip
\noindent With $A=\Phi_h|_{\mathcal A}$ and
$B=e^{h\Lind}$, consider the telescoping identity
\[
A^{n}-B^{n}=\sum_{k=0}^{n-1}A^{k}\,(A-B)\,B^{\,n-1-k},
\]
which is valid in any associative algebra and requires no commutativity. Since $A$ is a quantum channel, it is a trace-norm
contraction, so $\|A^k\|_{1\to1}\le1$. Furthermore, by Lemma \ref{lem:Ltilde}, $B$ is the restriction to $\mathcal{A}$ of the quantum channel $e^{h\tilde\Lind}$, so $\|B^{j}\|_{1\to1}\le1$ as well. Since $h=t/n$ and $B^n=e^{t\Lind}$, we get, for $n\ge t/h_*$, 
\[
\big\|A^{n}-B^{n}\big\|_{1\to1}\le n\,\|A-B\|_{1\to1}\le nKh^2=\frac{Kt^2}{n},
\]
which is \eqref{eq:rate}. As $K$ does not depend on $t$, the bound $Kt^2/n\le KT^2/n$ holds uniformly for
$t\in[0,T]$ and $n\ge T/h_*$.
\end{proof}

\subsection{Invariant State and Consistency with the Discrete Model}
\label{sec:lindbladsteady}

We now show that the invariant state of $\Lind$ is the $h\to0$ limit of the discrete invariant
state: the continuous coupling matrices are the limits of the discrete ones, and the whole construction of
the invariant state carries over.

The building blocks are the maps on $\B(\Hs)$
\[
\mathcal J_P=i[H_P,\cdot\,]+\gamma_P\id_P,\qquad \mathcal R_P=\gamma_P\,\mathcal J_P^{-1},
\]
and the maps $\mathcal J_Q$, $\mathcal R_Q$ on $\B(\Ho)$, defined in the same way with $H_Q$ and $\gamma_Q$. They are the continuous analogues of the discrete resolvent channels
\[
\mathcal R_P^h=(1-\alpha)(\id_P-\alpha\Wcal_P)^{-1},\qquad \mathcal R_Q^h=(1-\beta)(\id_Q-\beta\Wcal_Q)^{-1}.
\]

\begin{lemma}
\label{lem:Dinv}
The maps $\mathcal J_P$ and $\mathcal J_Q$ are invertible, and $\mathcal R_P$ and $\mathcal R_Q$ are quantum channels, being the $h\to0$
limits of $\mathcal R_P^h$ and $\mathcal R_Q^h$ under the scaling \eqref{eq:scaling}.
\end{lemma}

\begin{proof}
We treat $\mathcal J_P$ and $\mathcal R_P$, the case of $Q$ being identical. Let $(e_n)$ be an orthonormal eigenbasis of $H_P$, with $H_Pe_n=E_ne_n$. Then $\mathcal J_P(|e_n\rangle\langle e_m|)=\big(\gamma_P+i(E_n-E_m)\big)|e_n\rangle\langle e_m|$, and these eigenvalues have modulus at least $\gamma_P>0$, so $\mathcal J_P$ is invertible. Expanding $\Wcal_P(X)=X-ih[H_P,X]+\|X\|O(h^2)$, with $1-\alpha=\gamma_Ph+O(h^2)$, the scaling \eqref{eq:scaling} gives
\[
\id_P-\alpha\Wcal_P=\id_P-(1-\gamma_Ph)\big(\id_P-ih[H_P,\cdot\,]+O(h^2)\big)=h\,\mathcal J_P+O(h^2),
\]
so that
\[
\mathcal R_P^h=(\gamma_Ph+O(h^2))\,\big(h\mathcal J_P+O(h^2)\big)^{-1}=(\gamma_P+O(h))\big(\mathcal J_P+O(h)\big)^{-1}\xrightarrow[h\to0]{}\gamma_P\mathcal J_P^{-1}=\mathcal R_P .
\]
Each $\mathcal R_P^h$ is a quantum channel, and the set of quantum channels is closed, so $\mathcal R_P$ is a quantum channel; see also \cite{HJ} for an alternative argument.
\end{proof}

Since $\mathcal R_P$ and $\mathcal R_Q$ are trace-preserving, the continuous coupling matrices 
\begin{equation}
(T_P^c)_{kl}=\Tr\big[P_k\,\mathcal R_P(\pi_l)\big],\qquad (T_Q^c)_{kl}=\Tr\big[Q_k\,\mathcal R_Q(\tau_l)\big],
\label{eq:Tc}
\end{equation}
are column stochastic without any further normalization, $\sum_k(T_P^c)_{kl}=\Tr\big[\mathcal R_P(\pi_l)\big]=1$, using $\sum_kP_k=P$, and likewise for $T_Q^c$. By Lemma~\ref{lem:Dinv}, these are the $h\to0$ limits of the discrete coupling matrices $T_P$ and $T_Q$.

Let $\varrho$ be an invariant state of $\Lind$, with population vectors $\tilde\varrho_P=\big(\Tr(P_k\varrho)\big)_k$ and $\tilde\varrho_Q=\big(\Tr(Q_k\varrho)\big)_k$. As in Section~\ref{sec:fixed}, projecting the equation $\Lind(\varrho)=0$ onto $\Hs$ gives $\mathcal J_P(P\varrho P)=\gamma_Q\sum_k\tilde\varrho_Q^k\,\pi_k$, and similarly onto $\Ho$, that is
\begin{equation}
P\varrho P=\frac{\gamma_Q}{\gamma_P}\sum_k\tilde\varrho_Q^k\,\mathcal R_P(\pi_k),\qquad
Q\varrho Q=\frac{\gamma_P}{\gamma_Q}\sum_k\tilde\varrho_P^k\,\mathcal R_Q(\tau_k).
\label{eq:rhoss}
\end{equation}
Tracing against $P_k$ and $Q_k$ gives $\tilde\varrho_P=\frac{\gamma_Q}{\gamma_P}T_P^c\tilde\varrho_Q$ and $\tilde\varrho_Q=\frac{\gamma_P}{\gamma_Q}T_Q^c\tilde\varrho_P$, so that $\tilde\varrho_P=T_P^cT_Q^c\tilde\varrho_P$. Since $T_P^cT_Q^c$ is column stochastic, it admits $1$ as an eigenvalue, whose eigenvectors give the populations, and \eqref{eq:rhoss} reconstructs $\varrho$ from them.

The same holds for the discrete channel $\Phi_h$. By \eqref{eq:rhoinv}, an invariant state $\varrho_h$ of $\Phi_h$, with population vectors $\tilde\varrho_P(h)$ and $\tilde\varrho_Q(h)$, reads
\[
\varrho_h=\frac{1-\beta}{1-\alpha}\sum_k\tilde\varrho_Q^k(h)\,\mathcal R_P^h(\pi_k)+\frac{1-\alpha}{1-\beta}\sum_k\tilde\varrho_P^k(h)\,\mathcal R_Q^h(\tau_k).
\]
To leading order, this corresponds to \eqref{eq:rhoss} with $\mathcal R_P^h,\mathcal R_Q^h$ in place of $\mathcal R_P,\mathcal R_Q$, since $\frac{1-\beta}{1-\alpha}=\frac{\gamma_Q}{\gamma_P}+O(h)$ under the scaling \eqref{eq:scaling}.

Assume that $\varrho$ is unique, which, by the reconstruction \eqref{eq:rhoss}, amounts to $\ker(T_P^cT_Q^c-\Id)$ being one-dimensional. Since $T_P^cT_Q^c$ is column stochastic, its eigenvalue $1$ is semi-simple. By \cite[Thm.~II.5.1]{K}, for $h$ small enough, $T_PT_Q$ has a single eigenvalue close to $1$, which is $1$ itself since $T_PT_Q$ is column stochastic, and its eigenprojection, of rank one, converges to that of $T_P^cT_Q^c$. Hence $\varrho_h$ is unique for $h$ small (the trace condition fixes the normalization), $\tilde\varrho_P(h)\to\tilde\varrho_P$, and, together with the convergence of the resolvent channels, $\varrho_h\to\varrho$. The invariant state of $\Lind$ is thus the limit of the invariant state of $\Phi_h$, which confirms the consistency of the two models.

\medskip 
\noindent {\bf Acknowledgments.} The work of T.A. is supported by the QuanTEdu-France program (ANR-22-CMAS-0001), funded by France 2030 and by the European Union through NextGenerationEU. The work of A.J. is partially supported by the French National Research Agency (ANR) through the grant Dynacqus (ANR-24-CE40-5714-02). \medskip

\noindent {\bf AI Declaration.} This manuscript, including all mathematical proofs, was written entirely by the authors. Large Language Models were used for language editing, proofreading, reference finding and code assistance. The authors take full responsibility for the content of the paper.

\section{Appendix}

\subsubsection*{Proof of Lemma \ref{lem:LR}}
\label{app:LR}

Let $\xi_k=\langle p_m|P_k|p_n\rangle$ and $K_{\mathrm{src}}=\frac{1-\alpha}{\lambda_0-\beta}\,\xi$. We verify that the coefficients of $R$ are
\begin{equation}
K=\big(\Id-s(\lambda_0)\,T_P\,T_Q\big)^{-1}K_{\mathrm{src}},\qquad
C=\frac{1-\beta}{\lambda_0-\alpha}\,T_Q\,K ,
\label{eq:val}
\end{equation}
which are well defined by \eqref{eq:a} and the remark following it.

\smallskip
We first check $L$. Apply the adjoint \eqref{eq:phistar} to $L=|p_n\rangle\langle p_m|$:
\[
\Phi_0^*(L)=\alpha\,e^{-i(\theta_n-\theta_m)}\,L+(1-\beta)\sum_iQ_i\,\Tr\big(\pi_i|p_n\rangle\langle p_m|\big).
\]
Each $\pi_i$ commutes with $W_P$, so it leaves the eigenspaces of $W_P$ invariant. Since $\theta_n\neq\theta_m$, this gives $\langle p_m|\pi_i|p_n\rangle=0$, so that $\Tr(\pi_i|p_n\rangle\langle p_m|)=0$ and the reset term vanishes. Thus $\Phi_0^*(L)=\overline{\lambda_0}\,L$ and $L$ is a left eigenoperator.

\smallskip
Now we check $R$. Since $\pi_j$ commutes with $W_P$ and $\tau_l$ with $W_Q$, one has $\Wcal_P(\pi_j)=\pi_j$ and
$\Wcal_Q(\tau_l)=\tau_l$. Moreover, one has $\Tr(P_k|p_n\rangle\langle p_m|)=\xi_k$,
$\Tr(P_k\pi_j)=(T_P)_{kj}$ and $\Tr(Q_j\tau_l)=(T_Q)_{jl}$, while $\Tr(Q_j|p_n\rangle\langle p_m|)$,
$\Tr(Q_j\pi_{j'})$ and $\Tr(P_k\tau_l)$ vanish. Hence \eqref{eq:phi} gives
\begin{align*}
\Phi_0(R)=\lambda_0|p_n\rangle\langle p_m|&+\sum_j\Big[\alpha C_j+(1-\beta)(T_QK)_j\Big]\pi_j%
+\sum_k\Big[\beta K_k+(1-\alpha)\big(\xi+T_PC\big)_k\Big]\tau_k .
\end{align*}
By \eqref{eq:val}, $(1-\beta)T_QK=(\lambda_0-\alpha)C$, so the first bracket is $\lambda_0C_j$. Also by \eqref{eq:val},
$K=K_{\mathrm{src}}+s(\lambda_0)\,T_PT_QK$. Multiplying by $\lambda_0-\beta$ and using the expression of $C$ gives
\[
(\lambda_0-\beta)K=(1-\alpha)\xi+\frac{(1-\alpha)(1-\beta)}{\lambda_0-\alpha}\,T_PT_QK=(1-\alpha)\big(\xi+T_PC\big),
\]
so the second bracket is $\lambda_0K_k$. Hence $\Phi_0(R)=\lambda_0R$.
\qed

\subsubsection*{Proof of Lemma \ref{lem:affdim}}
\label{app:affdim}
Since $\Vcal$ is $*$-stable, every $X\in\Vcal$ writes uniquely as $X=A+iB$ with
$A=\frac{X+X^*}2$ and $B=\frac{X-X^*}{2i}$ in $\Vcal_h$. Hence $(A,B)\mapsto A+iB$ is an $\R$-linear isomorphism
from $\Vcal_h\times\Vcal_h$ onto $\Vcal$, and $2\dim_\R\Vcal_h=\dim_\R\Vcal=2m$. Thus $\dim_\R\Vcal_h=m$.
 
Let $\varrho_0=\Id_{\HH}/\dim\HH$. The trace is a real linear form on $\Vcal_h$,
nonzero since $\Tr\Id_{\HH}=\dim\HH$. Hence $\Vcal_h^0=\{X\in\Vcal_h:\Tr X=0\}$ has real dimension
$m-1$, and $\mathcal T=\varrho_0+\Vcal_h^0=\{X\in\Vcal_h:\Tr X=1\}$ is an affine hyperplane of $\Vcal_h$ of real dimension
$m-1$. Since states are self-adjoint, $\mathcal S(\Vcal)\subseteq \mathcal T$, and since $\mathcal T$ is affine,
$\operatorname{aff}\mathcal S(\Vcal)\subseteq \mathcal T$, where $\operatorname{aff}$ denotes the affine hull.
 
Conversely, $\varrho_0\in\mathcal S(\Vcal)$, so $\operatorname{aff}\mathcal S(\Vcal)=\varrho_0+\mathcal N$, where
$\mathcal N=\{Y-Z:\ Y,Z\in\operatorname{aff}\mathcal S(\Vcal)\}$ is the direction of the affine hull. Let
$X\in\Vcal_h^0\setminus\{0\}$ and $0<|t|\leq1/(\dim\HH\,\|X\|)$. Then $\varrho_0+tX\in\Vcal$ is self-adjoint with
trace $1$, and for every $x\in\HH$, by Cauchy--Schwarz,
\[
\langle x,(\varrho_0+tX)x\rangle\ge\frac{\|x\|^2}{\dim\HH}-|t|\,\|X\|\,\|x\|^2\geq0 .
\]
So $\varrho_0+tX\in\mathcal S(\Vcal)$. Since $\varrho_0$ and $\varrho_0+tX$ both lie in $\mathcal S(\Vcal)$, their
difference $tX$ lies in $\mathcal N$, hence $X\in\mathcal N$ as $t\neq0$. Therefore $\Vcal_h^0\subseteq\mathcal N$, and
\[
\mathcal T=\varrho_0+\Vcal_h^0\subseteq\varrho_0+\mathcal N=\operatorname{aff}\mathcal S(\Vcal).
\]
Together with $\operatorname{aff}\mathcal S(\Vcal)\subseteq \mathcal T$, this gives $\operatorname{aff}\mathcal S(\Vcal)=\mathcal T$,
of real dimension $m-1$.
\qed

\begin{spacing}{.8}

\end{spacing}

\end{document}